hep-ph/9505366   23 May 95

# PRECISION TESTS
# OF THE STANDARD MODEL:
# EVIDENCE FOR RADIATIVE
# CORRECTIONS AND
# HIGHER ORDER EFFECTS

**Paolo Gambino**

May 1995

PH.D. DISSERTATION
Physics Department
New York University
4 Washington Place, New York
NY 10003, USA[1]

Principal Advisor: Alberto Sirlin

[1] E-mail: gambino@acf2.nyu.edu. Address after September 1, 1995: Max Planck Institut für Physik, Werner Heisenberg Institut, Föhringer Ring 6, D-80805 München, Germany.

# Abstract

Recent developments in the field of high precision calculations in the Standard Model are illustrated, with particular emphasis on the evidence for radiative corrections and on the estimate of the theoretical error in perturbative calculations. The most important high energy observables and the basic features of the renormalization program are discussed in the on-shell and $\overline{\text{MS}}$ frameworks, and a calculation of the relation between the electroweak mixing angle $\sin^2 \hat\theta_w(M_z)$ and the effective coupling $\sin^2 \theta_{eff}^{lept}$ measured at LEP and SLD is considered. I present strong indirect evidence for the contribution of bosonic electroweak corrections in the Standard Model, and argue that at the present level of experimental accuracy the full one-loop electroweak corrections are needed to describe the data. A complete calculation of $O(\alpha\alpha_s)$ effects in bosonic vacuum polarization functions is discussed in the most general case of arbitrary quark masses and momentum transfer. Compact exact formulas are derived for the self-energies of electroweak vector bosons and Higgs scalars, including the case of some extensions of the Standard Model; the connection between the calculations based on dimensional regularization and on dispersion relations is also investigated. I illustrate some applications to electroweak observables, giving an estimate of the error due to the truncation of the perturbative series, which appears to be well under control. The first complete calculation of the $O(G_\mu^2 m_t^2 M_w^2)$ contributions to the $\rho$ parameter in the Standard Model is then described for the case of neutrino-lepton scattering, and the result used to estimate the theoretical error in the prediction of $M_w$ and $\sin^2 \theta_{eff}^{lept}$. In view of the experimental precision expected at future colliders, this suggests that a complete two-loop calculation of the electroweak corrections is needed. Finally, in the appendices I list useful two-loop integrals, asymptotic formulas for the QCD corrections to two-point functions, and Ward identities connecting electroweak self-energies.

# Acknowledgements

It is a pleasure to thank Alberto Sirlin for his patience, his support, and for innumerable and instructive conversations. I benefited immensely also from the collaboration with my friends Giuseppe Degrassi and Abdel Djouadi. A large part of the work described here has been done with them. I am grateful to all my collaborators, Mauro Anselmino, Sergio Fanchiotti, Ferruccio Feruglio, Jan Kalinowski, and Alessandro Vicini, for sharing with me their insight, the excitement of research, and their friendship, and to Enrico Predazzi for encouragement. I am happy to acknowledge significant technical help from Sergio Fanchiotti, and Caterina and Ruben Levi. Joannis Papavassiliou kindly helped me with some of the figures. The financial support of the Physics Department (Meyer Fellowship) and of the Graduate School of Arts and Science (James Arthur Fellowship) of New York University is gratefully acknowledged.

# Contents

# Chapter 1

# Introduction.

High energy experiments have reached in the last few years an impressive accuracy, which has called for an analogous precision in the predictions provided by theorists. The mass of the neutral gauge boson, for example, is now measured within few parts in $10^5$, and the electroweak mixing angle at the Z mass scale is known at the level of two parts in a thousand, with the prospect to reach a relative uncertainty of $5 \times 10^{-4}$ in the very near future.

As in the electroweak sector of the Standard Model (SM) the coupling constants are small, perturbation theory is generally valid: perturbative calculations beyond the lowest order in quantum field theory involve a considerable amount of conceptual and calculational complexity, but at present they are the most valuable tool to test the existing theories and to explore possible alternatives, in particular in a future of increasing precision.

The SM has indeed been very successful in describing a large amount of data over a wide range of energies, from zero momentum transfer up to the Z mass scale. There are a few discrepancies, that need to be carefully considered, but they are at the level of 2-3$\sigma$, and certainly not compelling. At present there are no glaring contradictions with SM predictions. On the other hand, there are strong theoretical arguments indicating that there must be physics beyond the SM; the current situation suggests that we need to investigate the very fine details of the SM structure before being able to open a window on the physics beyond. Precision calculations are therefore of crucial importance: they serve the twofold purpose of verifying the SM as a full-fledged Quantum Field Theory, uncovering the evidence for quantum corrections, and of allowing the detection of possible deviations from its predictions.

In this framework, the recent reports of evidence for a top quark by the CDF and D∅ Collaborations at Fermilab [1,2] hit remarkably on the mark: the average of the values for the top quark mass given by CDF and D∅ ($m_t = 180 \pm 12$ GeV) agrees surprisingly well with a recent global fit of experimental data ($m_t = 178 \pm 11^{+18}_{-19}$ GeV), obtained including all known radiative corrections of the SM [3]; these are often sensitive functions of the top mass, and the comparison of different observables allows us to set limits on this unknown parameter. As the precision in the measurement of $m_t$ improves, the focus is inevitably going to shift to the other standing enigma of the SM: the Higgs boson and the symmetry breaking mechanism, and the challenge to experimentalists and theorists would become more difficult, as the dependence of most observables on this sector of the theory is much weaker.

In this dissertation I will try to illustrate some of the recent developments in the field of high precision calculations in the SM of the electroweak and strong interactions. In particular, I will address two complementary questions: what is the present experimental evidence for quantum corrections beyond the leading fermionic contributions; what is the possible role of higher order (beyond one-loop) contributions to electroweak observables. Specifically, I will consider the cases of $M_W$ and $\sin^2\theta_{eff}^{lept}$, that appear to be most promising, among high-energy observables, for a significant improvement of the experimental accuracy.

In Ch. 2 I introduce to two of the renormalization procedures most frequently employed in precision physics, and present a detailed calculation of the effective mixing angle measured on the $Z$-resonance by experiments at LEP and SLC. In Ch. 3 I will discuss the phenomenological evidence for radiative corrections in the SM. A simple example will illustrate that there is very strong indirect evidence from high-energy experiments for the bosonic corrections of the theory. Ch. 4 deals with the universal QCD corrections, i.e. the perturbative QCD corrections to vector boson self-energies that enter the determination of most precision observables. I will present a complete calculation of these effects in bosonic vacuum polarization functions at first order in $\alpha_s$, for arbitrary values of the momentum transfer and of the quark masses involved, and discuss some of the Ward identities that connect them. In Ch. 5 I will consider potentially large two-loop contribution of electroweak origin, focussing on a calculation of the leading and next-to-leading contribution to the $\rho$ parameter in a heavy top expansion. The result is used to estimate the theoretical error in the determination of $M_W$ and $\sin^2\theta_{eff}^{lept}$. Finally, Ch. 6 contains my conclusions.

# Chapter 2

# Basic Observables and Renormalization

In this chapter I will briefly introduce the basic features of the renormalization of the Standard Model (SM), the most important precision observables, and two renormalization procedures most frequently employed in the analysis of high-energy physics, the on-shell and $\overline{\text{MS}}$ schemes[1]. Finally, in Sec. 2.4, I will present a detailed calculation of the effective mixing angle measured at LEP and SLC in the $\overline{\text{MS}}$ scheme.

## 2.1 Input parameters and tests of the theory

### 2.1.1 Input parameters

The electroweak lagrangian is characterized by a number of free parameters: the fermion masses, the Higgs boson mass $M_H$, the mixing angles and phase of the Kobayashi-Maskawa quark matrix[2], and three independent parameters that describe the gauge sector of the SM. Two possible choices for the latter are

$$g, \quad g', \quad v, \qquad \text{or} \qquad g, \ M_W, \ M_Z, \tag{2.1}$$

where $g$ and $g'$ are the $SU(2)_L$ and $U(1)_Y$ couplings, $v$ the vacuum expectation value of the Higgs field, and $M_W$ and $M_Z$ the vector boson masses. As the theory is renormalizable, this minimal set of parameters can in principle describe the physics at any order in perturbation theory. For what concerns the gauge sector of the SM, we therefore need only three input quantities, to be chosen among the most precise experimental observables; any additional measurement will test the theory. As can be seen from Table 2.1, based on Refs. [1–3, 7], at present the four most precise high-energy observables are the masses of the $Z$ and $W$ vector bosons, the total decay width of the $Z$ boson, and the leptonic asymmetries. In addition to these high-energy quantities, and to the masses of the fermions except the top, two other electroweak observables are measured at low-energy with great accuracy: the fine structure constant $\alpha$, which is most precisely determined from the anomalous magnetic moment of the electron, and the Fermi constant $G_\mu$, which is extracted from the muon lifetime. In Table 2.2 I show this set of six observables with their experimental errors.

---

[1] Among the many reviews on the SM, the procedures to renormalize it, and on precision tests, one can refer for instance to Refs. [4–6].

[2] In the event that neutrinos are massive, additional parameters have to be considered, masses and mixing angles, but neutrino masses are not expected to contribute significantly to precision physics, and I will disregard them in the following.

| observable | experimental value | SM fit Ref. [3] |
|---|---|---|
| $M_Z(\text{GeV})$ | 91.1888(44) | 91.1887 |
| $\Gamma_Z^{\text{tot}}(\text{GeV})$ | 2.4974(38) | 2.4973 |
| $M_W(\text{GeV})$ | 80.23(18) | $80.32 \pm 0.06 \pm 0.01$ |
| $R_h$ | 20.795(40) | 20.786 |
| $A_{FB}^{0,\ell}$ | 0.0170(16) | 0.0153 |
| $\mathcal{A}_\tau$ | 0.143(10) | 0.143 |
| $\mathcal{A}_e$ | 0.135(11) | 0.143 |
| $R_b$ | 0.2202(20) | 0.2158 |
| $R_c$ | 0.1583(98) | 0.172 |
| $A_{FB}^{0,b}$ | 0.0967(38) | 0.1002 |
| $A_{FB}^{0,c}$ | 0.0760(91) | 0.0714 |
| $1 - M_W^2/M_Z^2(\nu N)$ | 0.2253(47) | 0.2242 |
| $\sin^2 \theta_{eff}^{lept}(A_{LR})$ | 0.2294(10) | $0.2320 \pm 0.0003^{+0.0000}_{-0.0002}$ |
| $m_t(\text{GeV})$ | 180(12) | $178 \pm 11^{+18}_{-19}$ |
| $M_H(\text{GeV})$ | $\gtrsim 65.1$ | $\lesssim 730$ |
| $\alpha_s$ | 0.118(6) | $0.125 \pm 0.005 \pm 0.002$ |

Table 2.1: High-energy observables compared to their SM expectations. The SM predictions are based on a fit including all LEP, SLD, $\nu N$, and $p\bar{p}$ collider data, but not the new $m_t$ measurements; the first error and the central value refer to $M_H = 300\,\text{GeV}$, and the second error corresponds to the variation of the central value when $M_H$ varies between 60 GeV and 1 TeV.

The most obvious choice in the present situation is to use

$$\alpha \qquad G_\mu \qquad M_Z \qquad (2.2)$$

together with the fermion masses and mixing parameters as inputs to derive precise predictions for the other observables $\sin^2 \theta_{eff}^{lept}$, $M_W$, $\Gamma_Z$, and in principle for any electroweak observable. I will try to estimate the theoretical errors involved in this process, including the effects of the uncertainties of the inputs, which can be however expected to be small. In addition to these specifically electroweak input parameters, I will use the value

$$\alpha_s(M_Z) \equiv \alpha_s(M_Z)^{(5)}_{\overline{\text{MS}}} = 0.118 \pm 0.006 \qquad (2.3)$$

for the strong interaction coupling constant, corresponding to the 1994 world average [8] for the $\overline{\text{MS}}$ definition, with five active quark flavors.

| observable | experimental value [Ref.] | uncertainty |
|---|---|---|
| $\alpha^{-1}$ | 137.0359895(61) [14] | $4.5 \times 10^{-9}$ |
| $G_\mu$ | $1.16639(1) \times 10^{-5} \text{GeV}^{-2}$ [14] | $8.5 \times 10^{-6}$ |
| $M_z$ | 91.1888(44) GeV [3] | $4.8 \times 10^{-5}$ |
| $\sin^2 \theta_{eff}^{lept}$ | 0.2316(4) [3] | $2 \times 10^{-3}$ |
| $M_W$ | 80.23(18)GeV [15] | $2 \times 10^{-3}$ |
| $\Gamma_z^{tot}$ | 2.4974(38)Gev [3] | $1.5 \times 10^{-3}$ |

Table 2.2: The six most precise electroweak observables. The last column shows the relative uncertainties. The values for $M_z$ and $\Gamma_z$ are the 1994 LEP average, $\sin^2 \theta_{eff}^{lept}$ is given as the average of LEP and SLC asymmetry values, and $M_W$ is the 1994 collider world average. It does not include the value $M_W = 80.24(25)\text{GeV}$ from $s^2$ measurements in $\nu N$ low-energy scattering.

A number of natural relations link the couplings and the masses of the gauge bosons occurring in the bare SM Lagrangian. Most notably, we see that, although initially defined in terms of the gauge couplings, after spontaneous symmetry breaking the electroweak mixing angle $\theta_W^0$ relates both bare masses and couplings among themselves:

$$\tan \theta_W^0 = \frac{g_0'}{g_0}, \quad e_0 = g_0 \sin \theta_W^0, \quad M_W^0 = \cos \theta_W^0 M_z^0. \tag{2.4}$$

At the tree level, using $\sqrt{2}G_\mu = g^2/(4M_W^2)$ and $\alpha = e^2/(4\pi)$, the above relations can be re-written in terms of the input parameters (2.2)

$$\sin^2 \theta_W \cos^2 \theta_W = \frac{\pi\alpha}{\sqrt{2}G_\mu M_z^2}, \tag{2.5a}$$

$$M_W^2 = \cos^2 \theta_W M_z^2. \tag{2.5b}$$

Radiative corrections modify the natural relations of the bare Lagrangian, and it is convenient to summarize their effect in a compact way [9–11]

$$\frac{\pi\alpha}{\sqrt{2}G_\mu M_z^2} = s^2 \ c^2 \left(1 - \Delta r\right) = \hat{s}^2 \hat{c}^2 \left(1 - \Delta \hat{r}\right)$$
$$M_W^2 = c^2 \ M_z^2 = \hat{c}^2 \ \hat{\rho} \ M_z^2. \tag{2.6}$$

The two different possibilities for the right-hand side refer to the fact that, when quantum corrections are incorporated, the relations of Eqs.(2.5) cannot hold true simultaneously, and a precise definition of the parameters involved is needed. The connection between the renormalized parameters of the Lagrangian and the experimental inputs is precisely the content of a renormalization scheme. It seems reasonable to define the masses as the physical masses of the gauge bosons, as they are the quantities that are presently being measured at experiments. In particular, one considers the complex-valued position of the pole of the propagator, $\bar{s} = m_0^2 + A(\bar{s})$, where $m_0$ is the bare mass and $A(s)$ the vector-boson self-energy. This is certainly a gauge invariant quantity, as it is based on the fundamental properties of the S-matrix. Writing $\bar{s} = m_2^2 - im_2\Gamma_2$, two frequently used definitions are $m_{phys}^2 = \text{Re}\,\bar{s} = m_2^2$ and $m_{phys}^2 = m_1^2 \equiv m_2^2 + \Gamma_2^2$. It turns out that, in a large class of renormalizable gauges, $m_1^2$ differs from the on-shell definition $m^2 = m_0^2 + \text{Re}\,A(m^2)$ by gauge-dependent terms of $\mathcal{O}(\alpha^3)$ which lie beyond current and expected accuracies (in the complementary class of gauges, a gauge-dependent difference arises in $\mathcal{O}(\alpha^2)$ but it is bounded and very small). Furthermore, $m_1$ can be identified with the mass parameter measured at LEP [16], and can therefore be used in precision calculations. A similar pattern is expected for the mass of the $W$ boson.

Concerning the mixing angle, different choices are possible. Unlike the vector boson masses, the definition of a renormalized $\sin^2\theta_W$ is purely conventional, and one can choose either of Eq.(2.5) to relate a renormalized mixing angle to observables[3], or a different path, using the $\overline{\text{MS}}$ (Modified Minimal Subtraction) scheme [21]. All definitions differ by finite calculable radiative corrections.

In Eq.(2.6) $s$, $c$, and $\hat{s}$, $\hat{c}$, are therefore abbreviations for, respectively, the "on-shell" $\sin\theta_W$, $\cos\theta_W$, and the $\overline{\text{MS}}$ quantities $\sin\hat{\theta}_W$, $\cos\hat{\theta}_W$. The first mixing angle, $\theta_W$, is defined by the requirement that the third relation of Eq.(2.4) be valid at any order in perturbation theory [9],

$$\sin^2\theta_W \equiv 1 - \frac{M_W^2}{M_Z^2}. \qquad (2.7)$$

while the second one, $\hat{\theta}_W$, is defined by [20]

$$\sin^2\hat{\theta}_W(\mu)_{\overline{\text{MS}}} \equiv \frac{\hat{\alpha}(\mu)_{\overline{\text{MS}}}}{\hat{\alpha}_2(\mu)_{\overline{\text{MS}}}}. \qquad (2.8)$$

Here $\hat{\alpha}(\mu)_{\overline{\text{MS}}}$ and $\hat{\alpha}_2(\mu)_{\overline{\text{MS}}}$ are the electromagnetic and $SU(2)_L$ couplings defined at a mass scale $\mu$ in the $\overline{\text{MS}}$ scheme, a renormalization procedure that I will introduce in the next sections. The $\overline{\text{MS}}$ angle $\hat{\theta}_W(\mu)$ is therefore a scale dependent quantity; for $Z$-physics the natural choice is to evaluate it at $\mu = M_Z$.

## 2.1.2 The electromagnetic coupling

Although the fine structure constant has practically no experimental uncertainty, the use of $\alpha$ to describe physics on the $Z$-resonance is somewhat anomalous: among the three input parameters of (2.2), $\alpha$, defined at $q^2 = 0$, is the only one that involves long-distance dynamics. This is witnessed by the appearance of mass singularities in the conventional renormalization of the electromagnetic coupling [9]. In fact, physics at the $Z$-scale is naturally described by short-distance parameters defined at the $M_Z$ scale, like $\alpha(M_Z)$, the conventional electromagnetic running coupling evaluated at $M_Z^2$

$$\alpha(M_Z^2) = \frac{\alpha}{1 + \text{Re}\Pi_{\gamma\gamma}^{(f)}(M_Z^2) - \Pi_{\gamma\gamma}^{(f)}(0)} = \frac{\alpha}{1 - \Delta\alpha}, \qquad (2.9)$$

---

[3]The first solution (or improvements thereof) is used in the so-called $G_\mu$ scheme [17–19], and the second in the on-shell scheme, that I will discuss in detail.

where $\Pi_{\gamma\gamma}^{(f)}(q^2)$ is the fermionic photon two-point function[4], defined at squared transfer momentum $q^2$. Numerically, $\alpha(M_Z^2) \approx 1/128$. Because of the importance of this input parameter, the precise knowledge of the denominator of Eq.(2.9) is crucial. Note that any other definition of $\alpha$ at the $M_Z$ scale would be appropriate; another possibility is the $\overline{\text{MS}}$ definition that I will discuss in the next section. Unfortunately, the evaluation of $\text{Re}\Pi_{\gamma\gamma}(M_Z^2) - \Pi_{\gamma\gamma}(0)$ involves long-distance QCD dynamics in the light quark contribution for which perturbation theory cannot be used. The standard procedure [9, 12, 13] is to use experimental input from $e^+e^- \to$ hadrons and numerically integrate the spectral function obtained in this way (see Sec. 4.5). This method introduces a substantial uncertainty in the determination of $\alpha(M_Z^2)$ stemming from low-energy data, a fact that has important consequences in the analysis of the precision electroweak data. In the following, I will use a very recent state-of-the-art determination [13] for the five light quark contribution to the running of $\alpha$, which gives $\Pi_{\gamma\gamma}^{(5)}(0) - \text{Re}\Pi_{\gamma\gamma}^{(5)}(M_Z^2) = 0.0280 \pm 0.0007$[5]. The leptonic contribution to the denominator of Eq.(2.9), $\Delta\alpha_l$, keeping all leptonic masses [14] in the one-loop QED expressions, is $\Delta\alpha_l = 0.031419$. This implies $\Delta\alpha = 0.0594 \pm 0.0007$, and $\alpha^{-1}(M_Z^2) = 128.90 \pm 0.10$. The uncertainty is quite substantial, $\approx 7 \times 10^{-4}$, and affects any determination of the e.m. coupling at a high mass scale based on the fine structure constant (i.e. it is scheme independent). In fact, we will see that it is a real bottleneck to achieve greater theoretical accuracy. The increasing experimental precision indeed requires new accurate measurements of the cross-section $e^+e^- \to$ hadrons at low energies. The problem could, in principle, be circumvented by using high-energy experiments to determine $\alpha(M_Z^2)$, but this entails a severe loss of information and is quite premature at present.

### 2.1.3   $M_W$ and $\sin^2\theta_{eff}^{lept}$

Once the input parameters are specified, as we have done in the preceding sections, predictions for SM observables can be obtained and serve as tests of the theory. In the following, I will focus on two pseudo-observables[6]: the mass of the $W$ boson and $\sin^2\theta_{eff}^{lept}$. They are among the most precise quantities measured in high-energy experiments, and their experimental precision is expected to improve significantly in the next decade, as can be seen in Table 2.3 (see Ref. [24]; the estimate of the error on $\sin^2 \theta_{eff}^{lept}$ at LHC is taken from Ref. [25]). They are less affected by QCD corrections than $\Gamma_Z^{\mathrm{tot}}$; the QCD corrections to $M_W$ and $\sin^2 \theta_{eff}^{lept}$ will be studied in Ch. 4.

---

[4]The convention adopted for the vacuum polarization functions throughout this work is the following: $A_{ab}(q^2)$ is the transverse component of the vacuum polarization tensor for the transition between two vector bosons $a$ and $b$, defined as $-i$ times the standard Feynman amplitude. For the photon, $A_{\gamma\gamma}(q^2) = -q^2\Pi_{\gamma\gamma}(q^2)$, and in general $\Pi_T^{ab}(q^2) = -A_{ab}(q^2)$, in the notation introduced in Ch. 4.

[5]Recently, two other re-evaluations of the hadronic contribution to the running of alpha have appeared [22, 23], and differ from Ref. [13] by 1-1.5$\sigma$. However, one of them [23] relies on three-loop perturbative QCD up to very low energies, instead of fully exploiting the experimental data, and it is not yet clear whether the procedure of Ref. [22] is completely consistent. The discrepancy is potentially significant, but its clarification lies beyond the scope of this work.

[6]$M_W$ and $\sin^2\theta_{eff}^{lept}$, as well as a number of other high-energy quantities, are sometimes called *pseudo-observables*, in contrast to *real* observables such as cross-sections and asymmetries. The former are related to the latter by some well-defined set of specific assumptions that constitutes a sort of deconvolution procedure and involves a certain amount of theoretical input. The distinction is not unimportant, as this procedure introduces a large theoretical error, mainly connected to the initial state QED corrections, that is generally taken into account by the experimental groups.

| observable | present uncertainty [exp.] | $\approx$ year 2000 | far future |
|---|---|---|---|
| $\delta M_W(\mathrm{MeV})$ | $180[\mathrm{w.a.}]$ | $50 \pm 20[\mathrm{Tevatron}]$ | $20[\mathrm{NLC}]$ |
| | $230[\mathrm{CDF}]$ | $40[\mathrm{LEP200}]$ | |
| $\delta \sin^2 \theta_{eff}^{lept}$ | $4 \times 10^{-4}[\mathrm{SLD + LEP}]$ | $1 \times 10^{-4}[\mathrm{SLD + LEP}]$ | $6 \times 10^{-5}$ |
| | $10 \times 10^{-4}[\mathrm{SLD}]$ | $2.5 \times 10^{-4}[\mathrm{SLD}]$ | $[\mathrm{LHC}]$ |
| $\delta m_t(\mathrm{GeV})$ | $12[\mathrm{CDF + D\emptyset}]$ | $4[\mathrm{Tevatron}]$ | $2[\mathrm{LHC}]$ |

Table 2.3: Precision goals at future colliders. NLC stands for the Next Linear Collider, a 500GeV $e^+e^-$ collider under study. "Tevatron" implies extensive upgrades of the collider and detectors, and the start of Main Injector.

The effective weak interaction angle employed by the LEP groups is defined by

$$1 - 4\sin^2 \theta_{eff}^{lept} \;=\; \mathrm{Re} \, \frac{g_V^\ell}{g_A^\ell}, \tag{2.10}$$

where $g_V^\ell$ and $g_A^\ell$ are the effective vector and axial couplings in the $Z^0 \to \ell\bar{\ell}$ amplitude at resonance (with the exclusion of photon exchange effects), and $\ell$ denotes a charged lepton [3,26]. $\sin^2 \theta_{eff}^{lept}$ is extracted from the on-resonance left-right and forward-backward asymmetries. In the determination of the effective angle, it is common practice to use the $b$ and $c$ quark asymmetry data together with the leptonic ones, as this procedure is practically model independent. In order to establish the connection with the $\overline{\mathrm{MS}}$ parameter $\sin^2 \hat{\theta}_W(M_Z)$, or with $s$, the on-shell sine, we note that this amplitude is proportional to [27]

$$< \ell\bar{\ell}|J_Z^\lambda|0 > \;=\; -\bar{u}_\ell \gamma^\lambda \left[ \frac{1 - \gamma_5}{4} - \hat{k}_\ell(q^2) \, \hat{s}^2 \right] v_\ell, \tag{2.11}$$

where the electroweak form factor $\hat{k}(q^2)$ and $\hat{s}^2$ can be replaced by $k(q^2)$ and $s^2$ in the on-shell scheme, and $v_\ell$ and $\bar{u}_\ell$ are the lepton spinors[7]. The actual radiative corrections that enter into the determination of these two pseudo-observables therefore depend on the renormalization scheme. In the following I will present the case of the on-shell and $\overline{\mathrm{MS}}$ schemes.

## 2.2   On-shell scheme

In the on-shell scheme [9, 28], in analogy with the traditional renormalization of QED, the fundamental parameters are chosen to be $\alpha$ and the physical masses of all the particles of the SM and, in particular for the gauge sector, $M_Z$ and $M_W$. From the experimental values of the vector boson masses, we find $(\delta s^2/s^2)_{\rm exp} \approx -2(c^2/s^2)(\delta M_W/M_W)_{\rm exp} \approx 1.6\%$, a very poor precision for a fundamental coupling. As the experimental precision on $M_W$ is not very good at present, and even after LEP200 will not be better than the one on $G_\mu$, one has to resort to the relation, Eq.(2.6),

$$s^2 c^2 = \frac{M_W^2}{M_Z^2}\left(1 - \frac{M_W^2}{M_Z^2}\right) = \frac{\pi\alpha}{\sqrt{2}G_\mu M_Z^2[1 - \Delta r(M_W, m_t, M_H)]}, \qquad (2.12)$$

where $\Delta r(M_W, m_t, M_H)$ is a finite radiative correction that includes the term $\Delta\alpha$ defined above and the non-QED electroweak corrections to the muon decay [9][8]. The inclusion of $(1 - \Delta r)$ in the denominator of Eq.(2.12) effectively resums all leading logarithms of $\mathcal{O}(\alpha^n \log^n \frac{m_t}{M_Z})$ (in this way replacing $\alpha$ with $\alpha(M_Z^2)$), and to good approximation the terms of $\mathcal{O}(\alpha^2 \log \frac{m_t}{M_Z})$ [29][9]. Eq.(2.12) can be solved iteratively for $M_W$ and yields a prediction for $M_W$ or $\sin^2\theta_W$ for any given value of $m_t$ and $M_H$. The values for $M_W$ and $s^2$ derived in this way are then used in the calculation of any other observable in the on-shell scheme. It is interesting to show the asymptotic behavior of the function $\Delta r(m_t, M_H)$, as the comparison between the $M_W(m_t, M_H)$ obtained from Eq.(2.12) and $M_W^{\rm exp}$ allows us to set limits on the possible range of $m_t$ and $M_H$. The leading asymptotic behaviors for large $m_t$ and $M_H$ are

$$\Delta r \sim -\frac{3\alpha}{16\pi s^4}\frac{m_t^2}{M_Z^2} + \cdots \qquad \left(\frac{m_t}{M_Z} \gg 1\right)$$

$$\Delta r \sim \frac{11\alpha}{48\pi s^2}\ln\left(\frac{M_H^2}{M_Z^2}\right) + \cdots \qquad \left(\frac{M_H}{M_Z} \gg 1\right) \qquad (2.13)$$

---

[7]Alternatively, one could define effective couplings directly through the so-called *bare* asymmetries $A^0$, derived from the *experimental* asymmetries after subtraction of soft photon and photon exchange effects. For example, $A_{LR}^{0,\ell} = 2x/(1 + x^2)$, with $x = g_V^\ell/g_A^\ell$. The difference comes from the imaginary part of $\hat{k}(q^2)$, but we will see in Sec. 2.4 that the it is numerically very small.

Note that the dependence is quadratic in $m_t$ and logarithmic in $M_H$, and that the effects have opposite signs. This partially accounts for the fact that, in global analyses of the electroweak data, small $M_H$ values favor relatively small values of $m_t$; it also explains why it is much more difficult to set bounds on $M_H$ than on $m_t$. More generally, the whole set of precision observables can be calculated as a function of $m_t$ and $M_H$, and the result fitted to the existing data, yielding the bounds for $m_t$ and $M_H$ shown in Table 2.1.

In the on-shell framework it is natural to define the renormalized mixing angle according to Eq.(2.7). That definition is manifestly based on short-distance parameters, and it therefore absorbs large photon vacuum polarization effects that would be present adopting, for instance, the first of the relations in Eq.(2.5). The use of the on-shell mixing angle in Born amplitudes will not induce large loop effects of that kind. However, Eq.(2.7) does not absorb in $\sin^2\theta_W$ large isospin violation effects that occur in the on-resonance amplitudes, and are due to the mass splitting in the top-bottom isodoublet. These manifest themselves in the counterterm of

---

[8] As $M_Z$, $M_W$, $G_\mu$ and $\alpha$ are physical observables, it follows that the same is true of $\Delta r$.

[9] In the on-shell scheme the incorporation of leading higher order reducible contributions of $\mathcal{O}(G_\mu^2 m_t^4)$ is subtle. It can be implemented following the procedure outlined in Ref. [30].

$\sin^2 \theta_W$

$$\delta s^2 \;=\; -c^2\, Y, \qquad Y \;=\; \frac{\left(\frac{\delta M_W^2}{M_W^2} - \frac{\delta M_Z^2}{M_Z^2}\right)}{\left(1 - \frac{\delta M_Z^2}{M_Z^2}\right)}, \qquad (2.14)$$

where $\delta M_W^2 = \mathrm{Re}A_{WW}(M_W^2)$ and $\delta M_Z^2 = \mathrm{Re}A_{ZZ}(M_Z^2) + (\gamma, Z)$ mixing terms, are the mass counterterms of the vector bosons. At one-loop level we find $Y = x_t + \ldots$, where $x_t = 3\frac{G_\mu m_t^2}{8\pi^2\sqrt{2}}$ is the leading top correction to the $\rho$ parameter [71], and the ellipses stand for the rest of the one-loop contribution. This problem is frequently circumvented by introducing an effective parameter $\bar{s}^2 \equiv [1 + (c^2/s^2)x_t]s^2$ or, equivalently, $\bar{c}^2 = c^2(1 - x_t)$.

The determination of $M_W$ through Eq.(2.12) is affected by the parametric uncertainty on the inputs $G_\mu$, $\alpha$, and $M_Z$, and by the theoretical uncertainty on $\Delta r$, due to the truncation of the perturbative series. For instance, a change $\delta\Delta r$ induces a shift in the determination of $M_W$ through the following relation

$$(\delta M_W)_{\mathrm{th}} = -\frac{M_W}{2}\frac{s^2}{(c^2 - s^2 - 2c^2 x_t)}\frac{\delta\Delta r}{(1 - \Delta\alpha)}, \qquad (2.15)$$

where some higher order effects have been retained [36]. Using the result of Ref. [13], we see that the uncertainty on the hadronic contribution to the running of $\alpha$ implies $\delta M_W^{\mathrm{hadr.}} \approx$ 13MeV. This is the ultimate limit on the precision of a theoretical determination of $M_W$ in the present situation. In the next chapters, however, we will see that the theoretical error due to electroweak higher order effects in this determination can be of the same size. We can compare this uncertainty with the one coming from the experimental error for the input parameters. For instance, in the case of $M_Z$, $\delta M_Z \simeq 4.4$MeV. From Eq.(2.12) it is easy to see that this induces on $M_W$ an uncertainty $\delta M_W \approx 5.4$MeV, smaller than the one induced by $\delta\Delta\alpha^{(hadr)}$.

## 2.3 $\overline{\mathrm{MS}}$ scheme

The on-shell scheme has the privilege of simplicity, but as we have seen above, it induces potentially large higher order corrections in the on-resonance amplitudes. The on-shell couplings do not seem therefore adequate expansion parameters unless one employs the effective couplings $\alpha(M_Z^2)$ and $\bar{s}^2$. A very good alternative is provided by the $\overline{\mathrm{MS}}$ scheme, which is the prevalent framework for QCD studies.

The on-shell definition of the mixing angle is based on physical quantities; the renormalized parameter is set to be equal to an observable quantity, the ratio of the physical masses of the vector bosons. On the other hand, it is possible to define a renormalized parameter by choosing the counterterm associated to it. The $\overline{\mathrm{MS}}$ definition of a renormalized parameter [21] is based on a theoretical prescription for the associated counterterm and requires that the calculation be performed in one particular regularization framework, i.e. dimensional regularization. The idea is to subtract only the divergent part (poles in $n - 4$) and the associated constants $\gamma$ and $\ln 4\pi$ from the bare parameter[10], as these are fixed exclusively by the bare Lagrangian, unlike the finite parts of the counterterm, that depend on the renormalization condition. The result is a scale dependent quantity, and a first example is $\hat{\alpha}(\mu)$, the $\overline{MS}$ e.m. coupling constant, defined by

$$\hat{\alpha}(\mu)_{\overline{MS}} \equiv \frac{\alpha}{1 - \Delta_\gamma(\mu)} \tag{2.16}$$

where $\Delta_\gamma(\mu) = \Pi_{\gamma\gamma}^{(f)}(0)_{\overline{MS}} + \frac{\alpha}{\pi}(7/2\ln(M_W/\mu) - 1/6)$, with the subscript $\overline{MS}$ to indicate that $1/(n-4)$ poles have been subtracted with their related constants. In addition to the fermionic self-energies, $\Delta_\gamma$ contains the contribution of $W$ bosons. The hadronic part is evaluated adding and subtracting $\Pi_{\gamma\gamma}^{(5)}(M_Z^2)$, which can be calculated perturbatively, and using the value $\Pi_{\gamma\gamma}^{(5)}(0) - \mathrm{Re}\Pi_{\gamma\gamma}^{(5)}(M_Z^2) = 0.0280 \pm 0.0007$ [13]. For $M_Z$ and $M_W$ as given in Table 2.2, $m_t = 175\,\mathrm{GeV}$, and including two-loop QCD effects following the results of Ch. 4, we obtain $\Delta_\gamma(M_Z) = 0.0653 \pm 0.0007$ and $\hat{\alpha}^{-1}(M_Z) = 128.09 \pm 0.10$. However, $m_t > M_Z$, and according to the common practice in the $\overline{MS}$ scheme, we may subtract the top effects from Eq.(2.16), absorbing them in the definition of $\hat{\alpha}$, as it would be the case for another heavy unknown charged particle. Decoupling the top quark from the definition of $\hat{\alpha}$, we therefore have $\Delta_\gamma(M_Z) = 0.0666 \pm 0.0007$, and $\hat{\alpha}^{-1}(M_Z) = 127.91 \pm 0.10$, which differs from $\alpha(M_Z)$, the conventional running coupling constant of Sec. 2.1.2 by $0.73\%$, and has a similar uncertainty of $\approx 7 \times 10^{-4}$.

An analogous definition can be employed for the other couplings and masses. The bare relations Eq.(2.4) hold for all $\overline{MS}$ couplings and masses at a given scale $\mu$. In particular, we can relate the bare sine to the $\overline{MS}$ definition of Eq.(2.8) through (the ellipses indicate higher order contributions)

$$\sin^2\hat{\theta}_W(\mu)_{\overline{MS}} = \sin^2\theta_W^0 \left[1 + \frac{\alpha_0}{4\pi}\left(\frac{11}{3} + \frac{19}{6\sin^2\theta_W^0}\right)\frac{1}{\epsilon} + \ldots\right]. \tag{2.17}$$

Here I have subtracted only the poles and the related constants (not indicated), but one can think of implementing the decoupling of heavy particles as done above for $\hat{\alpha}$. However, in this case a subtlety arises, as $\alpha_2(\mu)_{\overline{MS}}$ can be defined by charged or neutral current couplings, and the top affects them differently. According to the Marciano-Rosner [M-R] convention [38], adopted also in Ref. [36], the logarithmic term is subtracted in the evaluation of $\mathrm{Re}\ A_{\gamma Z}^{(top)}(M_Z^2)/M_Z^2$, which enters the form factor $\hat{k}_\ell(M_Z^2)$ of Eq.(2.11) as

$$-\frac{\hat{c}}{\hat{s}}\frac{\mathrm{Re}\ A_{\gamma Z}^{(top)}(M_Z^2)}{M_Z^2} = -\frac{\hat{\alpha}}{6\pi\hat{s}^2}\left(1 - \frac{8}{3}\hat{s}^2\right)\left[\left(1 + \frac{\alpha_s}{\pi}\right)\log\xi_t - \frac{15}{4}\frac{\alpha_s}{\pi}\right] + D(\frac{1}{\xi_t}). \tag{2.18}$$

Here $\xi_t \equiv m_t^2/M_Z^2$ and $D(1/\xi_t)$ represents small terms that decouple in the limit $\xi_t \to \infty$. We see that at two-loop there is also an $m_t$-independent term that should be subtracted as well. This additional finite counterterm keeps the $\gamma - Z$ mixing independent of $m_t$ in neutral current amplitudes and allows $\sin^2 \hat{\theta}_W(\mu)_{\overline{MS}}$ to be continuous at $\mu = m_t$, up to small $\mathcal{O}(\hat{\alpha}\alpha_s)$ terms. The aim of the prescription is to make the value of $\sin^2 \hat{\theta}_W(M_Z)$, as extracted from the on-resonance asymmetries, very insensitive to heavy particles of mass $M > M_Z$. As can be seen from Eq.(2.18), the numerical difference between $\sin^2 \hat{\theta}_W(M_Z)$ evaluated with or without the M-R decoupling convention is very small, but not negligible, $\approx 0.0002$ in the present $m_t$ range.

---

[10]This is consistently done at any order in perturbation theory by rescaling the 't Hooft mass scale $\mu \to \mu'e^{\gamma/2}(4\pi)^{-1/2}$.

The use of $\overline{MS}$ couplings would suggest the use of $\overline{MS}$ masses $\hat{M}_Z(\mu)$ and $\hat{M}_W(\mu)$ for the vector bosons[11]. However, as we have seen, the experiments provide the values of the physical or pole masses of the vector bosons, and the relation between $M$ and $\hat{M}(\mu)$ involves large radiative corrections of $\mathcal{O}(G_\mu m_t^2 \ln m_t/\mu)$. At high orders in perturbation theory, it can be seen that a whole class of potentially large $\mathcal{O}(G_\mu^n m_t^{2n-2} M_W^2)$ contributions, corresponding to fermionic loop insertions on vector boson propagators, can be absorbed in the definition of the on-shell masses. The on-shell mass counterterms cancel completely the leading quadratic dependence of the heavy fermion loop. Thus, the use of $\overline{MS}$ couplings together with on-shell masses in the propagators seems preferable in order to keep higher order radiative corrections small [11]. An explicit example of this kind will be presented in Ch.5, in the case of the two-loop $\mathcal{O}(G_\mu^2 m_t^2 M_Z^2)$ contribution to the $\rho$ parameter, which turns out to be smaller when employing this prescription.

If we choose this route, the auxiliary parameter $\sin^2 \hat{\theta}_W(M_Z)$ is derived from the inputs setting the mass scale equal to $M_Z$, and using

$$\hat{s}^2 \hat{c}^2 = \frac{\pi \alpha}{\sqrt{2} G_\mu M_Z^2 [1 - \Delta \hat{r}(M_W, m_t, M_H, \hat{s})]}, \tag{2.19a}$$

and $M_W$ is determined through Eq.(2.6)

$$M_W^2 = \hat{\rho}(M_W, m_t, M_H, \hat{s}) \; \hat{c}^2 \; M_Z^2. \tag{2.19b}$$

Eqs.(2.19) can be solved simultaneously by iteration. The radiative correction $\hat{\rho}(m_t, M_H, M_W, \hat{s})$, which now governs the $M_W - M_Z$ interdependence, incorporates the custodial symmetry violation effect, quadratic in $m_t$. $\hat{\rho}$ can be readily obtained from the difference between the two definition of $\sin^2 \theta_W$. Using $s_0^2 = s^2 - \delta s^2 = \hat{s}^2 - \delta \hat{s}^2$, we have $\hat{c}^2 = c^2 (1 - Y_{\overline{MS}})$, where $\overline{MS}$ indicates that ultraviolet poles and related constants have been subtracted from $Y$ in Eq.(2.14), and $\mu$ has been set equal to $M_Z$. It then follows that $\hat{\rho} = (1 - Y_{\overline{MS}})^{-1}$. The function $\Delta \hat{r}$ is obtained calculating the electroweak corrections to the muon decay in the $\overline{MS}$ framework [10, 11]. The asymptotic behavior of these two functions is similar to that of $\Delta r$, but the top dependence of $\sin^2 \hat{\theta}_W(M_Z)$ is milder, as part of it has been absorbed in the definition [10]

$$\Delta \hat{r} \sim -\frac{3 \hat{\alpha}}{16 \pi \hat{s}^2 \hat{c}^2} \; \frac{m_t^2}{M_Z^2} + \dots \quad (m_t \gg M_Z). \tag{2.20}$$

Explicit formulae including some higher order effect can be found in Refs. [11, 36]. The use of $\alpha$ instead of $\hat{\alpha}$ in the numerator of Eq.(2.19a) implies the appearance in $\Delta \hat{r}$ of the mass singularities contained in $\Delta_\gamma$, and the corresponding hadronic uncertainty. The latter affects the determination of $\hat{s}^2$ according to

$$(\delta\hat{s}^2)_{th} \approx \frac{\hat{c}^2\hat{s}^2}{(\hat{c}^2 - \hat{s}^2)} \frac{\delta\Delta\hat{r}}{1 - \Delta_\gamma(M_z)}, \tag{2.21}$$

which leads to an uncertainty $\approx 2.5 \times 10^{-4}$ due to the hadronic contribution to the running of $\alpha$; this is one order of magnitude larger than the one induced by the uncertainty in $M_z$. The determination of $M_W$ through Eq.(2.19b) differs by the one through Eq.(2.12) by higher order subleading contributions, when the leading $m_t$ terms are consistently implemented. A numerical analysis [36] of this scheme dependence has shown a maximum discrepancy of $\approx 4 \text{MeV}$.

---

[11] This approach is illustrated in the first article of Ref. [6].

An important part in the choice of the renormalization framework is played by the simplicity of the formulation. For instance, the use of $\sin^2\theta_{eff}^{lept}$ as definition for a renormalized coupling would complicate beyond reason the analytic computation of electroweak radiative corrections, as this parameter is based on a very specific physical amplitude. If this may appear as an aesthetic point of view in one-loop calculations, it becomes a vital matter in higher order calculations. The $\overline{MS}$ framework does provide a clear and simple environment for such involved computations[12].

Finally, it is worth mentioning that a precise determination of the $\overline{MS}$ parameters at the $Z$ scale is a very convenient basis for applications to Grand Unified Theories (GUT), and for new physics in general. In particular, using the renormalization group, it is possible to relate the values of these parameters at $M_z$ to the ones at a very high energy scale $\mu$, and in this way to test different scenarios of Grand Unification. For example, the minimal non-supersymmetric SU(5) GUT [31] predicts $\hat{s}^2 = 3/8$ at the unification scale, which extrapolates to $\hat{s}^2 = 0.2100 \pm 0.0025 \pm 0.0007$ at $M_z$ [32] using $\alpha$ and $\alpha_s$ as inputs. Together with proton decay data, this prediction rules out the simplest GUTs. In contrast, in the minimal supersymmetric SM (MSSM) Grand Unification yields $\sin^2\hat{\theta}_W(M_z) = 0.2334 \pm 0.0025 \pm 0.0025$, which is clearly compatible with present data. In practice, as the precision on $\sin^2\hat{\theta}_W(M_z)$ is now much better than the one on $\alpha_s(M_z)$, it is standard procedure to use the MSSM determination of $\sin^2\hat{\theta}_W(M_z)$ together with $\hat{\alpha}(M_z)$, and to predict $\alpha_s(M_z)$, assuming Grand Unification. Present data are compatible with this hypothesis [32]. This is just one illustration of the importance of precision calculation for uncovering new physics.

## 2.4 The leptonic effective sine in $\overline{MS}$

As we have seen above, the $\overline{MS}$ definition of $\sin^2\theta_w$ seems particularly suitable for the description of physics on the $Z$-resonance. It has been known for some time, for example, that the use of $\overline{MS}$ parameters in the Born amplitudes provides a very good approximation of the neutral current complete amplitude[13], and in particular $\sin^2\hat{\theta}_W(M_z)$ is very close numerically to the parameter $\sin^2 \theta_{eff}^{lept}$ extracted from the asymmetries. However, the reason and extent of this coincidence and the precise numerical relation between the two has not been spelled out in the literature. In view of the very accurate experimental determination of $\sin^2 \theta_{eff}^{lept}$, it is important to know the precise relation between the two mixing angle. In this section, I present a detailed calculation [34] of the form factor $\hat{k}(M_Z^2)$ which controls this relation.

---

[12] A comprehensive comparison of different renormalization schemes for the electroweak sector of the SM can be found in the second article of Ref. [6].

[13] An exception is the $Zb\bar{b}$ vertex [35], where large non-universal top terms are also present.

Up to terms of order $\mathcal{O}(\hat{\alpha})$ in the $\overline{\text{MS}}$ framework we have [27]

$$\hat{k}_\ell(q^2) = 1 - \frac{\hat{c}}{\hat{s}} \frac{[A_{\gamma Z}(q^2) - A_{\gamma Z}(0)]_{\overline{\text{MS}}}}{q^2 - A_{\gamma\gamma\overline{\text{MS}}}^{(f)}(q^2)} + \frac{\hat{\alpha}}{\pi \hat{s}^2} \hat{c}^2 \log c^2 - \frac{\hat{\alpha}}{4\pi \hat{s}^2} V_\ell(q^2), \tag{2.22}$$

where $A_{\gamma Z}(q^2)$ is the $\gamma - Z$ mixing self-energy, the subscript $\overline{\text{MS}}$ means that the $\overline{\text{MS}}$ renormalization has been carried out (i.e. the pole terms have been subtracted and the 't Hooft scale has been set equal to $M_Z$), the superscript $f$ stands for fermionic part, $\hat{\alpha}(M_Z) = [127.91 \pm 0.10]^{-1}$, $c^2 \equiv M_W^2 / M_Z^2$ and $V_\ell(q^2)$ is a finite vertex correction. Explicitly,

$$V_\ell(q^2) = \frac{1}{2} f(\frac{q^2}{M_W^2}) + 4\hat{c}^2 \; g(\frac{q^2}{M_W^2}) - \frac{1 - 6\hat{s}^2 + 8\hat{s}^4}{4\hat{c}^2} f(\frac{q^2}{M_Z^2}), \tag{2.23}$$

where $f(x)$ and $g(x)$ are defined in Eqs. (6d) and (6e) of Ref. [27]. I have included the photon self-energy $A_{\gamma\gamma\overline{\text{MS}}}^{(f)}(q^2)$ in the second term of Eq.(2.22) because, as it will be explained later, it gives rise to relatively large $\mathcal{O}(\alpha^2)$ terms.

It is clear from Eq.(2.11) that the ratio of the vector and axial vector couplings at resonance is given by $1 - 4 \; \hat{k}_\ell(M_Z^2) \; \hat{s}^2$. We can now discuss the various contributions to $\hat{k}_\ell(M_Z^2)$.

To $\mathcal{O}(\hat{\alpha})$ the fermionic contribution to the self-energy of Eq. (2.22) can be written in the form

$$-\frac{\hat{c}}{\hat{s}} \frac{A_{\gamma Z}^{(f)}(M_Z^2)}{M_Z^2} = \frac{\hat{\alpha}}{\pi \hat{s}^2} \sum_i \left( \frac{Q_i C_i}{4} - \hat{s}^2 Q_i^2 \right) \frac{\Pi^V(M_Z^2, m_i, m_i)}{M_Z^2}, \tag{2.24}$$

where $Q_i$, $C_i$, and $m_i$ are the charge, third component of weak isospin (with eigenvalues $\pm 1$), and mass of the $i$-th fermion, the summation includes the color degree of freedom, $\Pi^V$ is the vacuum polarization function involving vector currents, according to the normalization of Ch. 4, and henceforth the $\overline{\text{MS}}$ renormalization is not indicated explicitly. For the leptons we can safely neglect the masses and using the result in App. B find that the contribution to Eq.(2.24) is $(\hat{\alpha}/\pi\hat{s}^2)(5/12 + i\pi/4)(1 - 4\hat{s}^2) = (3.3 + 6.2i)10^{-4}$. In this calculation and henceforth I employ $\hat{s}^2 = 0.2317$, which corresponds to the central values $m_t = 178$ GeV and $m_H = 300$ GeV in the global fit of Ref. [3], and $\hat{\alpha} = (127.9)^{-1}$.

For the first five quark flavors I set again $m_i = 0$ (it can be verified that this is an excellent approximation) and, including $\mathcal{O}(\hat{\alpha}\alpha_s)$ corrections available from the expressions in App. B, obtain a contribution $(\hat{\alpha}/\pi\hat{s}^2) (7/12 - 11 \; \hat{s}^2/9) [5/3 + (\alpha_s/\pi) (55/12 - 4\zeta(3) + i\pi)] = (5.35 + i10.51) \times 10^{-3}$, where I have used $\alpha_s = \alpha_s(M_Z) = 0.118$ and $\zeta(3) = 1.20206....$

The top quark contribution to Eq.(2.24) can be derived from Eq.(2.18), and the complete expression can be again gleaned from App. B. For the current range $150 \text{GeV} \lesssim m_t \lesssim 210$ GeV [3], $D(1/\xi_t)$ varies from $6 \times 10^{-5}$ to $3 \times 10^{-5}$ and is of the same order of magnitude as neglected two-loop contributions $\sim (\hat{\alpha}/\pi \hat{s}^2)^2 \approx 10^{-4}$ to Eq.(2.22). We reach the conclusion that when the M-R prescription is applied, the top quark contribution to Eq.(2.24) is very small.

The other contributions to $\hat{k}_\ell(M_Z^2)$ in Eq.(2.22) can be readily obtained from the literature. This form factor is gauge invariant, but several individual components are not. I evaluate them in the 't Hooft-Feynman gauge, using $M_W = 80.23$ GeV [3]: $(i)$ the bosonic contributions $-(\hat{c}/\hat{s}) [A_{\gamma Z}^{(b)}(M_Z^2) - A_{\gamma Z}^{(b)}(0)]/M_Z^2$ can be extracted from Ref. [39] and amount to $-5.94 \times 10^{-3}$; $(ii)$ $-(\hat{\alpha}/4\pi \hat{s}^2) V_\ell(M_Z^2)$ can be obtained from Eq.(2.23) above and Eqs.(6d,e) of Ref. [27], and gives $+(3.33 + 2.78i) \, 10^{-3}$; $(iii)$ $(\hat{\alpha}/\pi \hat{s}^2)\hat{c}^2 \log c^2 = -2.11 \times 10^{-3}$; $(iv)$ although two-loop effects have not been fully calculated, I include the $\mathcal{O}(\hat{\alpha}^2)$ contribution arising from the product of $\mathrm{Im} A_{\gamma Z}(M_Z^2)$ and $\mathrm{Im} A_{\gamma \gamma}(M_Z^2)$ in the second term of Eq.(2.22). The ratio $M_Z^2/(M_Z^2 - A_{\gamma \gamma}^{(f)}(M_Z^2))$ is given by $0.992 - 0.0175i$ for the current range of $m_t$, and the interference between the two imaginary parts gives a contribution to Eq.(2.24) of $\approx 1.9 \times 10^{-4}$. It is quite sizable, relative to typical $\mathcal{O}(\hat{\alpha}^2)$ contributions, because these imaginary parts involve several additive terms. On the other hand, the large logarithmic $\mathcal{O}(\alpha^2)$ corrections associated with the running of $\alpha$ are already taken into account, in the $\overline{\mathrm{MS}}$ scheme, by employing $\hat{\alpha}$ in the evaluation of $A_{\gamma Z}$.

Combining all the above results we have $\hat{k}_\ell(M_Z^2) = 1 + (0.33 + 0.62i + 5.32 + 105.1i - 5.94) \times 10^{-3}(0.9921 - 0.0178i) + D(1/\xi_t) + (3.33 + 2.78i - 2.11) \times 10^{-3}$, which to good approximation becomes

$$\hat{k}_\ell(M_Z^2) = 1.0012 + i\, 0.0139, \tag{2.25}$$

where the real part slightly decreases for a heavy top. It is clear, on the basis of Eq.(2.25), that at the one-loop level the ratio of effective vector and axial vector couplings in the $Z \to \ell \bar{\ell}$ amplitude is a complex number. This is also expected from general principles. On the other hand, LEP groups interpret both sides of Eq.(2.10) as real quantities. This can be justified on the grounds that the imaginary component of $\hat{k}_\ell(M_Z^2)$ gives a negligible contribution to the leptonic T-even bare asymmetries and partial widths [14]. For instance, the bare left-right asymmetry is given by $A_{LR}^{0,\ell} = 2 \, \mathrm{Re}(g_V/g_A)/[1 + |g_V/g_A|^2]$ and one readily verifies that the inclusion of $\mathrm{Im} \, \hat{k}_\ell(M_Z^2)$ decreases its value by only $-0.02\%$, and $\delta \sin^2 \theta_{eff}^{\ell} \approx -2\% \delta A_{LR}^{0,\ell}/A_{LR}^{0,\ell}$ for the current value of $\sin^2 \theta_{eff}^{lept}$. Similarly, $A_{FB}^{0,\ell}$ is modified by $\approx -0.03\%$. Therefore we identify

$$\sin^2 \theta_{eff}^{lept} = \hat{s}^2 \, \mathrm{Re} \, \hat{k}_\ell(M_Z^2). \tag{2.26}$$

Using Eq.(2.25) we have

$$\sin^2 \theta_{eff}^{lept} - \sin^2 \hat{\theta}_W(M_Z) = 2.8 \times 10^{-4} \approx 3 \times 10^{-4}, \tag{2.27}$$

which tends to approximate to $2.7 \times 10^{-4}$ for a heavy top $(m_t \gtrsim 185 \mathrm{GeV})$. The following observations are appropriate at this stage: $(a)$ because the Higgs boson does not contribute at the one-loop level to Eq.(2.22), the results of Eqs.(2.25, 2.27) are independent of $m_H$; $(b)$ it is clear that the closeness of Re $\hat{k}_\ell(M_Z^2)$ to unity and, correspondingly, the small difference in Eq.(2.27) are due to the cancellation of significantly larger terms. For instance, the light quark and bosonic contributions to the $\gamma$–$Z$ mixing self-energy are of the roughly expected order of magnitude $\sim \hat{\alpha}/(2\pi\hat{s}^2) \approx 5.4 \times 10^{-3}$, but they largely cancel against each other. On the other hand, the $\mathcal{O}(\hat{\alpha})$ contributions to the Im $\hat{k}_\ell(M_Z^2)$ are $\approx 1\%$, an order of magnitude larger; (c) as the relation between $\sin^2 \hat{\theta}_W(M_Z)$ and $\sin^2 \theta_W \equiv 1 - M_W^2/M_Z^2$ is well-known (see Sec. 2.3), Eq.(2.27) determines the connection between the three parameters.

---

[14] The imaginary part of $\hat{k}_\ell(m_Z^2)$, however, gives important contributions to T-odd leptonic asymmetries. See Ref. [40].

If the Marciano-Rosner decoupling convention is *not* applied, so that in the $\overline{\text{MS}}$ renormalization one only subtracts poles and sets the 't Hooft scale equal to $M_Z$, there is a further contribution to Re $\hat{k}_\ell(M_Z^2)$ arising from the first term in Eq.(2.18). Using $\alpha_s(m_t) \approx 0.11$, this amounts to $-6.1 \times 10^{-4}$, $-8.7 \times 10^{-4}$, $-1.09 \times 10^{-3}$, for $m_t = 150$, 180 and 210 GeV, respectively. Correspondingly, Re $\hat{k}_\ell(M_Z^2)$ becomes 1.0006, 1.0003, 1.0001, even closer to unity. As a consequence, although the difference between $\sin^2 \theta_{eff}^{lept}$ and $\sin^2 \hat{\theta}_W(M_Z)$ depends more on $m_t$ when the M-R prescription is not applied, it is actually smaller for the current range $150 \lesssim m_t \lesssim 210$ GeV. In fact, we find that it is $1 \times 10^{-4}$ for $150 \leq m_t \leq 191$ GeV, and there is no difference in the fourth decimal for $192 \leq m_t \leq 210$ GeV.

One can obtain a rough consistency check of the order of magnitude of Eq.(2.27) by comparing the fits of Ref. [3] (Table 2.1) with the calculations of Ref. [36]. Using the LEP, SLD, collider and $\nu$ data, Ref. [3] finds $m_t = 178 \pm 11\,^{+18}_{-19}$ and $\sin^2 \theta_{eff}^{lept} = 0.2320 \pm 0.0003\,^{+0.0000}_{-0.0002}$. Their central values assume $m_H = 300$ GeV, the first error represents experimental and theoretical uncertainties, while the second reflects changes corresponding to the assumptions $m_H = 60$ GeV and $M_H = 1$ TeV. According to Eq.(2.27), the corresponding central value for $\hat{s}^2$ should be 0.2317. On the other hand, from Ref. [36] one finds $\hat{s}^2 = 0.2317$ for $m_t = 178$ GeV and $m_H = 300$ GeV. Thus, the comparison of the conclusions of Ref. [3] with the calculations of Ref. [36] is roughly consistent with Eq.(2.27). Of course, such consistency checks are not a substitute for precise, ab initio calculations, like the one leading to Eq.(2.27).

Two additional comments can be useful at this stage: (*a*) Consistently with Eq.(2.10), $\sin^2 \theta_{eff}^{lept}$ can be defined in terms of a bare forward-backward asymmetry $A_{FB}^{0,\ell}$, which is obtained from the physical asymmetry $A_{FB}^\ell$ after extracting the effect of photon-mediated contributions and other radiative correction effects [26]. Therefore, we should not attempt to find the numerical relation between $\sin^2 \theta_{eff}^{lept}$ and $\sin^2 \hat{\theta}_W(M_Z)$ by comparing detailed $\overline{\text{MS}}$ calculations of the physical asymmetry $A_{FB}^\ell$, as those in Ref. [27], with theoretical expressions for $A_{FB}^{0,\ell}$ expressed in terms of $\sin^2 \theta_{eff}^{lept}$. The point is that $A_{FB}^\ell$ contains electroweak effects not contained in $A_{FB}^{0,\ell}$. (*b*) Global analyses often cite the value of $\sin^2 \theta_{eff}^{lept}$ as extracted only from the on-resonance asymmetries, while they give the prediction for $m_t$ derived from the complete data base. Current asymmetry results lead to determinations of $\sin^2 \theta_{eff}^{lept}$ that are somewhat smaller than the $\sin^2 \hat{\theta}_W(M_Z)$ numbers corresponding to the central $m_t$. This, however, is not a contradiction with Eq.(2.27), because the on-resonance asymmetries represent only a part of the experimental information. This is quite visible in the detailed report of Ref. [3], in which one finds $\sin^2 \theta_{eff}^{lept} = 0.2317 \pm 0.0004$ from the on-resonance asymmetries and, as mentioned before, larger values from the global fits.

In summary, I have attempted to clarify the connection between $\sin^2 \theta_{eff}^{lept}$ and $\sin^2 \hat{\theta}_W(M_z)$ and obtained the value of their difference by means of a detailed calculation, both with and without the M–R decoupling convention [34]. I recall that $\sin^2 \theta_{eff}^{lept}$ is defined in terms of a physical amplitude, while this is not the case for $\sin^2 \hat{\theta}_W(M_z)$, the running $\overline{\text{MS}}$ parameter evaluated at the $M_z$-scale. Thus, it is clear that the two parameters are conceptually very different. On the other hand, we have found that their numerical values are very close (cf. Eq.(2.27)). As illustrated in the discussion after Eq.(2.27), this fact is probably due to a fortuitous cancellation of radiative corrections.

## 2.5 Higher order corrections

In the preceding sections I have outlined the main features of the renormalization program for electroweak high-energy physics. Although the bulk of the radiative corrections is doubtless represented by the one-loop contributions, it is necessary to consider all possible sources of large higher order contributions. In some case, as for the electromagnetic coupling, renormalization group techniques help to resum large contributions at any order. Whenever they are possible, however, explicit multi-loop calculations are the best way to investigate the convergence of the electroweak perturbative series. In fact, what we really need is an estimate of the theoretical error due to the truncation of the perturbative series. There exist heuristic methods for estimating the higher order contribution in QCD, but in general no such method applies to the more complicated situation of electroweak interactions. In the last few years, multi-loop calculations in electroweak physics have become increasingly important, as the lower bound on the top mass increased, and the experimental accuracy improved. One of the effects of the activity in this field has been the realization that the convergence of the electroweak series is not as smooth as it was believed: large higher order effects do exist, and must be taken into account. In the following I will restrict the analysis to the two observables considered in this chapter, $M_W$ and $\sin^2 \theta_{eff}^{lept}$. We can distinguish two kinds of possibly large higher order effects that can affect the predictions for these observables.

- **Large mass effects** $M_H$ and $m_t$ are heavy, of the order of the spontaneous symmetry breaking scale. Heavy masses do not decouple in the SM and therefore we can expect loop contributions that are not bounded as $m_t$ and $M_H$ go to infinity. In particular, the so-called decoupling theorem of Appelquist and Carazzone [43] does not apply in this context. If we exclude the case of amplitudes involving external Higgs bosons, we can see that at the one-loop level the radiative corrections scale as the square of the top mass, and as the logarithm of $M_H$, as in Eq.(2.13)[15]. At two-loop level, the leading heavy mass effects are of $\mathcal{O}(G_\mu^2 m_t^4)$ and $\mathcal{O}(\alpha G_\mu M_H^2)$. Since complete two-loop calculation do not exist at present, a good starting point would be to have all power-like heavy mass

---

[15]The logarithmic dependence on $M_H$ of the one-loop radiative corrections in the gauge sector of the SM has been first shown in Ref. [41]. In the limit of large $M_H$ the SM can be viewed as a gauged non-linear $\sigma$-model and the mass of the Higgs boson can be interpreted as an ultraviolet regulator. It is then easy to see that at one-loop the non-linear $\sigma$-model contains only logarithmic divergences [42].

corrections under control. This kind of effects will be considered in Ch. 5.

- **QCD effects** Because of the large coupling constant, perturbative QCD corrections
  can be sizable, even at a high energy scale. In particular, the determination of $M_W$
  and $\sin^2 \theta_{eff}^{lept}$ proceeds through physical amplitudes (muon decay and $Z\ell\bar{\ell}$ vertex) that
  involve only *oblique* QCD corrections, i.e. QCD corrections to quark loop insertions on
  vector boson propagators. We have seen that large quadratic top effects can arise as a
  consequence of the $SU(2)_L$ symmetry breaking in the $(t, b)$ isodoublet; we can therefore
  expect relevant QCD corrections of $\mathcal{O}(G_\mu \alpha_s m_t^2)$. This kind of effects will be studied in
  Ch. 4.

As we have seen, in the $\overline{\text{MS}}$ framework, the determination of $\sin^2 \theta_{eff}^{lept}$ proceeds through
Eq.(2.27) and Eq.(2.19a). The relation between $\sin^2 \theta_{eff}^{lept}$ and $\sin^2 \hat{\theta}_W(M_Z)$, however, has
already been evaluated keeping leading QCD effects in the hadronic self-energies, and, if
the Marciano-Rosner prescription is employed, should be free of large power-like heavy mass
effects of $\mathcal{O}(G_\mu^2 m_t^2 M_Z^2)$. Therefore, the precise determination of $M_W$ and $\sin^2 \hat{\theta}_W(M_Z)$ through
Eq.(2.19) becomes the most important task in precision physics. Some aspects of this effort
will be illustrated in the following.

But first, in the next chapter, I will present a simple method to attain indirect evidence
that a full one-loop calculation is needed to describe current data.

# Chapter 3

# Evidence for radiative corrections

We know that a renormalizable quantum field theory includes quantum corrections of $\mathcal{O}(\hbar)$, and we know that the SM is very successful in describing current experimental data. One might then wonder what do we know phenomenologically about the quantum field structure of the SM. In this chapter I will discuss the current phenomenological evidence for radiative corrections. A simple example [33] will illustrate that there is very strong indirect evidence from high-energy experiments for the leading contributions due to fermionic loops, as well as for the numerically subleading contributions that involve virtual boson exchange.

## 3.1   Evidence from low-energy

The fundamental importance of radiative corrections in the electroweak theory has been known for a long time, and has been established first by low-energy experiments [47]. For example, in the absence of radiative corrections, a large violation of the unitarity of the Cabibbo-Kobayashi-Maskawa matrix (which is mainly determined from low-energy experiments) would appear. When all the one-loop electroweak corrections are included and the nuclear overlapping is appropriately taken into account, an analysis [44, 45] of the $^{14}O$ super-allowed Fermi transitions leads to $V_{ud} = 0.9745 \pm 0.0005 \pm 0.0004$. $V_{us}$ and $V_{ub}$ are much smaller, and are extracted from $\Delta S = 1$ and $B$ decays, yielding [45]

$$|V_{ud}|^2 + |V_{us}|^2 + |V_{ub}|^2 = 0.9983 \pm 0.0015, \qquad (3.1)$$

very close to unity. The radiative corrections affecting the derivation of Eq.(3.1) are quite sizable, about 4.1%, and involve vector boson exchanges. As the momentum scale is small, no large electromagnetic corrections due to the running of $\alpha$ are present, and the sensitivity to $m_t$ is cancelled in the ratio between the amplitudes for $\beta$-decay and muon decay, so that we can ascribe this large contribution to bosonic effects. When we analyze the data employing only tree level expressions and the Fermi-Coulomb function, we obtain $1.0386 \pm 0.0013$ [47], which is about $30\sigma$ away from unity. In fact, the appearance of a large radiative correction in the determination of $V_{ud}$ can be traced back to the observation that the photonic corrections to the $\beta$-decay in the local V-A electroweak theory are not convergent, unlike the ones to the muon decay[1]. They involve a logarithmic divergence $\sim \ln \frac{\Lambda}{m_p}$, where $m_p$ is the mass of the nucleon. In the one-loop SM calculation [44], the cutoff is simply replaced by $M_Z$, yielding a sizable contribution.

On the other hand, one could argue that the effect of SM loop corrections that go beyond the QED corrections can be mimicked by simply setting the cutoff equal to $M_Z$. However, in order to match the experimental precision, an accurate determination of the cutoff $\Lambda$ in the Fermi effective field theory would be necessary. Although the order of magnitude of the cutoff can be easily guessed (the symmetry breaking scale $v \sim G_\mu^{-1/2}$), without a precise determination the local V-A theory result for Eq.(3.1) can still be very different from unity [50]. Such a precise determination can only be provided in the framework of the full renormalizable SM. In conclusion, it appears that low-energy experiments provide a very strong evidence for electroweak radiative corrections of bosonic origin.

## 3.2    Evidence from high-energy

While the low-energy corrections in the problem of universality involve virtual fermions and $W^\pm$, $\gamma$, and $Z$ bosons, at high-energy the dominant corrections come from fermionic loops (an exception is the $Z \rightarrow b\bar{b}$ vertex, where large corrections are induced by loops involving virtual bosons). As we have seen in the previous chapter, one of the main sources of large corrections is the running of $\alpha$, with the associated large logarithms. Evidence for radiative corrections beyond the running of $\alpha$ has been analyzed in Ref. [47]. Direct evidence of such contributions comes from the comparison of the experimental data for specific observables with the predictions of a Born approximation in which $\alpha$ has been replaced by $\alpha(M_Z)$ or $\hat{\alpha}$. This kind of direct evidence is not particularly compelling, and presently reaches $\approx 1.5 - 2\sigma$ [48]. However, it has been argued [47] that very strong (at the level of 5–6$\sigma$) indirect or inferred evidence can be uncovered when the comparison is made between the Born approximation and the results of a global fit in the full SM to the whole body of electroweak data. The global fit provides more precise values than experiments for single observables, and incorporates all the interlocking relations that highly constrain the SM.

It is also interesting to note that no Born approximation can accurately describe all the available experimental electroweak data. Indeed, we have seen in the previous chapter that two very different values for $\sin^2 \theta_W$ can be extracted from experiment: $\sin^2 \theta_{eff}^{lept}$ and $1 - M_W^2/M_Z^2$, respectively $0.2317 \pm 0.0004$ and $0.2259 \pm 0.0035$; they differ by $1.7\sigma$. No Born approximation for the electroweak mixing angle can accommodate both values. Again, the discrepancy is much stronger if we compare the values for these observables in the SM fit of Table 2.1: we find $\sin^2 \theta_{eff}^{lept} = 0.2320 \pm 0.0004$ and $1 - M_W^2/M_Z^2 = 0.2242 \pm 0.0012$, which differ by $5.6\sigma$. Note also that the fit of Table 2.1 does not include the latest CDF and DØ values for $m_t$.

---

[1]A beautiful and simple explanation of this difference, based on Fierz transformations, can be found in Ref. [46].

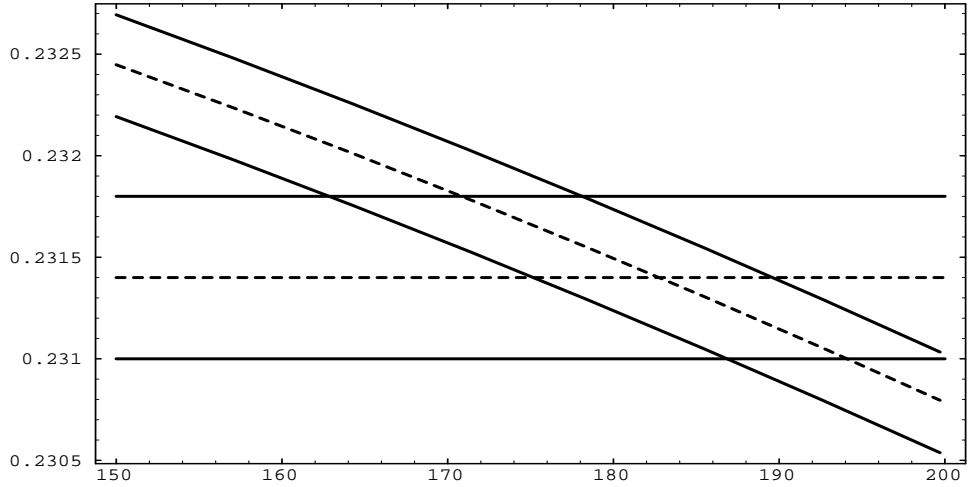

Figure 3.1: Determination of $\sin^2 \hat{\theta}_W(M_Z)$ from asymmetries (horizontal lines), and from $M_Z$ (slanted curves), in the full SM with $M_H = 300 \,\mathrm{GeV}$. The $1\sigma$ errors are indicated.

### 3.2.1 Bosonic radiative correction

Besides the running of $\alpha$, another source of large fermionic contributions is the sharp breaking of isospin symmetry in the $t - b$ quark doublet, which introduces terms that are quadratic in the top quark mass, and allows the determination of $m_t$ through radiative corrections. For instance, the radiative correction $\Delta r$, introduced in the previous chapter (Eq.(2.12)), can be decomposed in

$$\Delta r = \Delta \alpha - \frac{c^2}{s^2} x_t + \Delta r_{rem} \tag{3.2}$$

where $\Delta \alpha$ and $x_t = 3 \frac{G_\mu m_t^2}{8\pi^2 \sqrt{2}}$ are the large fermionic contributions coming from the running of the e.m. coupling constant and from the $t - b$ doublet, while $\Delta r_{rem}$ collects all other contributions, mostly bosonic in origin. The typical size of $\Delta r_{rem}$ is 1%, for example in the case $m_t = 180 \,\mathrm{GeV}$, $\Delta r \approx 3.8\%$, $x_t \approx 1.0\%$, and $\Delta r_{rem} \approx 1.3\%$.

Especially after the experimental observation of the top quark, we can safely argue that the fermionic sector of the SM is phenomenologically very well established. Fermionic loops are unambiguous theoretically, and the couplings have been tested in many high-energy precision experiments. On the other hand, we know that, in addition to these fermionic contributions, there are a number of conceptually very important but numerically "subleading" corrections involving virtual bosons ($Z$, $W^{\pm}$, $\gamma$, and $H$). They involve the plethora of bosonic couplings of the SM, including the well-known tri-linear vertices, which have never been tested independently[2], and affect self-energies, vertices, and box diagrams (in four fermion processes, bosonic vertex and box diagrams also include virtual fermions). Since the separation into fermionic and bosonic contribution is gauge invariant and finite at the one-loop level, and therefore unambiguous, it is possible to regard these two subsets as completely independent.

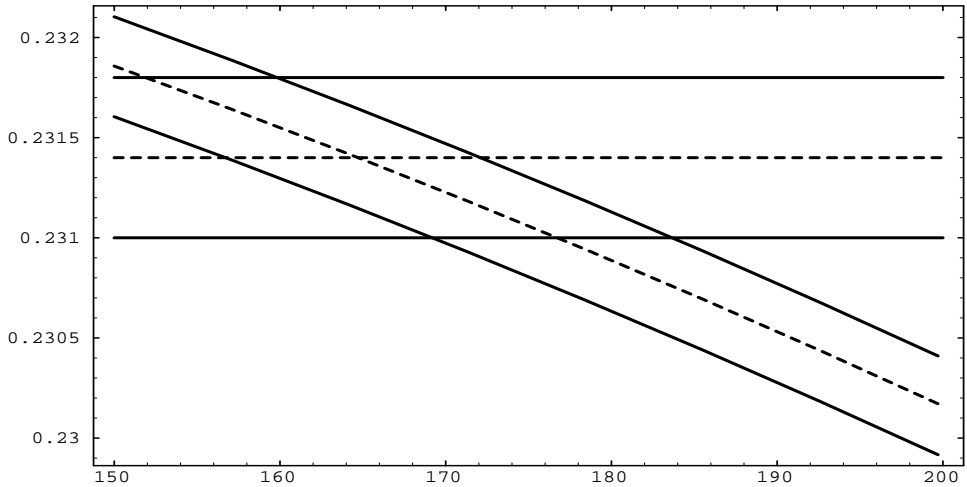

Figure 3.2: Determination of $\sin^2 \hat{\theta}_W(M_Z)$ in the full SM with $M_H = 100$GeV.

With the increasing experimental accuracy the question arises of what is the sensitivity of present high-energy experiments to subleading contributions as the ones mentioned before, and to what extent present data provide evidence for the virtual structure of the theory. In this respect, the idea is to analyze the effect of the bosonic part of the theory on basic observables and their relations, and show that the current accuracy already allows us to recognize the presence of the associated virtual corrections. As an illustration, I will consider two different precise determinations of the $\overline{\text{MS}}$ parameter $\sin^2 \hat{\theta}_W(M_Z) \equiv \hat{s}^2$, which conveniently describes physics at the Z-mass scale, as discussed in the previous chapter. Alternatively, this procedure can be viewed as a determination of $\sin^2 \theta_{eff}^{lept}$ in the $\overline{\text{MS}}$ framework.

The first determination of $\sin^2 \hat{\theta}_W(M_Z)$ is obtained from the effective weak mixing angle $\sin^2 \theta_{eff}^{lept}$ which is now known with a very good precision. As mentioned above, the average of LEP and SLC asymmetry measurements gives $\sin^2 \theta_{eff}^{lept} = 0.2317 \pm 0.0004$. The two parameters $\hat{s}^2$ and $\sin^2 \theta_{eff}^{lept}$ are related by Eq.(2.27) which, to very good accuracy, is independent of both $m_t$ and $M_H$. The parameter $\hat{s}^2$ can also be obtained using $M_Z$, $G_\mu$, $\alpha$ through Eq.(2.19a),

$$\hat{s}^2 \hat{c}^2 = \frac{\pi \alpha}{\sqrt{2} G_\mu M_Z^2 (1 - \Delta \hat{r})} \tag{3.3}$$

---

[2]The present experimental resolution on tri-boson couplings is very poor, but is expected to improve significantly at LEP200.

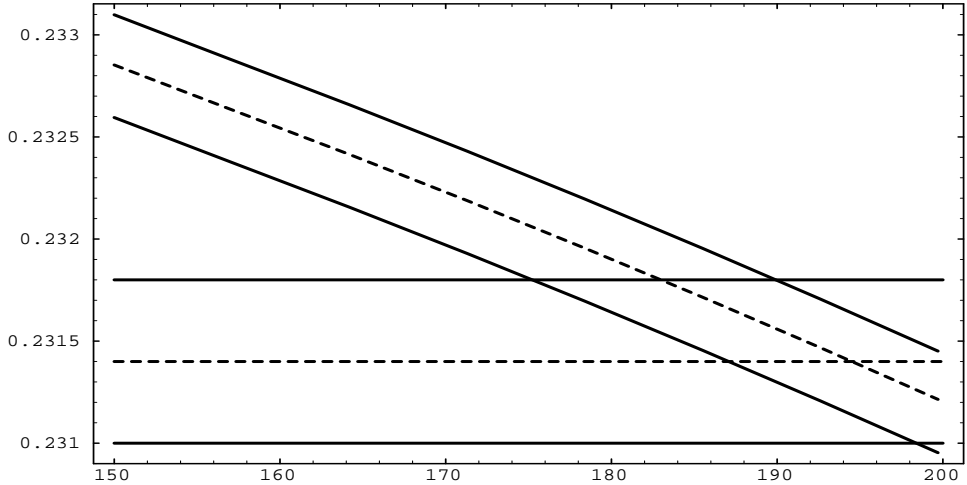

Figure 3.3: Determination of $\sin^2 \hat{\theta}_W(M_Z)$ in the full SM with $M_H = 600$GeV.

and it has been shown that the radiative correction $\Delta \hat{r}$ is a sensitive function of $m_t$ and $M_H$. The results of the two determinations of $\hat{s}^2$ in the SM are shown in Figs. 3.1, 3.2, and 3.3, as a function of $m_t$, in the case of $M_H$ respectively equal to 300, 100, and 600GeV, with the corresponding $1\sigma$ errors. The error on the determination from Eq.(3.3) is almost completely due to the uncertainty on the light hadron contribution to the running of $\alpha$. The comparison between the two determinations of $\sin^2 \hat{\theta}_W(M_Z)$ forces a bound on the top quark mass. In the case of the SM with $M_H = 300$ GeV this bound is in good agreement with the quoted fit of Ref. [3], and from Figs. 3.2 and 3.3 we can see that there is a slight preference for a light Higgs boson. In the evaluation of $\Delta \hat{r}$ I have neglected small $\mathcal{O}(\hat{\alpha}^2)$ and $\mathcal{O}(\hat{\alpha}\alpha_s)$ corrections, except for the leading two-loop effects of $\mathcal{O}(\alpha_s G_\mu m_t^2)$, which I have incorporated using a simple method recently discussed by Sirlin [51] and illustrated in Chap.4. The values of the inputs are as reported in Chap. 2.

In Fig. 3.4 the situation is analyzed in a truncated version of the theory, where all contributions involving virtual bosons (vertices, boxes, and bosonic self-energies) have been removed. From Sec. 2.4, removing the bosonic contributions, one obtains $\mathrm{Re}\hat{k}_\ell(M_Z^2) = 1.0060$, so that

$$(\hat{s}^2)_{tr} = 0.2303 \pm .0004 \qquad (Asymmetries). \tag{3.4}$$

Henceforth the subscript $tr$ reminds us that this value corresponds to a "truncated" version of theory, with bosonic contributions removed in the electroweak corrections. In evaluating $\mathrm{Re}\hat{k}_\ell(M_Z^2)$ we have neglected small $\mathcal{O}(\hat{\alpha}\alpha_s)$ corrections that were retained in Sec. 2.4. On the other hand, removing all the bosonic contributions from $\Delta \hat{r}$ leads to

$$(\hat{s}^2)_{tr}(\hat{c}^2)_{tr} = \frac{\pi \alpha}{\sqrt{2} G_\mu M_Z^2 (1 - (\Delta \hat{r})_{tr})}, \tag{3.5}$$

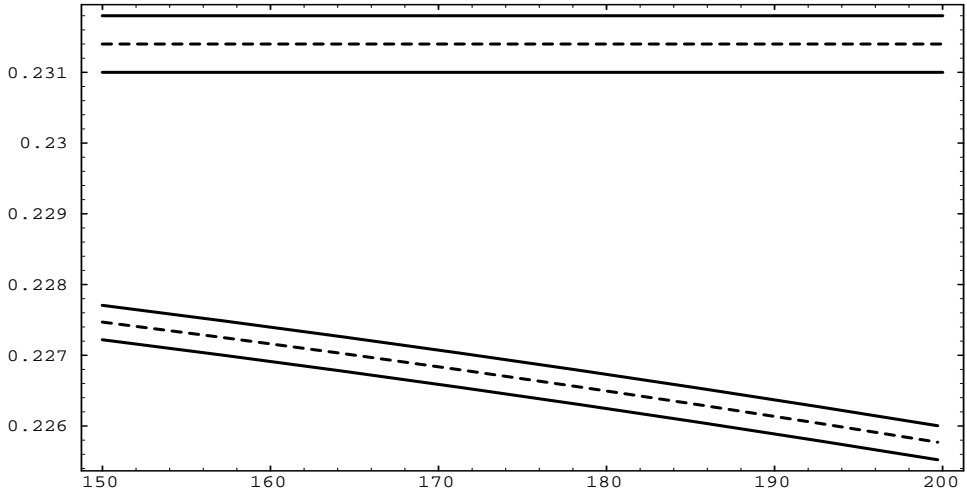

Figure 3.4: Determination of $\sin^2 \hat{\theta}_W(M_Z)$ as a function of $m_t$ from asymmetries and from $M_Z$ without bosonic contributions.

Here I have also removed the irreducible two-loop effects of $\mathcal{O}(\alpha^2 m_t^4)$ [115] as they involve virtual bosons. It is apparent from Fig. 3.4 that the removal of the bosonic component of the electroweak corrections leads to a sharp disagreement. The two determinations do not overlap for any value of $m_t$ in the present experimental range $m_t = 180 \pm 12\,\mathrm{GeV}$ [1, 2]. At the very conservative lower bound of $m_t = 150$ GeV, the value obtained from Eq.(3.5) is

$$(\hat{s}^2)_{tr} = 0.2275 \pm 0.0003 \qquad (G_\mu, \alpha, M_Z, m_t = 150\,\mathrm{GeV}) \qquad (3.6)$$

and we see that the discrepancy with Eq.(3.4) amounts to $5.6\sigma$ (if the SLC value were not included, the top curve in Fig.3.4 would be shifted upwards by $\approx 0.0004$, the error would be slightly increased, and the difference would be $4.5\sigma$). As shown in Fig. 3.4, the discrepancy rapidly increases with $m_t$. For instance, it is more than $7\sigma$ for $m_t = 180$ GeV. Comparing Figs. 3.4 and 3.1, we see that the removal of the bosonic corrections leads to lower values of $\hat{s}^2$. The discrepancy arises because the effect is much more pronounced in the $(G_\mu, \alpha, M_Z)$ determination. An analogous investigation of the effects of bosonic contributions on the determination of $M_W$ gives much less interesting results; the prediction derived using Eq.(2.12) is in acceptable agreement with the experimental value over most of the accepted $m_t$ range, and no glaring contradiction is uncovered. Somewhat similar conclusions have been drawn in Ref. [52], where the effect of bosonic contributions was studied in $\sin^2 \theta_{eff}^{lept}$, $M_W$, and the decay width $\Gamma_z$.

We conclude that the accuracy currently reached in high-energy experiments is such that there is strong indirect evidence for the presence in the SM of these important subleading corrections. In the next few years precision experiments are likely to become sensitive to the structure of the gauge and spontaneous symmetry breaking sectors of the SM. The whole set of one-loop radiative corrections of the SM is definitely needed to describe current high-energy data.

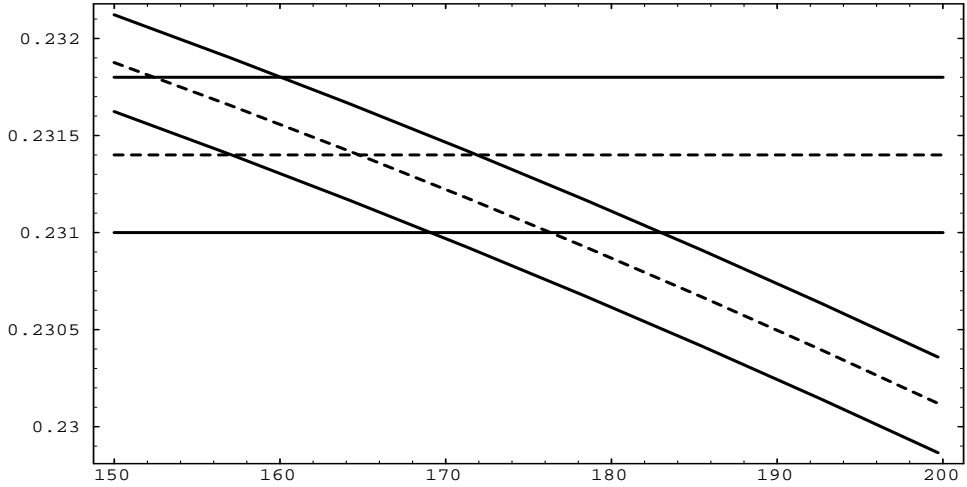

Figure 3.5: Determination of $\sin^2 \hat\theta_W(M_Z)$ as a function of $m_t$ from asymmetries and $M_Z$ without Higgs contributions.

### 3.2.2  The Higgs contribution

Given the sharpness of the signal, it is natural to ask whether one can use the same approach to probe specific components of the bosonic corrections. For instance, can one search for signals of Higgs boson contribution by removing them from the electroweak corrections, retaining the rest? As $H$ does not contribute, at one-loop level, to $\hat k_\ell(M_Z^2)$, the value of $\hat s^2$ derived in this case from $\sin^2 \theta_{eff}^{lept}$ is that of the full SM:

$$\hat s^2 = 0.2314 \pm 0.0004 \qquad (Asymmetries), \qquad (3.7)$$

instead of Eq.(3.4). In order to be physically meaningful, the removal of the Higgs contribution from $\Delta\hat r$ must be done in a gauge-invariant and finite manner. Fortunately, in the SM the diagrams involving $H$ in the self-energies contributing to $\Delta\hat r$ form a gauge-invariant subset. On the other hand, they are divergent. Therefore, one must specify the renormalization prescription and the scale at which these partial contributions are evaluated. As $\hat s^2$ is the $\overline{\rm MS}$ parameter and the electroweak data are dominated by information at the $Z^0$-peak, it is natural to subtract the $\overline{\rm MS}$-renormalized Higgs boson contribution evaluated at the $M_Z$ scale.

Neglecting non-leading $\mathcal{O}(\alpha^2)$ terms, the latter is given by

$$(\Delta\hat{r})_{H.B.} = \frac{\alpha}{4\pi\hat{s}^2}\left\{\frac{1}{\hat{c}^2}H(\xi) - \frac{3}{4}\frac{\xi\ln\xi - \hat{c}^2\ln\hat{c}^2}{\xi - \hat{c}^2} + \frac{19}{24} + \frac{\hat{s}^2}{6\hat{c}^2}\right\}_{\overline{\rm MS}} \tag{3.8}$$

where $\xi = M_H^2/M_Z^2$, $H(\xi)$ is a function studied in Ref. [9, 37], and the subscript $\overline{\rm MS}$ reminds us that the $\overline{\rm MS}$ renormalization has been carried out and the scale $\mu = M_Z$ chosen. The need to specify the scale in defining the Higgs boson contribution can be most easily understood in the on-shell method of renormalization, where one employs $\sin^2\theta_W = 1 - M_W^2/M_Z^2$ instead of $\hat{s}^2$. In that case, the relevant radiative correction is $\Delta r$, rather than $\Delta\hat{r}$. Although $\Delta r$ is a physical observable and is therefore $\mu$-independent, the Higgs-boson contribution is $\mu$-dependent. Thus, a specification of the scale is necessary in its definition.

Subtracting then the Higgs boson contribution one obtains a new truncated version of $\Delta\hat{r}$, independent of $M_H$, from which we can compute the corresponding $(\hat{s}^2)_{tr}$ via Eq.(3.5). The comparison with Eq.(3.3) is given in Fig. 3.5. In contrast with Fig.3.4, where the complete block of bosonic contributions was removed, there are no signals of inconsistency. This is easily understood by noting that $(\Delta\hat{r})_{H.B.}$ vanishes for $M_H \approx 113$ GeV. Thus, the subtraction of Eq.(3.8) is equivalent to a SM model calculation with a relatively light Higgs scalar, $M_H \approx 113$ GeV, and this is consistent with current electroweak data. Similar results, from a different viewpoint, have been found in Ref. [53].

In summary, I have presented strong indirect evidence for the presence in the SM of bosonic electroweak corrections (Fig.3.4). If one probes just the Higgs component of these corrections, no evidence has been uncovered in our very simple analysis. However, it is likely that the signals will become sharper as the precision increases and $m_t$ is measured. Finally, the approach I have illustrated, namely the removal of the bosonic electroweak corrections and their components from the relevant corrections, could be also extended systematically to global analyses.

# Chapter 4

# QCD corrections

*What is common to all men of goodwill is this: that in the final analysis our work disappoint us,*
*that we must always start again at the beginning, that the sacrifice must always be made afresh.*

H. HESSE, Narcissus and Goldmund

In the last few years many interesting results have been obtained in the field of perturbative QCD corrections to the electroweak phenomena, and the inclusion of leading two and even three-loop effects has become routine in most phenomenological applications. Indeed, in most cases relevant for precision physics the effective scale of the processes involved is high enough for perturbation theory to be reliable. A notable exception is the hadronic contribution to $\Delta\alpha$ discussed in Ch. 2.

In this chapter I will first illustrate two examples of two-loop calculations, concerning the QCD corrections to electroweak bosonic two-point functions. The idea behind these calculations is to provide a complete set of exact and compact $\mathcal{O}(\alpha\alpha_s)$ expressions for electroweak bosonic vacuum polarization functions (VPF), in the most general possible case, including some extensions of the SM. From this set of general expressions it is possible to obtain asymptotic expansions for some special case of physical interest (very large and vanishing external momentum and/or masses), as well as the imaginary parts, which in turn are related to the gluonic corrections to the hadronic decay widths of the corresponding boson. These calculations generalize and extend previous work on the subject [54–57], which was limited to the transverse parts of the vector boson self-energies, and relied on some approximation on the masses involved, destined to cover only the present most important phenomenological cases. The final goal of this kind of efforts should be to place analytical calculations in the Standard Model at a level comparable to what is known in QED where, in a pioneering work, the authors of Ref. [58] have derived exactly the VPF of the photon at two-loop order.

As we will see in Sec. 4.6, the universal part of the QCD corrections to four-fermion electroweak processes (and the whole QCD correction in the case of no external hadrons) comes from the transverse parts of the electroweak vector boson self-energies. In particular this is the case of the most important precision observables, based on the muon decay and the leptonic asymmetries. The longitudinal parts of the vector boson VPF's can however prove to be important, for example, in problems involving the production of heavy particles, and the scalar VPF is necessary in the case of Higgs decay, and to study radiative corrections to the masses of the Higgs particles involved in the symmetry breaking, both in the SM and in many of its extensions. Furthermore, an exact calculation without any assumption on the quark masses might turn out to be mandatory in the case of a fourth generation of fermions, the existence of which is still allowed by present experimental data if the associated neutrino is heavy enough [3].

Several checks of the cumbersome calculations are possible and will be illustrated. A nice example is provided by the use of Ward identities, described in Sec. 4.3, while the comparison of the dimensional regularization method with a dispersive approach will be discussed in Sec. 4.4.

QCD calculations, because of the strong scale dependence of the coupling constant $\alpha_s(\mu^2)$, are very sensitive on the choice of mass scale $\mu$ to be considered. As this scale dependence is unavoidable in perturbative calculations, where the perturbative series is truncated, this choice is crucial; a brief discussion of the subject will be given in Sec. 4.5.

Finally, in the last section of this chapter, I will briefly review the incorporation of these QCD perturbative effects in the most important electroweak corrections, and illustrate the first and most natural applications of the results of the preceding sections. In App. A and B the integrals used in the calculation, some asymptotic formulae for special situations of physical interest, and the imaginary parts are reported for completeness.

## 4.1 QCD corrections to electroweak vector boson self-energies: complete calculation

In this section, I present exact and compact analytical expressions for the $\mathcal{O}(\alpha\alpha_s)$ contributions of quark doublets to the vacuum polarization functions of the electroweak gauge bosons in the most general case: real and imaginary parts of both the transverse and longitudinal components, for different and non-zero quark masses and for arbitrary momentum transfer [59].

The contribution of a fermionic loop to the vacuum polarization tensor of a vector boson $i$, or to the mixing amplitude of two bosons $i$ and $j$, denoted $\Pi_{\mu\nu}^{ij}$, can be written as

$$\Pi_{\mu\nu}^{ij}(q^2) = -i \int d^4x \, e^{iq\cdot x} < 0|\mathrm{T}^*\left[J_\mu^i(x)J_\nu^{j\dagger}(0)\right]|0 > \tag{4.1}$$

where $\mathrm{T}^*$ is the covariant time ordered product and $q$ the four–momentum transfer; $J_\mu^i, J_\nu^j$ are fermionic currents coupled to the vector bosons $i, j$ and constructed with spinor fields whose corresponding masses are $m_a, m_b$. The definition of $\Pi_{\mu\nu}^{ij}$ corresponds to $+i$ times the standard Feynman amplitude. The vacuum polarization tensor can then be decomposed into a transverse and a longitudinal part,

$$\Pi_{\mu\nu}^{ij}(q^2) = \left(g_{\mu\nu} - \frac{q_\mu q_\nu}{q^2}\right)\Pi_T^{ij}(q^2) + \frac{q_\mu q_\nu}{q^2}\Pi_L^{ij}(q^2) \tag{4.2}$$

and the two components can be directly extracted by contracting $\Pi_{\mu\nu}^{ij}(q^2)$ by the two projectors $g_{\mu\nu} - q_\mu q_\nu/q^2$ and $q_\mu q_\nu/q^2$; in $n = 4 - 2\epsilon$ dimensions one has

$$\Pi_T^{ij}(q^2) = \frac{1}{3 - 2\epsilon}\left(g^{\mu\nu} - \frac{q^\mu q^\nu}{q^2}\right)\Pi_{\mu\nu}^{ij}(q^2)$$

$$\Pi_L^{ij}(q^2) = \frac{q^\mu q^\nu}{q^2}\Pi_{\mu\nu}^{ij}(q^2) \tag{4.3}$$

To set the notation it seems convenient to discuss briefly the well-known one-loop expressions of the fermionic vector boson self-energies: the one-loop transverse and longitudinal components can be written as (with $s \equiv q^2$)

$$\Pi_{T,L}^{ij}(s) = \frac{\alpha}{\pi}N_c\left[\left(v^i v^j + a^i a^j\right)s\,\Pi_{T,L}^+(s)\ +\ \left(v^i v^j - a^i a^j\right)m_a m_b\,\Pi_{T,L}^-(s)\right] \tag{4.4}$$

with $N_c$ the number of colors of the fermionic doublet and $v^i$ and $a^i$ the vector and axial-vector couplings of the gauge boson $i$ to the fermions expressed in units of the proton charge $e = \sqrt{4\pi\alpha}$. The vector and axial-vector components of the vacuum polarization function, with coupling constants factored out, are then simply given by

$$\Pi_{T,L}^{V,A}(s) = s\,\Pi_{T,L}^+(s)\ \pm\ m_a m_b\,\Pi_{T,L}^-(s) \tag{4.5}$$

which exhibits the fact that $\Pi^{A,V}(s)$ can be obtained from $\Pi^{V,A}(s)$ by making the substitution $m_a \to -m_a$ (or $m_b \to -m_b$) as expected from $\gamma_5$ reflection symmetry.

At the one-loop level the vacuum polarization amplitude receives contributions only by the graph of Fig. (4.1). The regularized amplitude must be multiplied by $\mu^{2\epsilon}$ for each momentum integration, where $\mu$ is the 't Hooft renormalization mass scale introduced to make the coupling constant dimensionless in $n = 4 - 2\epsilon$ dimensions; I have also introduced an extra term $(e^\gamma/4\pi)^\epsilon$, where $\gamma$ is the Euler constant, to prevent uninteresting combinations of $\ln 4\pi$, $\gamma\ldots$, in the final result. After contracting the amplitude with the projectors $(g^{\mu\nu} - q^\mu q^\nu/q^2)$ and $q^\mu q^\nu/q^2$, the resulting integrals can be expressed in terms of the scalar one-loop integrals that are given in App. A.1 for completeness.

The expressions of $\Pi_{T,L}^\pm(s)$ in the general case $m_a \neq m_b \neq 0$ are then

$$\Pi_T^+(s) = \frac{2 + 3\alpha + 3\beta}{6\epsilon} - \frac{1}{4}\left(\frac{2}{3} + \alpha + \beta\right)(\rho_a + \rho_b) + \frac{5}{9} + \frac{\alpha + \beta}{3} - \frac{(\alpha - \beta)^2}{6}$$
$$-\ \frac{1}{12}(\alpha - \beta)^3\ln\frac{\alpha}{\beta} + \frac{1}{12}\lambda^{\frac{1}{2}}[3(1 + \alpha + \beta) - \lambda](\ln x_a + \ln x_b),$$

$$\Pi_L^+(s) = \frac{\alpha + \beta}{2\epsilon} - \frac{1}{4}(\alpha + \beta)(\rho_a + \rho_b) + \alpha + \beta + \frac{1}{2}(\alpha - \beta)^2$$
$$+\ \frac{1}{4}(\lambda - 1)(\alpha - \beta)\ln\frac{\alpha}{\beta} + \frac{1}{4}\lambda^{\frac{1}{2}}(\lambda - 1 - \alpha - \beta)(\ln x_a + \ln x_b),$$

$$\Pi_T^-(s) = \Pi_L^-(s) = \frac{1}{\epsilon} - \frac{\rho_a + \rho_b}{2} + 2 + \frac{\alpha - \beta}{2}\ln\frac{\alpha}{\beta} + \frac{1}{2}\lambda^{\frac{1}{2}}(\ln x_a + \ln x_b), \tag{4.6}$$

where we use the variables

$$\alpha = -\frac{m_a^2}{s}\ ,\ \ \rho_a = \ln\frac{m_a^2}{\mu^2}\ ,\ \ \ x_a = \frac{2\alpha}{1 + \alpha + \beta + \lambda^{\frac{1}{2}}}, \tag{4.7}$$

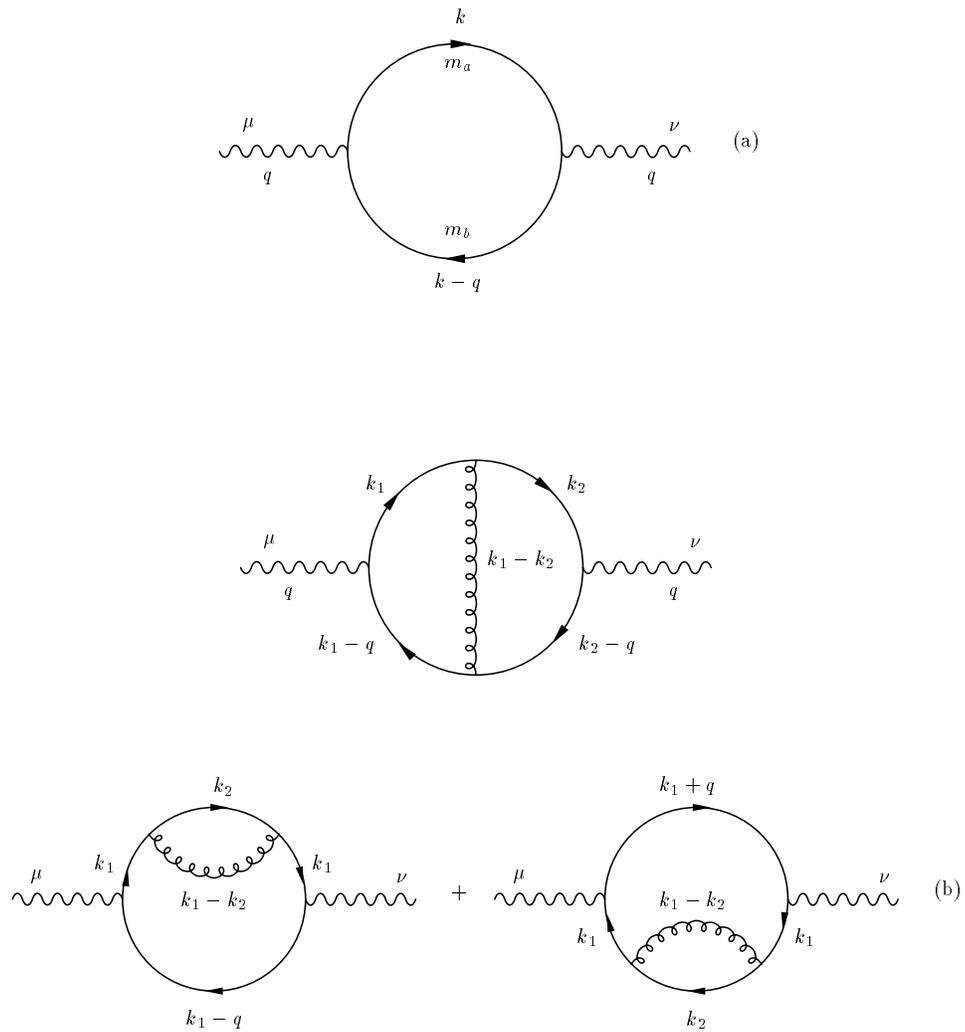

Figure 4.1: Feynman diagrams for the contribution of a quark isodoublet to the vacuum polarization function of a gauge boson at the one-loop (a) and two-loop (b) levels.

$$\lambda = 1 + 2\alpha + 2\beta + (\alpha - \beta)^2, \tag{4.8}$$

and $\beta$, $\rho_b$, and $x_b$ are defined accordingly, exchanging $m_{a,b}$.

In all the previous expressions the momentum transfer has been defined to be in the space-like region, $q^2 < 0$. When continued to the physical region above the threshold for the production of two fermions, $s \geq (m_a + m_b)^2$, the vector boson self-energies acquire imaginary parts, which are related to the decay widths of the vector bosons into fermions. Adding an infinitesimal imaginary part $-i\epsilon'$ to the squared fermion masses, the analytical continuation is consistently defined. The imaginary parts of the vector boson VPF's are discussed in Appendix B.2.

The knowledge of the imaginary part of the polarization function, which can be calculated directly using Cutkosky rules [64], allows an alternative way for obtaining the real part $\mathcal{R}e\Pi(s)$, which can be expressed as a dispersive integral of $\mathcal{I}m\Pi(s)$. The connection between the dispersive approach and the results which are derived using dimensional regularization, will be discussed in Sec. 4.4.

The imaginary parts of Eq. (4.1) are proportional to the decay width of the corresponding vector boson into quarks. This also explains why the expressions of Eq. (4.6) are free of mass singularities in the limit of vanishing quark masses: the real parts can be derived by dispersive integration of their imaginary parts, which are guaranteed to be free of these singularities by a theorem by Kinoshita, and Lee and Nauenberg [61, 62]. Clearly, this must hold true at any order in perturbation theory, and will provide a check of the two-loop expressions.

### 4.1.1  Mass definition and scheme dependence

The two-loop diagrams contributing to the VPF $\Pi_{\mu\nu}^{ij}(q^2)$ at $\mathcal{O}(\alpha\alpha_s)$ are shown in Fig.4.1b. In the 't Hooft-Feynman gauge of QCD, using the routing of momenta shown in the figure and following the notations introduced in the previous section, one can write the bare amplitude as

$$\Pi_{\mu\nu}^{ij}(q^2)\Big|_{\text{bare}} = -\frac{4}{3}\alpha\alpha_s(16\pi^2)\left(\frac{\mu^2 e^{\gamma}}{4\pi}\right)^{2\epsilon} \int \frac{d^n k_1}{(2\pi)^n} \int \frac{d^n k_2}{(2\pi)^n} \left[\mathcal{A}_{\mu\nu}^{ij} + \mathcal{B}_{\mu\nu}^{ij}\right] \tag{4.9a}$$

with

$$\mathcal{A}_{\mu\nu}^{ij} = \text{Tr}\frac{(\slashed{k}_1 + m_a)\gamma_\mu(v^i - a^i\gamma_5)(\slashed{k}_1 - \slashed{q} + m_b)\gamma_\lambda(\slashed{k}_2 - \slashed{q} + m_b)\gamma_\nu(v^j - a^j\gamma^5)(\slashed{k}_2 + m_a)\gamma^\lambda}{(k_1 - k_2)^2 \, (k_1^2 - m_a^2) \, (k_2^2 - m_a^2) \, [(k_1 - q)^2 - m_b^2] \, [(k_2 - q)^2 - m_b^2]}$$

$$\mathcal{B}_{\mu\nu}^{ij} = \text{Tr}\frac{(\slashed{k}_1 + m_a)\gamma_\mu(v^i - a^i\gamma_5)(\slashed{k}_1 - \slashed{q} + m_b)\gamma_\nu(v^j - a^j\gamma^5)(\slashed{k}_1 + m_a)\gamma_\lambda(\slashed{k}_2 + m_a)\gamma^\lambda}{(k_1 - k_2)^2 \, (k_1^2 - m_a^2)^2 \, (k_2^2 - m_a^2) \, [(k_1 - q)^2 - m_b^2]}$$

$$+ (m_a \leftrightarrow m_b) \tag{4.9b}$$

In addition to these amplitudes, the expansion of the bare masses and couplings in the one-loop amplitude provides the necessary counterterms. By virtue of a QED-like Ward identity, the vertex and fermion wave function counterterms cancel each other, so that the electroweak couplings do not get renormalized at $\mathcal{O}(\alpha_s)$, and only quark mass renormalization has to be included. The latter is obtained by considering the diagram shown in Fig.4.2, the amplitude of which reads in dimensional regularization

$$-i\Sigma(\not{p}) = -\alpha_s \frac{16\pi}{3} \left(\frac{e^\gamma \mu^2}{4\pi}\right)^\epsilon \int \frac{d^n k}{(2\pi)^n} \frac{\gamma^\lambda (\not{p} - \not{k} + m)\gamma_\lambda}{[(p-k)^2 - m^2] \, k^2} \tag{4.10}$$

where $p$ is the four-momentum of the quark and $m$ its bare mass. This expression can be decomposed into a piece proportional to $(\not{p} - m)$ which will enter the wave function renormalization and another piece proportional to $m$ which will give the mass counterterm. After integration over the loop momentum, the latter is given by

$$\Sigma_m(p^2) = \frac{\alpha_s}{\pi} \left(\frac{\mu^2 e^\gamma}{m^2}\right)^\epsilon \frac{m}{\epsilon} \frac{1 - 2\epsilon/3}{1 - 2\epsilon} \, \Gamma(1 + \epsilon) \left[1 + \left(1 - \frac{p^2}{m^2}\right) \mathcal{O}(\epsilon)\right]. \tag{4.11}$$

The mass counterterm depends now on the choice of the renormalization condition, i.e. on the definition of the quark mass. For instance, the fermion mass can be defined in a gauge invariant way as the real part of the complex pole position $\bar{s} = m_2 - i\Gamma_2$ of its propagator. In applications, however, it is usually more convenient to use the zero of the real part of the inverse propagator: $m = m_0 + \mathrm{Re}\Sigma(m)$. These two definitions differ by terms of $\mathcal{O}(\Gamma_2^2/m)$, which are completely negligible in view of the foreseen experimental precision on the mass of the top quark. This mass will be denoted simply by $m$, and it is generally referred to as the on-shell, pole, or physical mass; the mass counterterm is then given by

$$\delta m \equiv m - m_0 = \frac{\alpha_s}{\pi} \frac{m}{\epsilon} \left(\frac{\mu^2}{m^2}\right)^\epsilon \frac{1 - 2\epsilon/3}{1 - 2\epsilon} \left(1 + \frac{\pi^2}{12}\epsilon^2 + \dots\right), \tag{4.12}$$

where the ellipses stand for higher order terms in $\epsilon$. Pole masses imply the possibility of defining asymptotic states of the corresponding fermion fields. Although this is perfectly reasonable for leptons, the confinement of quarks makes this concept much less suitable to quark masses. A clear indication that in QCD the pole mass is not a well-defined concept will be described in Sec. 4.5. However, in the case of heavy quarks, the definition of mass as the pole of the propagator maintains a validity. Indeed, the lifetime of heavy quarks, and in particular of the top quark, is much smaller than the time-scale involved in strong interaction (of the order of $1/\Lambda_{QCD}$). Equivalently, the top decay width $\Gamma_t$, of the order of 1-2GeV, is much larger than $\Lambda_{QCD} \approx 200 - 300\mathrm{MeV}$. Moreover, the pole mass appears to be the natural parameter in the experimental determination of the top mass through kinematic reconstruction currently carried out at Fermilab [1,2]. From a formal point of view, the pole mass is the mass parameter implicitly used in the evaluation of VPF's through dispersion relations.

Another possibility is the use of the $\overline{\mathrm{MS}}$ scheme in which the mass is defined by simply removing the divergent term in the expression of $\Sigma_m(m^2)$ (the related constants $\ln 4\pi, \gamma, \dots$ have already been absorbed in the normalization). The $\overline{\mathrm{MS}}$ mass $\hat{m}(\mu)$ is related to the pole mass (considering only QCD effects) by [65,66]

$$m = \hat{m}(m) \left[1 + \frac{4}{3}\frac{\alpha_s}{\pi} + K \frac{\alpha_s^2}{\pi^2} + \mathcal{O}(\alpha_s^3)\right], \tag{4.13}$$

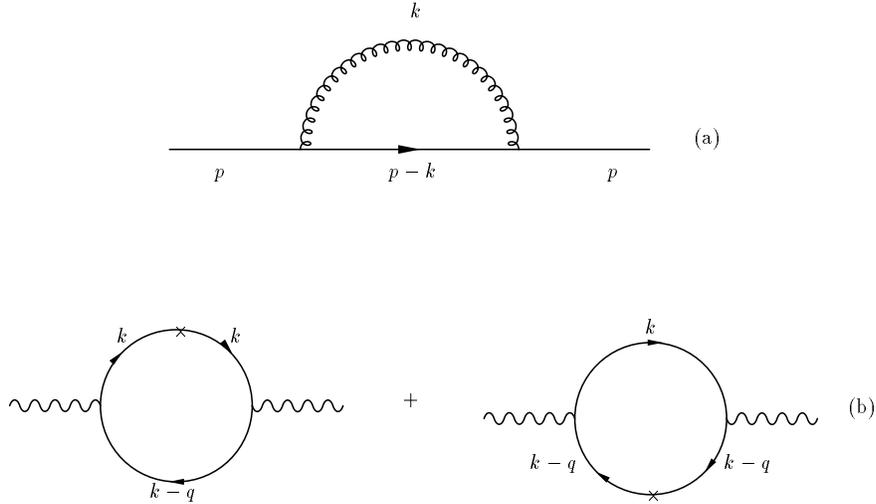

Figure 4.2: Feynman diagrams for the one-loop quark self-energy diagram (a) and for the contribution of the mass counterterm to the VPF at two-loop level (b).

where $K = 10.96$ for the top quark, 12.5 for the bottom, etc. A study of the relation between fermion pole mass and $\overline{\text{MS}}$ Yukawa coupling in the SM can be found in Ref. [63].

What is relevant here is that the definition of mass is completely arbitrary and there may be different legitimate choices, depending also on the problem at hand. Once the choice is made one inserts the counterterm in the one-loop self-energies, as depicted in the diagrams of Fig.4.2b, which is equivalent to calculate

$$\Pi_{\mu\nu}^{ij}(q^2)\Big|_{\text{CT}} = -\delta m_a \frac{\partial}{\partial m_a} \Pi_{\mu\nu}^{ij}(q^2)\Big|_{1-\text{loop}} - \delta m_b \frac{\partial}{\partial m_b} \Pi_{\mu\nu}^{ij}(q^2)\Big|_{1-\text{loop}} \qquad (4.14)$$

where the one-loop vacuum polarization function is given by Eq. (4.6). The renormalized two-loop self-energies will then read

$$\Pi_{\mu\nu}^{ij}(q^2) = \Pi_{\mu\nu}^{ij}(q^2)\Big|_{\text{bare}} + \Pi_{\mu\nu}^{ij}(q^2)\Big|_{\text{CT}} \qquad (4.15)$$

Having at hand the result of the vacuum polarization function in the on-shell scheme, one can obtain the polarization function in any other scheme X: one simply has to add to the expression of the two-loop self-energy Eq. (4.15) the quantity[1]

$$\Delta\Pi_{\mu\nu}^{ij}(q^2)\Big|_X = \Delta m_a^X \frac{\partial}{\partial m_a} \Pi_{\mu\nu}^{ij}(q^2)\Big|_{1-\text{loop}} + \Delta m_b^X \frac{\partial}{\partial m_b} \Pi_{\mu\nu}^{ij}(q^2)\Big|_{1-\text{loop}} \qquad (4.16)$$

---

[1] Note that since at this stage we are only discussing the difference between two renormalization schemes, we do not need the $\mathcal{O}(\epsilon)$ terms in the one-loop result of the vector boson self-energies. These terms are of course needed to evaluate the renormalized two-loop self-energies in a given scheme.

where $\Delta m_{a,b}^{X} = m_{a,b} - m_{a,b}^{X}$ is the difference between the pole mass $m_{a,b}$ and the quark mass $m_{a,b}^{X}$ defined in the scheme $X$.

Decomposing the one-loop transverse and longitudinal components as in Eq. (4.4), one obtains for the derivative of these components

$$
\begin{aligned}
m_a \frac{\partial}{\partial m_a} \left. \Pi_T^+(s) \right|_{1-\mathrm{loop}} = & \; \frac{\alpha}{\epsilon} - \frac{1}{2}\alpha(\rho_a + \rho_b) + \alpha(\beta - \alpha) - \frac{1}{2}\alpha(\alpha - \beta)^2 \ln\frac{\alpha}{\beta} \\
& + \frac{1}{2}\lambda^{\frac{1}{2}} \; \alpha(1 - \alpha + \beta)(1 - 2\beta\lambda^{-1})(\ln x_a + \ln x_b),
\end{aligned}
$$

$$
\begin{aligned}
m_a \frac{\partial}{\partial m_a} \left. \Pi_L^+(s) \right|_{1-\mathrm{loop}} = & \; \frac{\alpha}{\epsilon} - \frac{1}{2}\alpha(\rho_a + \rho_b) + \alpha(2 - 3\beta + 3\alpha) \\
& + \frac{1}{2}\alpha[3(\alpha - \beta)^2 + 4\alpha]\ln\frac{\alpha}{\beta} \\
& + \frac{\alpha}{2}\lambda^{\frac{1}{2}} \left[ 1 + 3(\alpha - \beta) + 2\frac{\beta}{\lambda}(1 + \beta - \alpha) \right] (\ln x_a + \ln x_b),
\end{aligned}
\tag{4.17}
$$

$$
\begin{aligned}
m_a \frac{\partial}{\partial m_a} \; m_a m_b \left. \Pi_{T,L}^-(s) \right|_{1-\mathrm{loop}} = & \; m_a m_b \left[ \frac{1}{\epsilon} - \frac{1}{2}(\rho_a + \rho_b) + 2 + \frac{1}{2}(3\alpha - \beta)\ln\frac{\alpha}{\beta} \right. \\
& \left. + \frac{1}{2}\lambda^{\frac{1}{2}} \left[ 1 + 2\alpha\lambda^{-1}(1 + \alpha - \beta) \right] (\ln x_a + \ln x_b) \right],
\end{aligned}
$$

and similarly for the piece involving $m_b$ which can be obtained by making the substitution $m_a \leftrightarrow m_b$. Once the complete result of the VPF is known in the on-shell scheme, it is therefore straightforward to obtain the corresponding results in any renormalization scheme using Eqs. (4.16,4.17). In practice, the most useful cases are the ones with equal quark masses, corresponding to the $Z$ boson, and with one nearly massless quark, corresponding to the $(t-b)$ doublet contribution to the $W$ boson VPF. The explicit expressions for these two cases are (up to a factor $N_c \alpha/\pi \; v^2(a^2)$)

$$
\Delta^X \Pi_{T, m_a = m_b}^{V} = 4 m_a^2 \; \frac{\Delta m_a^X}{m_a} \left( 1 + \frac{2\alpha}{\sqrt{1 + 4\alpha}} \ln\frac{4\alpha}{(1 + \sqrt{1 + 4\alpha})^2} \right)
$$

$$
\Delta^X \Pi_{T, m_a = m_b}^{A} = -4 m_a^2 \; \frac{\Delta m_a^X}{m_a} \left( \frac{1}{\epsilon} + 1 - \rho_a + \sqrt{1 + 4\alpha} \ln\frac{4\alpha}{(1 + \sqrt{1 + 4\alpha})^2} \right)
$$

$$
\Delta^X \Pi_{T, m_b = 0}^{A, V} = -m_a^2 \; \frac{\Delta m_a^X}{m_a} \left( \frac{1}{\epsilon} - \rho_a - \alpha + (1 - \alpha^2) \; \ln\frac{\alpha}{1 + \alpha} \right)
\tag{4.18}
$$

A few remarks are necessary at this stage.

First, one notices in the previous expressions the occurrence of terms that are inversely proportional to the velocity factor $\lambda^{1/2}$. In principle, after an analytical continuation to the physical region beyond the threshold for quark pair production, $s \geq (m_a + m_b)^2$, these terms would diverge for energy values near the production threshold, $\lambda \sim 0$. However, as we will see later, when evaluated in the on-shell scheme, which is the only scheme where the physical threshold is well defined, the renormalized two-loop polarization function is free of these $\lambda^{-1/2}$ factors. Near threshold, the dominant terms will be constant and would correspond, once the vector boson self-energy is normalized to its one-loop value, to the well known $\lambda^{-1/2}$ Coulomb singularities which require a non-perturbative treatment [62]. Of course, the $\lambda^{-1/2}$ terms can be present in the vacuum polarization function when it is evaluated in a different scheme but in this case the threshold is not well defined since the masses are not "physical" masses.

In principle one can even define the masses $m_a$ and $m_b$ in two completely different schemes; this could be useful if, for instance, different scales are involved in the evaluation of the VPF's. In the top-bottom isodoublet, one could employ the on-shell mass for the heavy top quark, and use a running mass evaluated at the scale $q^2$ for the relatively light $b$ quark, which in general avoids the appearance of large logarithms for $q^2 \gg m_b^2$ (see the example at the end of App. B). In the rest of this section, however, I will employ exclusively on-shell masses.

### 4.1.2  Exact two-loop results

In this section I give the expression of the vacuum polarization function at order $\mathcal{O}(\alpha\alpha_s)$ in the general case $m_a \neq m_b \neq 0$ and for arbitrary momentum transfer [59]. The result will be expressed in terms of on-shell quark masses. I will use the same notation previously introduced at the one-loop level, and $m_{a,b}$ will stand for the on-shell masses.

At $O(\alpha\alpha_s)$, the transverse and longitudinal components of the vacuum polarization function $\Pi_{\mu\nu}^{ij}(q^2)$ can be written as (the color factor $N_c = 3$ is now included[2])

$$\Pi_{T,L}^{ij}(s) = \frac{\alpha}{\pi}\frac{\alpha_s}{\pi}\left[\left(v^i v^j + a^i a^j\right) s\, \Pi_{T,L}^+(s) \;+\; \left(v^i v^j - a^i a^j\right) m_a m_b\, \Pi_{T,L}^-(s)\right], \quad (4.19)$$

where $\Pi_{T,L}^{\pm}$ are the sum of the corresponding components in the bare two-loop amplitude Eqs. (4.9) and the mass counterterm which can be obtained from the preceding section.

In analogy with the one-loop case, after contracting $\Pi_{\mu\nu}^{ij}$ in Eqs. (4.9) by the tensors $(g^{\mu\nu} - q^\mu q^\nu/q^2)$ and $q^\mu q^\nu/q^2$ and expressing the scalar products of momenta appearing in the numerators in terms of combinations of the polynomials in the denominators, one is led to the calculation of a set of scalar two-loop integrals. Most of these integrals have been first calculated by Broadhurst in Ref. [133]; the remaining integrals reduce after straightforward computations to the previous ones [55, 134] and are all listed in App. A.1. After a cumbersome computation, and taking advantage of the symmetry in the change $\alpha \leftrightarrow \beta$, we can write $\Pi_{T,L}^{\pm}(s)$ in a relatively simple and compact form[3]

$$\begin{aligned}
\Pi_T^+ = \Big\{ &-\frac{3\alpha}{2\epsilon^2} + \frac{1}{\epsilon}\left(\frac{1}{4} + 3\alpha\rho_a - \frac{11}{4}\alpha\right) + \frac{(\rho_a + \rho_b)}{4}\left(11\alpha - 1 - 9\alpha\rho_a + 3\alpha\rho_b\right) \\
&+ \frac{55}{24} - \frac{71}{24}\alpha - \frac{5}{6}\alpha^2 + \frac{11}{6}\alpha\beta + \frac{2}{3}\alpha[G(x_b) - G(x_a)] \\
&+ \frac{\ln x_a}{12}\left[\left(\alpha - \beta + \lambda^{\frac{1}{2}}\right)\left(11 + 19\alpha + 19\beta + 12\alpha\beta - 5\lambda\right) + (\alpha - \beta) \right.\\
&\left. \times \left(42 - 5\alpha - 5\beta\right)\right] + \frac{\alpha}{2}\ln x_a \ln x_b\left[3 + \beta(4 + 2\alpha)\right]
\end{aligned}$$

---

[2]The case of $SU(N)$ with $N \neq 3$ can be obtained by multiplying our result by $(N^2 - 1)/8$, while for QED it is sufficient to multiply by $1/4$ for leptons, and by $3/4\ Q_i Q_j$ for quarks, $Q_i$ being the charge of the $i$ quark in unit of the positron charge.

[3]Minor differences with the expressions in Ref. [59] are due to trivial algebraic simplifications.

$$+ \frac{\ln^2 x_a}{12} \left[ (1 - 3\alpha - 3\beta) \left( \lambda - 1 - \alpha - \beta + (\alpha - \beta)\lambda^{\frac{1}{2}} \right) - 9(\alpha + \beta) + 8\alpha\beta \right]$$

$$- \frac{\pi^2}{4}\alpha + \frac{1}{3} \left[ (\alpha - 1)\lambda - 6\alpha\beta \right] \mathcal{I} - \frac{1}{3} \left[ 3(1 + 2\alpha) - \lambda \right] \mathcal{I}' \Big\} \quad + \quad \{\alpha \leftrightarrow \beta\} \, ;$$

$$\Pi_T^- = \left\{ -\frac{3}{2\epsilon^2} + \frac{1}{\epsilon} \left( 3\rho_a - \frac{11}{4} \right) + \frac{11}{2}\rho_a - \frac{3}{4}(\rho_a + \rho_b)^2 + \frac{11}{8} + \alpha - \frac{\pi^2}{4} \right.$$

$$+ \frac{\ln x_a}{2} \left[ \alpha - \beta + (\alpha + \beta + 9)(\alpha - \beta + \lambda^{\frac{1}{2}}) \right] + \frac{\ln x_a \ln x_b}{2} \left[ 3 + 6\alpha + 2\alpha\beta \right]$$

$$+ \frac{\ln^2 x_a}{2} \left[ 1 + \alpha + \beta - \lambda + (\beta - \alpha)\lambda^{\frac{1}{2}} \right] - (1 + 2\alpha)\mathcal{I} - 2\mathcal{I}' \Big\} \quad + \quad \{\alpha \leftrightarrow \beta\} \, ;$$

$$\Pi_L^+ = \left\{ -\frac{3\alpha}{2\epsilon^2} + \frac{\alpha}{\epsilon} \left( 3\rho_a - \frac{11}{4} \right) + \frac{\alpha}{4}(\rho_a + \rho_b)\left( 11 - 9\rho_a + 3\rho_b \right) + \frac{3}{8}\alpha + \frac{7}{2}\alpha^2 \right.$$

$$+ \frac{9}{4}\ln x_a \left[ \left( \alpha - \beta + \lambda^{\frac{1}{2}} \right) \left( \lambda - 1 - \alpha - \beta - \frac{4}{3}\alpha\beta \right) + (\alpha - \beta)\left( \alpha + \beta + \frac{20}{9} \right) \right]$$

$$+ \frac{\ln^2 x_a}{4} \left[ \lambda(2\lambda - 2 - \alpha - \beta) - 3\alpha - 3\beta - 16\alpha\beta + (2\lambda + 1 + \alpha + \beta)(\alpha - \beta)\lambda^{\frac{1}{2}} \right]$$

$$- \frac{13}{2}\alpha\beta - \frac{\pi^2}{4}\alpha + \frac{3\alpha}{2}\ln x_a \ln x_b (1 - 2\alpha\beta) + \alpha\left[ 2\beta - \lambda \right] \mathcal{I} + (1 + 2\alpha - \lambda)\mathcal{I}' \right\}$$

$$+ \quad \{\alpha \leftrightarrow \beta\} \, ;$$

$$\Pi_L^- = \left\{ -\frac{3}{2\epsilon^2} + \frac{1}{\epsilon} \left( 3\rho_a - \frac{11}{4} \right) + \frac{11}{2}\rho_a - \frac{3}{4}(\rho_a + \rho_b)^2 + \frac{3}{8} - 3\alpha - \frac{\pi^2}{4} \right.$$

$$- \frac{3}{2}\ln x_a \left[ \alpha - \beta + (\alpha + \beta - 3)(\alpha - \beta + \lambda^{\frac{1}{2}}) \right] - (1 + 2\alpha)\mathcal{I} - 2\mathcal{I}'$$

$$+ \frac{\ln^2 x_a}{2} \left[ 1 + \alpha + \beta - \lambda + (\beta - \alpha)\lambda^{\frac{1}{2}} \right] + \frac{3}{2}\ln x_a \ln x_b \left[ 1 + 2\alpha - 2\alpha\beta \right] \right\}$$

$$+ \{\alpha \leftrightarrow \beta\} \, , \tag{4.20}$$

with $\mathcal{I}$ and $\mathcal{I}'$ given by

$$\mathcal{I} = F(1) + F(x_a x_b) - F(x_a) - F(x_b)$$

$$\mathcal{I}' = \lambda^{\frac{1}{2}} G(x_a x_b) - \frac{1}{2}(\beta - \alpha + \lambda^{\frac{1}{2}}) G(x_a) - \frac{1}{2}(\alpha - \beta + \lambda^{\frac{1}{2}}) G(x_b) \tag{4.21}$$

In terms of the polylogarithmic functions [135] $\mathrm{Li}_2(x) = -\int_0^1 y^{-1} \ln(1 - xy) \mathrm{d}y$ and $\mathrm{Li}_3(x) = -\int_0^1 y^{-1} \ln y \ln(1 - xy) \mathrm{d}y$, the functions $F$ and $G$ are given by

$$F(x) = 6\mathrm{Li}_3(x) - 4\mathrm{Li}_2(x)\ln x - \ln^2 x \ln(1 - x)$$

$$G(x) = 2\mathrm{Li}_2(x) + 2\ln x \ln(1 - x) + \frac{x}{1 - x}\ln^2 x \tag{4.22}$$

or also

$$F(x) = \sum_{n=1}^{\infty} [(2 - n \ln x)^2 + 2] \ x^n / n^3.$$

These two functions also admit a simple and useful integral representation [133]

$$F(x) = \int_0^x dy \left( \frac{\ln y}{1-y} \right)^2 \ln \frac{x}{y} \quad , \quad G(x) = x \ F'(x) = \int_0^x dy \left( \frac{\ln y}{1-y} \right)^2 \qquad (4.23)$$

The imaginary parts of the VPF are derived along the same lines as in the one-loop case, and are given in App B.2. Incidentally, we note that the analytic structure of the VPF is the same as at one-loop level, with a unique threshold at $(m_a + m_b)^2$. This is obviously due to the masslessness of the gluon. In general, at two-loop level the presence of new virtual massive particles in the loops prompts new thresholds, and it is this feature that complicates the evaluation of two-loop massive integrals. For instance, it has been noted [67] that two-loop scalar integrals can be expressed in terms of polylogarithmic functions only in a very restricted set of cases including the one we are considering (see also App. A).

The gluonic corrections to the hadronic widths of vector bosons in the case of arbitrary quark masses and for both transverse and longitudinal components have been known for some time [56, 68, 69] (in the first reference only the transverse part is given) and were obtained by directly calculating the QCD corrections to the flavor changing decay of a vector boson. These calculations were mostly motivated, in the context of QCD sum-rules, by the need to calculate at first order in perturbative QCD the Wilson coefficients of dimension two operators. The results presented here were obtained using a completely different method, and agree with those of Ref. [56, 68], and with Ref. [69], if some misprints in the integrals of their Appendix ($J_1$ and $J_2$) are corrected (see also Ref. [70]). This is a very important check of our calculation.

In App. B the asymptotic expressions for the most important special cases, corresponding to two equal mass quarks, one massless quark, and $q^2 \gg m_{a,b}^2$ are presented in detail. They represent excellent approximations of the case in which we are going to be mostly interested: the $(t-b)$ doublet, where the mass of the bottom can be safely set to zero as in most applications the neglected terms will be of $\mathcal{O}(m_b^2/q^2) = \mathcal{O}(m_b^2/m_W^2)$. The case of vanishing external momentum is discussed in the next section. We also verify that in the limit of vanishing quark mass, the self-energies are free of mass singularities as required by the mass singularity theorem [61, 62].

Even though the real parts of Eq. (4.20) are a new result, a number of partial results allow us several useful checks of the calculation. In the equal mass case, $\Pi_T^V(s)$ has been derived in the 1950s by the authors of Ref. [58] by means of a dispersive approach, and, more recently, $\Pi_T^{V,A}(s)$ have been calculated in Ref. [56] by the same method. Real parts of the transverse components in the cases $m_b = 0$ and $m_a = m_b$ had been previously derived using dimensional regularization in Ref. [54, 55] [4]. Finally, the expressions of $\Pi_T^{V,A}(s)$ in these special cases has also been obtained in Ref. [57], again by means of dispersive integration.

---

[4]As explained in Ref. [59], the expressions of Ref. [54, 55] differ from the ones reported in App. B by a constant term proportional to $\pi^2$, due to an omission in the mass counterterm used in that calculation. This does not affect most physical observables.

### 4.1.3 $\Delta\rho$ and a fourth generation of quarks

An immediate application of the expressions of the previous section is the calculation of the effects of a heavy quark isodoublet with SM couplings on the $\rho$ parameter, which measures the ratio of neutral and charged couplings at zero momentum transfer.

The fermionic one-loop contribution to the $\rho$ parameter can be expressed in terms of the difference between the transverse components of the $Z$ and $W$ bosons self-energies at zero-momentum transfer, $q^2 = 0$

$$\Delta\rho = \frac{A_{WW}(0)}{M_W^2} - \frac{A_{ZZ}(0)}{M_Z^2} = \frac{\Pi_T^{ZZ}(0)}{M_Z^2} - \frac{\Pi_T^{WW}(0)}{M_W^2}. \tag{4.24}$$

The contribution of a quark pair with different masses, $m_a \neq m_b \neq 0$, to $\Pi_{T,L}^{V,A}(0)$ in the one-loop approximation is

$$\Pi_T^{V,A}(0) = \Pi_L^{V,A}(0) = -\frac{1}{4}\left[(m_a^2 + m_b^2)\left(\frac{2}{\epsilon} + 1 - \rho_a - \rho_b\right) + \frac{m_a^4 + m_b^4}{m_a^2 - m_b^2}\ln\frac{m_b^2}{m_a^2}\right]$$
$$\pm\, m_a m_b\left[\frac{1}{\epsilon} + 1 - \frac{1}{2}(\rho_a + \rho_b) - \frac{1}{2}\frac{m_a^2 + m_b^2}{m_a^2 - m_b^2}\ln\frac{m_a^2}{m_b^2}\right], \tag{4.25}$$

where the upper (lower) sign refers to vector (axial) current. Note that, in this limit, the longitudinal and transverse components of the self-energies are equal. This is a trivial consequence of the analyticity of $\Pi^{\mu\nu}(q^2)$ at $q^2 = 0$; in particular, the coefficient of $q^\mu q^\nu$ has to be regular in the limit $q^2 \to 0$, thus forcing the identity $\Pi_T^{V,A}(0) = \Pi_L^{V,A}(0)$. Similarly, the conservation of equal mass vector current prompts a Ward identity which implies $\Pi^V(0) = 0$ for $m_a = m_b$. $\Delta\rho$ can therefore be written in terms of either the longitudinal or the transverse components of the VPF:

$$\Delta^{(1)}\rho = \frac{\sqrt{2}G_\mu}{8\pi^2}f_{L,T}^{(1)}(s = 0, m_a, m_b), \tag{4.26}$$

with

$$f_{L,T}(s, m_a, m_b) = \sum_{i=a,b}\sum_{j=A,V}\frac{\Pi_{L,T}^j(s, m_i, m_i)}{2} - \Pi_{L,T}^V(s, m_a, m_b) - \Pi_{L,T}^A(s, m_a, m_b), \tag{4.27}$$

where I have kept some $\Pi^V(s, m, m)$ terms even if they do not contribute to $\Delta\rho$ for future convenience. Using the previous expressions for $\Pi_{T,L}^{V,A}(0)$ one obtains the one-loop correction, which is quadratic in the fermion masses [71]

$$f^{(1)}(0, m_a, m_b) = \frac{N_c}{2}\left[m_a^2 + m_b^2 + \frac{2m_a^2 m_b^2}{m_a^2 - m_b^2}\ln\frac{m_b^2}{m_a^2}\right]. \tag{4.28}$$

At $\mathcal{O}(\alpha\alpha_s)$, the last equation gets modified in the case of virtual quarks. In the limit of zero momentum transfer, the expressions of $\Pi_{L,T}^{V,A}(s)$ simplify considerably, and one obtains from

Eqs. (4.20) $\left(\Pi_T^{V,A}(0) = \Pi_L^{V,A}(0) = \Pi^{V,A}(0)\right)$

$$\begin{aligned}
\Pi^{V,A}(0) = &\ (m_a^2 + m_b^2)\left(\frac{3}{2\epsilon^2} + \frac{11}{4\epsilon} + \frac{35}{8} + \frac{\pi^2}{4}\right) - \left(\frac{3}{\epsilon} + \frac{11}{2}\right)(m_a^2\rho_a + m_b^2\rho_b) \\
&+ 3m_a^2\rho_a^2 + 3m_b^2\rho_b^2 + \frac{(m_a^2 - m_b^2)}{4}\left[G\left(\frac{m_a^2}{m_b^2}\right) - G\left(\frac{m_b^2}{m_a^2}\right)\right] \\
&+ \frac{m_a^2 m_b^2}{m_a^2 - m_b^2}\ln\frac{m_a^2}{m_b^2} - m_a^2 m_b^2\frac{m_a^2 + m_b^2}{(m_a^2 - m_b^2)^2}\ln^2\frac{m_a^2}{m_b^2} \\
\pm &\ m_a m_b\left[-\frac{3}{\epsilon^2} + \frac{1}{\epsilon}\left(3\rho_a + 3\rho_b - \frac{11}{2}\right) - \frac{3}{2}(\rho_a + \rho_b)^2 + 4(\rho_a + \rho_b)\right. \\
&- \left.\frac{31}{4} + 3\frac{m_a^2\rho_b - m_b^2\rho_a}{m_a^2 - m_b^2} + 3\frac{m_a^2 m_b^2}{(m_a^2 - m_b^2)^2}\ln^2\frac{m_a^2}{m_b^2} - \frac{\pi^2}{2}\right],
\end{aligned} \tag{4.29}$$

where the function $G(x)$ has been introduced in the previous section, and again the upper (lower) sign refers to vector (axial) current.

In the two special cases $m_a = m_b$ and $m_b = 0$ the last equation simplifies further to

$$\Pi^A(0, m_a = m_b) = m_a^2\left[\frac{6}{\epsilon^2} + \frac{1}{\epsilon}(11 - 12\rho_a) + 12\rho_a^2 - 22\rho_a + \frac{31}{2} + \pi^2\right] \tag{4.30}$$

$$\Pi^{V,A}(0, m_b = 0) = \frac{m_a^2}{4}\left[\frac{6}{\epsilon^2} + \frac{1}{\epsilon}(11 - 12\rho_a) + 12\rho_a^2 - 22\rho_a + \frac{35}{2} + \frac{5}{3}\pi^2\right]$$

As expected, there is no singularity in the self-energies in this limit. In addition, besides the manifest symmetry in the exchange $m_a \leftrightarrow m_b$, the previous formulas exhibit the fact that $\Pi^{V,A}(0)$ can be obtained from $\Pi^{A,V}(0)$ by simply making the substitution $m_a \to -m_a$ (or $m_b \to -m_b$) as expected from $\gamma_5$ reflection symmetry.

Using the previous expression, one readily obtains the QCD corrections to the contribution of a heavy quark isodoublet to the $\rho$ parameter in the general case $m_a \neq m_b \neq 0$. Defining the contribution to the $\rho$ parameter in analogy with Eq. (4.26)

$$\Delta^{(2)}\rho = \frac{\sqrt{2}G_\mu}{8\pi^2}\frac{\alpha_s}{\pi}f^{(2)}(0, m_a, m_b) \tag{4.31}$$

the function $f^{(2)}$, defined as the two-loop analogue of $f^{(1)}$, is given for $N_c = 3$ by

$$\begin{aligned}
f^{(2)}(0, m_a, m_b) = -&\left\{m_a^2 + m_b^2 + 2\ \frac{m_a^2 m_b^2}{m_a^2 - m_b^2}\ln\frac{m_a^2}{m_b^2}\left[1 + \frac{m_a^2 + m_b^2}{m_a^2 - m_b^2}\ln\frac{m_b^2}{m_a^2}\right]\right. \\
&+ (m_a^2 - m_b^2)\left[2\text{Li}_2\left(\frac{m_a^2}{m_b^2}\right) + 2\ln\frac{m_a^2}{m_b^2}\ln\left(1 - \frac{m_a^2}{m_b^2}\right)\right. \\
&- \left.\left.\frac{m_a^2}{m_a^2 - m_b^2}\ln^2\frac{m_a^2}{m_b^2} - \frac{\pi^2}{3}\right]\right\}
\end{aligned} \tag{4.32}$$

Notice that $f^{(2)}(0, m_a, m_b)$ is free of ultraviolet divergences as it should be since $\Delta\rho$ is an observable physical quantity, and that it does not depend on the 't Hooft mass scale $\mu$. Note also that the cofactor of $m_a^2 - m_b^2$ in Eq.(4.32) is simply $(G(m_a^2/m_b^2) - G(m_b^2/m_a^2))/2$, which makes the symmetry in the interchange of $m_a$ and $m_b$ explicit (we have used the identity $G(1/x)+G(x) = 2\pi^2/3$). In the limit of large mass splitting between the two quarks, $m_a \gg m_b$, the QCD corrections to the $\rho$ parameter reduce to the known result for $m_b = 0$ [54,55]

$$\Delta\rho = \frac{3G_\mu\sqrt{2}}{16\pi^2}\ m_a^2\ \left[1 - \frac{\alpha_s}{\pi}\left(\frac{6+2\pi^2}{9}\right)\right], \qquad (4.33)$$

where it is apparent that the two-loop QCD correction significantly "screens" the one-loop effect (for $\alpha_s = 0.11$, which roughly corresponds to $\alpha_s$ at the top mass scale, the QCD correction amounts to about 10%).

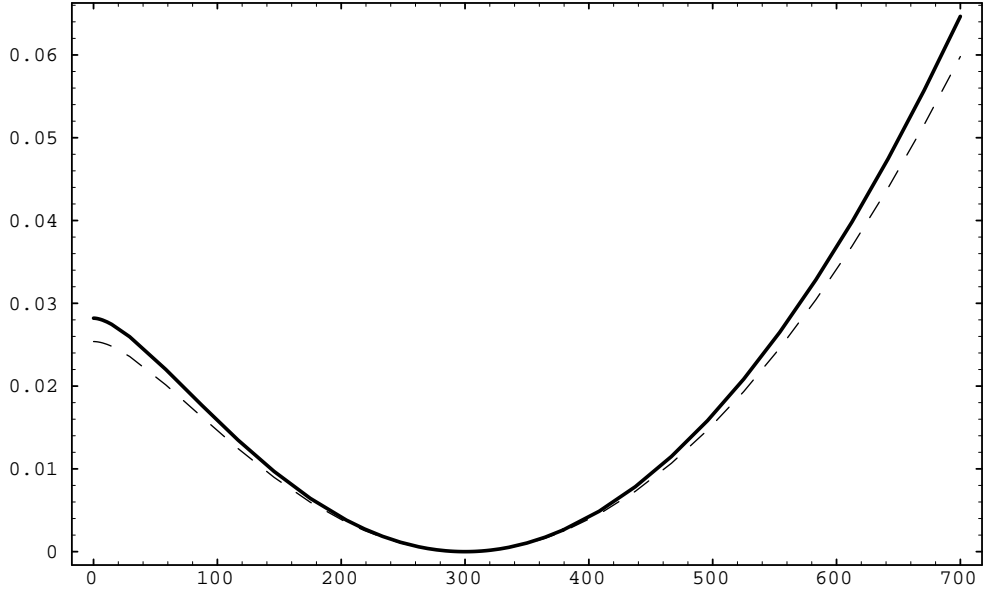

Figure 4.3: Contribution of a fourth generation of quarks to $\Delta\rho$, in the case of one quark mass equal to 300GeV and $\alpha_s = 0.11$, as a function of the other quark mass (GeV). The full line represents the one-loop contribution; the dashed line incorporates also two-loop $\mathcal{O}(\alpha\alpha_s)$ terms.

The contribution of a new quark isodoublet to the $\rho$ parameter is exemplified in Fig. 4.3 both in the one-loop approximation (full line) and including the QCD radiative corrections (dashed line). I have set the mass of one quark to 300 GeV and varied the mass of the other quark from 0 to 700 GeV; for the strong coupling constant I have used the numerical value $\alpha_s \simeq 0.11$ (see Sect. 4.4 for a discussion of the choice of scale). As one can see, the contribution of the heavy quark isodoublet to $\Delta\rho$ vanishes for degenerate quarks, $m_a = m_b$. This is expected, because the deviation of the $\rho$ from 1 is due to a violation of the SU(2) custodial symmetry [72] which can be induced only by hypercharge or by isospin violation in the Yukawa sector. In our case, only the latter actually contributes, and we see that the QCD corrections are always negative, thus screening the one-loop contribution to $\Delta\rho$. In fact, in this mass range the QCD corrections are practically constant and amount to a decrease of $\Delta\rho$ by approximately 7-8%, as can be seen from Fig.4.4. In the case of nearly degenerate quarks, Eqs.(4.31,4.32) reduce to

$$\Delta\rho = \frac{\sqrt{2}G_\mu}{4\pi^2} \left[ (m_a - m_b)^2 \ \left( 1 - \frac{2\alpha_s}{\pi} \right) + \ \mathcal{O}\left( \frac{(m_a - m_b)^4}{m_a^2} \right) \right]. \tag{4.34}$$

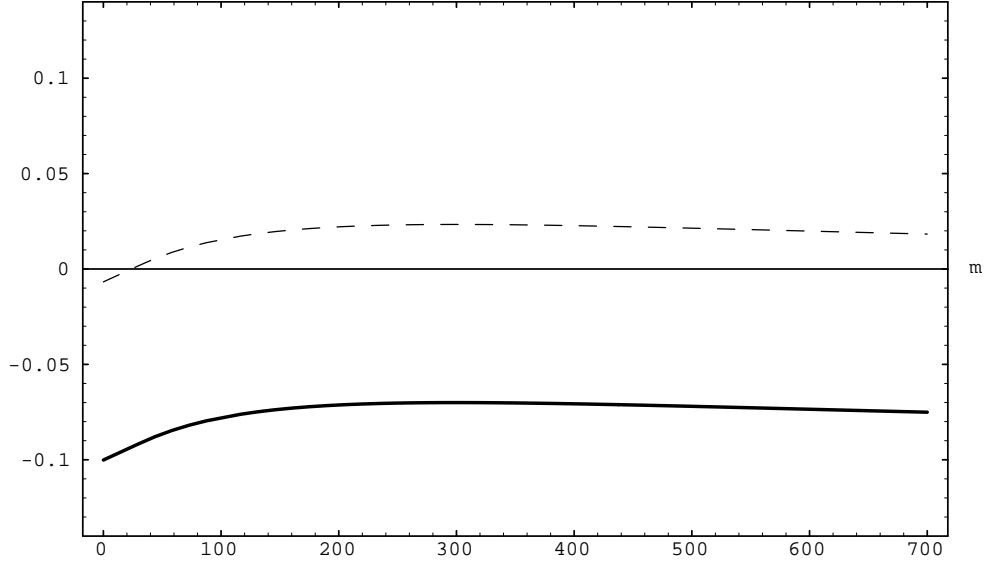

Figure 4.4: QCD correction to the contribution of a fourth generation of quarks to $\Delta\rho$, normalized to the one-loop contribution. One quark mass is set to 300GeV and the other one is varied from 0 to 700GeV; $\alpha_s = 0.11$. The dashed line represents the same correction when the one-loop contribution is expressed in terms of $\overline{\text{MS}}$ masses.

It is interesting to see what happens if one uses a different definition of mass: for example, if we express the one-loop correction in terms of $\overline{\text{MS}}$ masses $\hat{m}_{a,b}(m_{a,b})$, using Eq.(4.13) and neglecting higher orders in $\alpha_s$, we obtain an extra term in $\mathcal{O}(\alpha\alpha_s)$ equal to $\frac{8}{3}\frac{\alpha_s}{\pi}$ times the one-loop correction. In the case of one vanishing mass, Eq.(4.33) is modified into

$$\Delta\rho = \frac{3G_\mu\sqrt{2}}{16\pi^2} \ \hat{m}_a^2 \ \left[ 1 - 2\frac{\alpha_s}{\pi}\left( \frac{\pi^2}{9} - 1 \right) \right]. \tag{4.35}$$

In this way the QCD correction to $\Delta\rho$ is much smaller, about 0.7%; the comparison with the on-shell case is shown in Fig.4.4.

For what concerns the top-bottom doublet, I illustrate the effect of keeping $m_b \neq 0$ for $m_t = 200$GeV, and using $\alpha_s = 0.11$. From Eq.(4.33) we see that the one-loop contribution in the approximation of neglecting the mass of the bottom amounts to 1.2535%, while using Eqs.(4.26-4.28) with $m_b = 5$GeV we obtain 1.2427%. The $\mathcal{O}(\alpha_s)$ terms in Eq.(4.33) and (4.31-4.32) add to these results $-0.1255\%$ and $-0.1225\%$, respectively. We conclude that the finite bottom mass effect on the one-loop $\Delta\rho$ amounts to about $10^{-4}$, while the effect on the QCD corrections to $\Delta\rho$ is a mere $3{\times}10^{-5}$. In other words, a 5 GeV bottom mass decreases the QCD corrections to the one-loop formula by about $2.4\%$ for $m_t = 200$ GeV, roughly of the order $m_b/m_t$, as an effect of terms like $\frac{m_b^2}{m_t^2}\ln^2\frac{m_b^2}{m_t^2}$. These contributions are well beyond the accuracy of current analyses. The finite bottom mass effect is illustrated in Fig.4.5 , in the range $150\,\mathrm{GeV} < m_t < 220\,\mathrm{GeV}$.

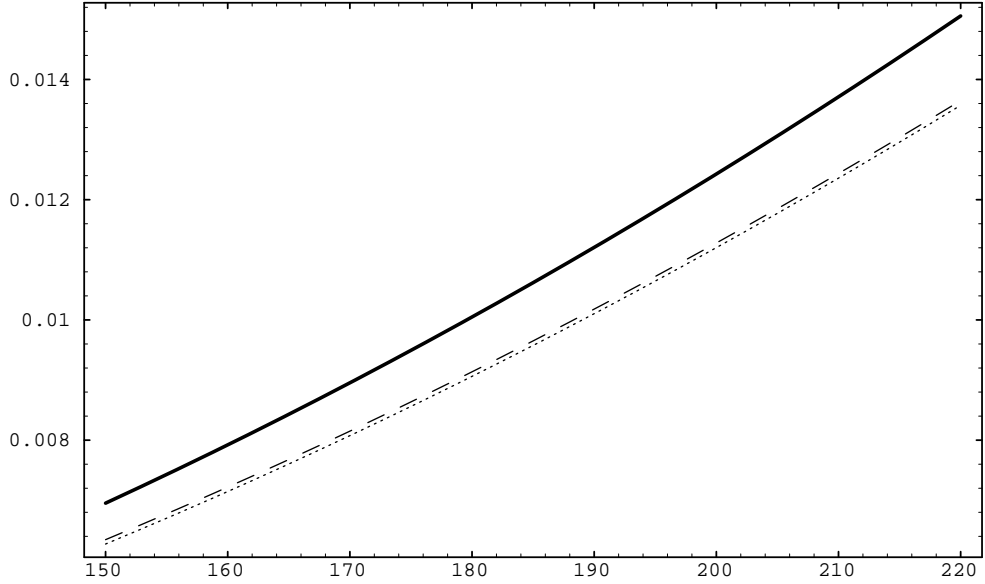

Figure 4.5: Contribution of the $(t, b)$ doublet to $\Delta\rho$ for $150 < m_t < 220$ GeV and $\alpha_s = 0.11$. The one-loop contribution with $m_b = 5$ GeV (full line), and the sum of one and two-loop contribution for $m_b = 0$ (dotted line) and for $m_b = 5$ GeV (dashed line) are shown.

## 4.2  QCD corrections to the Higgs vacuum polarization functions

The existence of at least one scalar particle, the Higgs boson, is required in the SM to generate the masses of the other fundamental particles, leptons, quarks and weak gauge bosons [75]. The discovery of this particle and the study of its fundamental properties will be the most important mission of future high-energy colliders.

The phenomenological properties of the unique Higgs particle of the SM have been studied in great details in the literature [75, 76]. In fact, because the precise knowledge of Higgs decay widths, branching fractions and production cross sections is mandatory, quantum corrections must be included and this subject has received much attention recently [77]. In particular, QCD corrections to Higgs decay and production processes are of the utmost importance. For instance, the Higgs decays into quark pairs and gluons, which together with the $H \to \tau^+\tau^-$ decays are the most important decay modes in the intermediate mass range, $M_W \lesssim M_H \lesssim 140\,\mathrm{GeV}$, receive very large QCD corrections. In the case of $H \to q\bar{q}$, they are known exactly to $\mathcal{O}(\alpha_s)$ [78, 79, 68] and up to $\mathcal{O}(\alpha_s^2)$ [80] in the approximation $m_q \ll M_H$; in the case of the gluonic decay, $H \to g\,g$, the QCD contributions are known up to next-to-leading-order [81], and the leading electroweak correction have also been calculated [82].

The electroweak radiative corrections to the SM Higgs decays [83] are of significance as well, as the leading contribution is proportional to the squared mass of the heavy top quark. In fact, a fourth generation of heavy fermions, the existence of which is still allowed by present experimental data with the proviso that the associated neutrino is heavy enough, would have a dramatic effect on the Higgs decay widths. Its contribution is universal in the sense that it does not depend on the final state particle, and will also increase quadratically with the heavy fermion masses. The universal part of the two-loop mixed $\mathcal{O}(\alpha_s G_F m_q^2)$ corrections, which have been calculated very recently [84, 85], will screen the leading one-loop contribution by a non negligible amount.

Many extensions of the Standard Model predict the existence of a larger Higgs sector. For instance, supersymmetric theories (SUSY) require the existence of at least two isodoublet scalar fields $\Phi_1$ and $\Phi_2$ to give masses separately to isospin up and down particles, thus extending the physical spectrum of scalar particles to five [75]. The physical Higgs bosons introduced by the minimal supersymmetric extension of the Standard Model (MSSM) are of the following type: two CP-even neutral bosons $h$ and $H$ (where $h$ will be the lightest particle), a CP-odd neutral boson $A$ (usually called pseudoscalar), and two charged Higgs bosons $H^\pm$. Besides the four masses $M_h$, $M_H$, $M_A$ and $M_{H^\pm}$, two additional parameters define the properties of the scalar particles and their interactions with gauge bosons and fermions: the ratio of the two vacuum expectation values $\tan\beta = v_2/v_1$ and a mixing angle $\alpha$ in the neutral CP-even sector. Note that, unlike a general two-Higgs doublet model where the six parameters are free, supersymmetry leads to several relations among these parameters and only two of them are independent. These natural relations also receive large radiative corrections [86].

In this section I present a calculation [74] of the hadronic contributions to the Higgs boson self-energies at $\mathcal{O}(\alpha\alpha_s)$. The most general case has been considered, leaving the momentum transfer arbitrary, with different quark flavors $u \neq d$ and consequently different quark masses $m_u \neq m_d$. The motivation for performing such a calculation is threefold: (i) As in the case of the SM Higgs boson, the strong [78, 79, 68, 87] and some of the electroweak [88] radiative corrections to SUSY Higgs boson decays are known at the one-loop level. In some limiting cases, as for the gluonic corrections for nearly massless quarks, SM two-loop results can be adapted to the SUSY case. Here, we provide the necessary material which allows to derive the universal part of the mixed $\mathcal{O}(\alpha_s G_F)$ radiative corrections to these Higgs decays. This is a generalization to a multi-Higgs doublet model of the recent SM calculation [84], which is just a special case (for a CP-even neutral Higgs boson) of the result presented here. (ii) The imaginary parts of the Higgs boson self-energies are related, through the optical theorem, to the partial decay widths of the Higgs bosons into quarks. Here I give exact analytical expressions of the QCD corrections to the Higgs decay widths in the most general case $m_u \neq m_d$, which is not available in the literature[5]. In the special cases $m_u = m_d$, the known results for the QCD corrections to CP-even and CP-odd [78, 79, 68] neutral Higgs bosons can be recovered. Since the expressions reported here have been obtained by a completely different method, this serves as an independent check. (iii) We can derive Ward identities that relate scalar VPF's to the longitudinal components of vector boson self-energies. In Sec. 4.3 I will show explicitly that it is possible to use these Ward identities to provide a powerful consistency check of the calculations described in the previous and in the present section.

### 4.2.1  Notation and one-loop results

In order to set the notation, I rederive here some one-loop results which will be relevant to our next discussion.

The contribution of a quark loop to the self-energy of a scalar Higgs boson $\Phi$ will be denoted by $\Pi^\Phi(s = q^2)$ where $q$ is the four-momentum transfer, and will correspond to $-i$ times the standard Feynman amplitude. To treat the cases of neutral CP-even, neutral CP-odd and charged Higgs bosons on the same footing, it is convenient to work in the general situation where the internal quarks in the loop are of different flavor, and have different masses. This corresponds to the case of a charged Higgs boson which couples to an up-type and down-type quark, with arbitrary masses $m_{u,d}$; the self-energies of neutral scalar and pseudoscalar Higgs bosons will be special cases of the previous one.

The coupling of charged Higgs bosons to fermions is a P-violating mixture of scalar and pseudoscalar couplings

$$g(H^+ u\bar{d}) = i \left( G_F/\sqrt{2} \right)^{1/2} [\, h_u(1 - \gamma_5) \; + \; h_d(1 + \gamma_5) \,]. \tag{4.36}$$

In the so-called type II two-Higgs doublet model [75], in which one doublet couples to the up quarks and neutrinos while the second doublet couples to down quarks and charged leptons,

| $\Phi$ | $h_u/m_u$ | $h_d/m_d$ |
|---|---|---|
| $H_{SM}$ | 1 | 1 |
| $h$ | $\cos\alpha/\sin\beta$ | $-\sin\alpha/\cos\beta$ |
| $H$ | $\sin\alpha/\sin\beta$ | $\cos\alpha/\cos\beta$ |
| $A$ | $1/\tan\beta$ | $\tan\beta$ |

Table 4.1: Neutral Higgs couplings to up-type and down-type fermions in the SM and MSSM.

we have

$$h_u = m_u/\tan\beta, \qquad h_d = m_d\,\tan\beta. \tag{4.37}$$

The MSSM belongs to this type. It is often convenient to use the scalar and pseudoscalar components of this coupling

$$v = h_d + h_u, \qquad a = h_d - h_u. \tag{4.38}$$

The couplings of scalar, that we will denote by $S$, and pseudoscalar $A$ Higgs bosons take the general form

$$g(Sq\bar{q}) = -i\left(G_F\sqrt{2}\right)^{1/2}\,h_q, \qquad g(Aq\bar{q}) = -\left(G_F\sqrt{2}\right)^{1/2}\gamma_5\,h_q \tag{4.39}$$

In the SM, the reduced couplings $h_q$ are just the quark masses; in the the MSSM these couplings for $\Phi = S, A$ are given in Table 4.2.1 for $u$ and $d$ quarks.

In the one-loop approximation, the contribution of a quark loop to the vacuum polarization amplitude of a charged Higgs boson, $\Pi^C(q^2)$, corresponds to the diagram of Fig. 4.2.1a. For arbitrary fermion masses $m_u \neq m_d \neq 0$ and momentum transfer $q^2$, this amplitude can be written as

$$\Pi^C(s) = \frac{N_c G_F}{2\sqrt{2}\pi^2}\,s\,\left[\,h_u^2\Pi_u^+(s) + h_d^2\Pi_d^+(s) + 2h_u h_d\frac{m_u m_d}{s}\Pi^-(s)\,\right], \tag{4.40}$$

where, using $\alpha = -m_u^2/q^2$, $\beta = -m_d^2/q^2$, $\rho_{a,b} = \ln\frac{m_{u,d}^2}{\mu^2}$, and all the other variables introduced in the last section, $\Pi^\pm(s)$ take the following form ($\Pi_u^+ = \Pi_d^+ = \Pi^+$)

$$\Pi^+(s) = \frac{1 + 2\alpha + 2\beta}{2\epsilon} - \frac{1}{4}(\rho_a + \rho_b)(1 + \alpha + \beta) - \frac{\alpha\rho_a + \beta\rho_b}{2} + 1 + \frac{3}{2}(\alpha + \beta)$$
$$+ \frac{1}{4}(1 + \alpha + \beta)\left[(\alpha - \beta + \lambda^{\frac{1}{2}})\ln x_a + (\beta - \alpha + \lambda^{\frac{1}{2}})\ln x_b\right],$$
$$\Pi^-(s) = -\frac{1}{\epsilon} - 2 + \frac{\rho_a + \rho_b}{2} - \frac{1}{2}\left[(\alpha - \beta + \lambda^{\frac{1}{2}})\ln x_a + (\beta - \alpha + \lambda^{\frac{1}{2}})\ln x_b\right]. \tag{4.41}$$

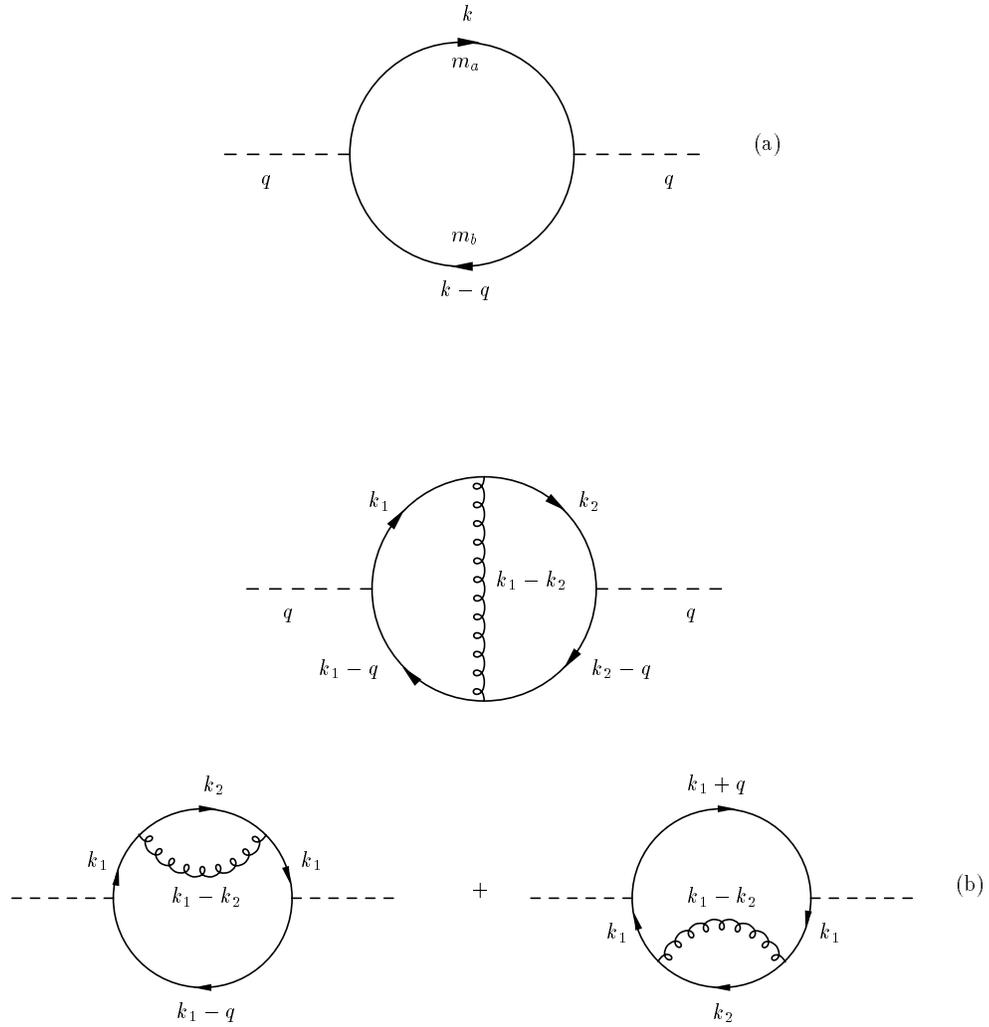

Figure 4.6: Feynman diagrams for the contribution of a quark isodoublet to the self-energy of a Higgs boson at the one-loop (a) and two-loop (b) levels.

As already done in the previous section, $\ln 4\pi$ and $\gamma$ factors have been absorbed in the normalization. The expressions of the self-energies for neutral scalar and pseudoscalar Higgs bosons can be simply obtained by setting $m_u = m_d = m_q$ in Eqs. (4.40) and (4.41), and by using the relevant couplings which are given in Tab. 4.2.1. The result is

$$\Pi^{S,A}(s) = \frac{3 G_F}{2\sqrt{2}\pi^2} s\, h_q^2 \left[\, \Pi_q^+(s) \;\pm\; (m_q^2/s)\; \Pi^-(s)\,\right], \qquad (4.42)$$

leading to

$$\Pi^S(s) = \frac{3 G_F}{2\sqrt{2}\pi^2} s\, h_q^2 \left[\frac{1}{2}\left(\frac{1}{\epsilon} - \rho_a\right)(1+6\alpha) + 1 + 5\alpha + \frac{1}{2}(1+4\alpha)^{\frac{3}{2}}\ln x\right],$$

$$\Pi^A(s) = \frac{3 G_F}{2\sqrt{2}\pi^2} s\, h_q^2 \left[\frac{1}{2}\left(\frac{1}{\epsilon} - \rho_a\right)(1+2\alpha) + 1 + \alpha + \frac{1}{2}(1+4\alpha)^{\frac{1}{2}}\ln x\right], \qquad (4.43)$$

where $x = 4\alpha/(1+\sqrt{1+4\alpha})^2$.

The imaginary parts of the scalar VPF's can be obtained along the same lines as described already for vector boson self-energies. The analytic structure of the VPF's is also identical. Considering the normalization we have chosen, the imaginary parts of a charged scalar self-energy are related to the partial decay widths of a charged Higgs boson $H^+$ into $u\bar{d}$ quark pairs through

$$\Gamma(H^+ \to u\bar{d}) = \frac{N_c G_F M_{H^+}}{2\sqrt{2}\pi^2} \left[h_u^2 \mathcal{I}m\Pi_u^+(s) + h_d^2 \mathcal{I}m\Pi_d^+(s)\right.$$

$$\left. + 2h_u h_d \frac{m_u m_d}{s} \mathcal{I}m\Pi^-(s)\right], \qquad (4.44)$$

and the case of neutral Higgs can be treated in a similar way.

## 4.2.2   Exact two-loop results

At $\mathcal{O}(\alpha\alpha_s)$, the two-loop diagrams contributing to the Higgs boson self-energies $\Pi^\Phi(q^2)$ are shown in Fig.4.2.1. In the 't Hooft-Feynman gauge, using the routing of momenta shown in the figure, and using[6] $N_c = 3$, one can write the bare amplitude as

$$\Pi^\Phi(q^2)\Big|_{\text{bare}} = \frac{16\pi G_F}{3\sqrt{2}} N_c \alpha_s \left(\frac{\mu^2 e^\gamma}{4\pi}\right)^{2\epsilon} \int \frac{d^n k_1}{(2\pi)^n} \int \frac{d^n k_2}{(2\pi)^n}\; [\mathcal{A} + \mathcal{B}], \qquad (4.45)$$

where

$$\mathcal{A} = \text{Tr}\frac{(\slashed{k}_1 + m_u)(v - a\gamma_5)(\slashed{k}_1 - \slashed{q} + m_d)\gamma_\lambda(\slashed{k}_2 - \slashed{q} + m_d)(v + a\gamma_5)(\slashed{k}_2 + m_u)\gamma^\lambda}{(k_1 - k_2)^2\,(k_1^2 - m_u^2)\,(k_2^2 - m_u^2)\,[(k_1 - q)^2 - m_d^2]\,[(k_2 - q)^2 - m_d^2]}$$

$$\mathcal{B} = \text{Tr}\frac{(\slashed{k}_1 + m_u)(v - a\gamma_5)(\slashed{k}_1 - \slashed{q} + m_d)(v + a\gamma_5)(\slashed{k}_1 + m_u)\gamma_\lambda(\slashed{k}_2 + m_u)\gamma^\lambda}{(k_1 - k_2)^2\,(k_1^2 - m_u^2)^2\,(k_2^2 - m_u^2)\,[(k_1 - q)^2 - m_d^2]}$$

$$+ \quad m_u \leftrightarrow m_d. \qquad (4.46)$$

---

[6]The case of different color group can be obtained as described in Sec. 4.1.2.

This bare amplitude has to be supplemented by counterterms; we can proceed in analogy with the case of vector boson VPF, with the difference that the Yukawa couplings of the quarks with the scalars are proportional to the quark masses, and they must be renormalized according to the mass definition. The result is that again only quark mass renormalization has to be considered, and the diagrams with quark mass insertions into the one-loop amplitude are the analogues of the ones in Fig.4.2b, with the addition of appropriate mass insertions in the vertices[7]. This is of course equivalent to calculate

$$\Pi^{\Phi}(q^2)\Big|_{\text{CT}} = -\frac{\delta m_u \partial}{\partial m_u} \Pi^{\Phi}(q^2)\Big|_{\text{1-loop}} - \frac{\delta m_d \partial}{\partial m_d} \Pi^{\Phi}(q^2)\Big|_{\text{1-loop}} \tag{4.47}$$

where the one-loop VPF is given by Eqs. (4.40, 4.41) and $h_{u,d}$ have been expressed in terms of $m_{u,d}$ before differentiation; $\mathcal{O}(\epsilon)$ terms not present in Eqs. (4.40,4.41) have to be included. The two-loop self-energies expressed in terms of renormalized masses then read

$$\Pi^{\Phi}(q^2) = \Pi^{\Phi}(q^2)\Big|_{\text{bare}} + \Pi^{\Phi}(q^2)\Big|_{\text{CT}} \tag{4.48}$$

The calculation proceeds exactly in the same way as for the vector boson VPF's, and involves the same integrals (see App. A.1).

Here I give the expression of the contribution of a $(u, d)$ isodoublet to the charged Higgs boson two-point function at order $\mathcal{O}(\alpha \alpha_s)$ in the case of arbitrary quark masses [74]. The result is expressed in terms of on-shell quark masses $m_{u,d}$. Using the same notations as in the one-loop case, the charged Higgs boson self-energy $\Pi^C(q^2)$ at the two-loop level is given by

$$\Pi^C(s) = \frac{G_F}{2\sqrt{2}\pi^2} \frac{\alpha_s}{\pi} s \left[ h_u^2 \Pi_U^+(s) + h_d^2 \Pi_D^+(s) + 2 h_u h_d \frac{m_u m_d}{s} \Pi^-(s) \right] \tag{4.49}$$

with $\Pi^{\pm}(s)$ given by the relatively simple and compact expressions[8]

$$\begin{aligned}
\Pi_u^+ = &-\frac{3}{2\epsilon^2}(1 + 4\alpha + 4\beta) - \frac{1}{\epsilon}\left[\frac{11}{4} + 14(\alpha + \beta) - 3\rho_a - 12\alpha\rho_a - 6\beta(\rho_a + \rho_b)\right] \\
&+ (\rho_a + \rho_b)\left[\frac{11}{4} + 14(\alpha + \beta) - 3(\rho_a + \rho_b)\left(\frac{1}{4} + \alpha + \beta\right) - 3(\alpha - \beta)\ln\frac{\alpha}{\beta}\right] \\
&+ \ln\frac{\alpha}{\beta}\left[5 + \frac{17}{2}(\alpha + \beta) - \frac{3}{2}(1 + 2\alpha + 2\beta)(\rho_a + \rho_b) + \frac{3}{4}(\alpha^2 - \beta^2)\ln\frac{\alpha}{\beta}\right] \\
&+ \frac{3}{8} - \frac{53}{2}(\alpha + \beta) + \frac{3}{4}\lambda^{\frac{1}{2}}(1 + \alpha + \beta)(\ln x_a + \ln x_b)\ln\frac{\alpha}{\beta} \\
&+ \frac{9}{4}\lambda^{\frac{1}{2}}(1 + \alpha + \beta)(\ln x_a + \ln x_b) + \frac{\alpha - \beta}{4}\ln\frac{\alpha}{\beta}(40 + 9(\alpha + \beta)) \\
&+ \frac{\ln^2 x_a + \ln^2 x_b}{4}\left[(1 - 2\alpha - 2\beta)(1 - \lambda + \alpha + \beta) - 3(\alpha + \beta) + 3(\alpha - \beta)^2\right]
\end{aligned}$$

$$+ \frac{\lambda^{\frac{1}{2}}}{2} \ln \frac{\alpha}{\beta} \left( \ln x_a + \ln x_b \right) \left( 1 + \alpha + \beta \right) - \frac{\pi^2}{4} (1 + 4\alpha + 4\beta)$$

$$+ \frac{3}{2} \ln x_a \ln x_b (\lambda + 2\alpha + 2\beta + 6\alpha\beta) - (1 + \alpha + \beta)^2 \mathcal{I} - 2(1 + \alpha + \beta)\mathcal{I}',$$

$$\Pi_d^+ \;\; = \;\; \Pi_u^+ \; [\; m_u \leftrightarrow \; m_d \;], \tag{4.50}$$

$$\Pi^- = \frac{6}{\epsilon^2} + \frac{1}{\epsilon} (14 - 6\rho_a - 6\rho_b) - 14(\rho_a + \rho_b) + 3(\rho_a + \rho_b)^2 + 20 + \pi^2$$

$$- 6 \ln x_a (\alpha - \beta + \lambda^{\frac{1}{2}}) - \ln^2 x_a \left[ \lambda - 1 - \alpha - \beta + (\alpha - \beta)\lambda^{\frac{1}{2}} \right]$$

$$- 6 \ln x_b (\beta - \alpha + \lambda^{\frac{1}{2}}) - \ln^2 x_b \left[ \lambda - 1 - \alpha - \beta + (\beta - \alpha)\lambda^{\frac{1}{2}} \right]$$

$$- 6 \ln x_a \ln x_b (1 + \alpha + \beta) + 2(1 + \alpha + \beta)\mathcal{I} + 4\mathcal{I}',$$

where the functions $F$ and $G$ and the integrals $\mathcal{I}$, $\mathcal{I}'$ are given in Sec. 4.1.2. One can derive the expressions of the self-energies for neutral scalar and pseudoscalar Higgs bosons from Eqs. (4.50) by setting $m_u = m_d = m_q$ and using the proper couplings. This is done in App. B.1, where the corresponding formulae are shown, together with the limits for infinite and vanishing momentum transfer. The expression of $\Pi_q^S$ recently derived in Ref. [84] in the case of the SM Higgs boson is in agreement with our result. Explicit $\mathcal{O}(G_\mu \alpha_s m_t^2)$ formulae for $\Pi_q^{S,A}(q^2)$ are crucial in order to calculate the leading radiative corrections to the Higgs boson fermionic decay widths. In the case of leptonic decays, for instance, there is only a universal factor coming from the Higgs wave function renormalization and the renormalization of the Higgs vacuum expectation value[9].

Analytic results for $\mathcal{I}m\Pi^S(s)$ [68,69,78,79] and $\mathcal{I}m\Pi^A(s)$ [79,68] have been obtained in the past by a number of authors by directly calculating the QCD corrections to the decay of a scalar Higgs boson into quark pairs. The results that we obtain from Eq. (4.50) using a completely different method agree with the previous ones; this serves as a check of our calculation in the general case. The expressions of the imaginary parts in those cases and for arbitrary quark masses are reported in App. B.2. Note also that for the value $\tan\beta = 1$, we recover the expression of the imaginary part for the longitudinal component of the electroweak vector bosons in the general case, which is given in Refs. [68,69]. Indeed, because of a Ward identity discussed in the next section, the imaginary part of the longitudinal component of the vector boson self-energy is the same as the one for the Higgs boson self-energy for this value of $\tan\beta$. This feature provides also a powerful check of the calculation presented here.

## 4.3   Ward Identities: a consistency check

It is well-known that there are Ward identities relating the longitudinal components of the electroweak vector bosons and the corresponding Goldstone bosons [89] (see also Sec 5.3).

---

[9]This factor has been calculated independently in Ref. [74,84]. For a more detailed discussion see [74] and references therein.

In this section, I will use the current algebra of the Standard Model to derive these Ward identities; I will then briefly show how to relate the Higgs boson self-energies calculated in the preceding section to the longitudinal parts of the electroweak vector boson self-energies, given in Sec. 4.1.2.

Defining the fermionic contribution to the VPF of the $W$ boson and of the corresponding Goldstone boson $\Phi$ as ($g$ is the $SU(2)_L$ coupling constant)

$$\Pi_{WW}^{\mu\nu}(q^2) = -i\frac{g^2}{2}\int d^n x e^{-iq\cdot x}\langle 0|T^*J_W^{\dagger\mu}(x)J_W^\nu(0)|0\rangle \tag{4.51}$$

$$\Pi_{\Phi\Phi}(q^2) = +i\frac{g^2}{2M_W^2}\int d^n x e^{-iq\cdot x}\langle 0|T^*S^\dagger(x)S(0)|0\rangle \tag{4.52}$$

where $J_W^\mu(x)$ and $S(x)$ are the charged fermionic currents coupled to the $W$ and to the $\Phi$ bosons and $T^*$ denotes the covariant time ordering product; for the notation and normalization of the currents, I follow Ref. [44] (see also Sec. 5.3, where I show them explicitly). Contracting $\Pi_{WW}^{\mu\nu}$ with the tensor $q^\mu q^\nu$, one obtains

$$q^\mu q^\nu \int_x \langle 0|T^*J_W^{\dagger\mu}(x)J_W^\nu(0)|0\rangle = \int_x \langle 0|T^*S^\dagger(x)S(0)|0\rangle - \frac{i}{2}\langle 0|S_1(0)|0\rangle \tag{4.53}$$

with $\int_x = \int d^n x e^{-iq\cdot x}$, $S_1$ the current coupled to the Standard Model Higgs boson, and where we have used

$$\partial_\mu J_W^\mu(x) = iS(x) \tag{4.54}$$

and

$$[J_W^0(x), S^\dagger(y)]_{x^0=y^0} = +\frac{1}{2}\,\delta^3(\vec{x}-\vec{y})[S_1(x) - iS_2(x)] \tag{4.55}$$

where $S_2$ is the current coupled to the neutral Goldstone boson. This, in turn, can be written as

$$q^2\Pi_{WW}^L(q^2) = -M_W^2\Pi_{\Phi\Phi}(q^2) - \frac{g^2}{4}\langle 0|S_1(0)|0\rangle \tag{4.56}$$

This equation relates the longitudinal part of the vacuum polarization function of the $W$ boson to the self-energy of the corresponding unphysical charged boson. One can see that the subtraction term $\langle 0|S_1(0)|0\rangle$ (a tadpole) is needed to cancel a spurious quartic dependence on the mass of the fermions.

Even though the previous derivation was at the one-loop level in the electroweak interactions, it is valid at any order in the strong interactions as the QCD generators commute with the ones of the electroweak group. To derive the self-energy of the charged Goldstone boson at $\mathcal{O}(\alpha)$ and $\mathcal{O}(\alpha\alpha_s)$, we therefore need only the expressions of the electroweak vector boson self-energies given in Ref. [59] in the general case and the one of the tadpole diagrams of Fig. 4.7 where both the two quarks of the same weak isodoublet are running in the loop. Using the same notation as in the previous sections, and for a single quark of mass $m_q$ (renormalized on-shell), we obtain for the tadpole amplitude up to order $\alpha_s$

$$\langle 0|S_1(0)|0\rangle = \frac{3m_q^4}{4\pi^2}\left[\frac{1}{\epsilon} + 1 - \ln\frac{m_q^2}{\mu^2} + \frac{\alpha_s}{3\pi}\left(-\frac{6}{\epsilon^2} - \frac{1}{\epsilon}\left(14 - 12\ln\frac{m_q^2}{\mu^2}\right)\right.\right.$$
$$\left.\left. -30 - \pi^2 + 28\ln\frac{m_q^2}{\mu^2} - 12\ln^2\frac{m_q^2}{\mu^2}\right)\right] \tag{4.57}$$

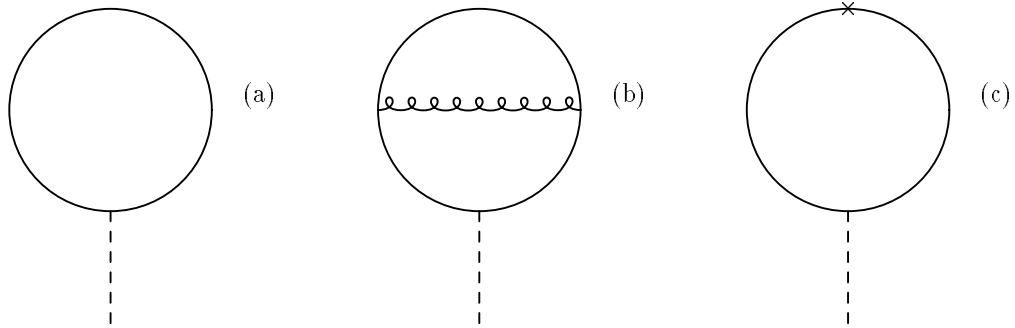

Figure 4.7: Tadpole diagrams relating the Higgs boson self-energies to the vector boson self-energies at the one-loop (a) and two-loop (b) levels, and mass counterterm (c) contributing at two-loop.

This equation, added to the one and two-loop expressions for the longitudinal part of the $W$ boson self-energies in the general case $m_u \neq m_d \neq 0$ given in Sec. 4.1.2, leads to the one and two-loop expressions of the charged Higgs boson self-energy given in Eqs. (4.41,4.50). This is just because for $\tan\beta = 1$, the charged Higgs boson couples to fermions exactly like the charged Goldstone of the SM, up to a relative minus sign for up-type and down-type quarks.

In a completely analogous manner, one can derive the Ward identity in the case of the neutral Goldstone boson $\Phi_2$

$$q^2\Pi_{ZZ}^L(q^2) = -M_Z^2\Pi_{\Phi_2\Phi_2}(q^2) - \frac{g^2}{4\cos^2\theta_W}\langle 0|S_1(0)|0\rangle \tag{4.58}$$

which allows to check the expressions of the self-energies of the pseudoscalar Higgs boson. For $\tan\beta = 1$, the latter has exactly the same couplings as the neutral SM Goldstone, again up to a relative minus sign for isospin up and down quarks.

This result provides a powerful consistency check of both the calculation of the electroweak vector boson self-energies described in Sec. 4.1, and the one of the neutral pseudoscalar and charged Higgs boson self-energies presented in Sec. 4.2.

## 4.4   Dispersive approach

In the preceding sections, I have discussed the calculation of vector boson and scalar self-energies using dimensional regularization. An alternative method consists in the use of dispersive integrals of the imaginary parts of the VPF's. The imaginary part of the self-energy of a vector boson or of a Higgs scalar is related to the partial decay width of the particle into quarks. This in turn can be easily calculated at one-loop by integrating in the phase space of the fermions the tree level decay amplitude. At two-loop, $\mathcal{O}(\alpha\alpha_s)$, the calculation of the decay amplitude must be pushed up to include virtual gluon corrections as well as real gluon production. This procedure is equivalent to the use of Landau-Cutkosky rules [64] on the relevant one or two-loop Feynman amplitudes, and implicitly assumes the use of on-shell fermion masses.[1] Once the imaginary part is known, the real part can be obtained solving a dispersive integral

$$\Pi(s,\Lambda) = \frac{1}{\pi} \int_{(m_a+m_b)^2}^{\Lambda^2} ds' \frac{\mathcal{I}m\Pi(s)}{s-s'-i\epsilon}, \tag{4.59}$$

where the cutoff scale $\Lambda \gg \sqrt{s}, m_a, m_b$ regulates the ultraviolet divergence, and a general dependence on the masses $m_{a,b}$ is understood. However, $\Pi(s) \approx s$ for large $s$, and in order for the previous integral to be meaningful, the dispersion relation of Eq. (4.59) needs to be subtracted. For a conserved current, as in the case of QED, the presence of spurious quadratic divergence is removed by a subtraction of the integral at $s = 0$, in this way restoring the Ward identity which requires the photon VPF to be transverse and proportional to $s$. The situation becomes more complicated for a non-conserved current, for which some logarithmic divergences are expected to be proportional to the squared fermion masses, and should not be removed in the subtraction procedure [56].

Subtraction procedures for Eq. (4.59) have been proposed in the past by Chang, Gaemers, and Van Neerven [56] and, more recently, by Kniehl and Sirlin [90,91], The dispersion relation proposed by the authors of Ref. [56] reads

$$\Pi^{V,A}(s,m_a,m_b) = \frac{1}{\pi} \int_{(m_a+m_b)^2}^{\Lambda^2} ds' \frac{\mathcal{I}m\Pi^{V,A}(s',m_a,m_b)}{s'-s-i\epsilon} \tag{4.60}$$
$$- \frac{1}{2\pi} \int_{4m_a^2}^{\Lambda^2} ds' \frac{\mathcal{I}m\Pi^V(s',m_a,m_a)}{s'} - \frac{1}{2\pi} \int_{4m_b^2}^{\Lambda^2} ds' \frac{\mathcal{I}m\Pi^V(s',m_b,m_b)}{s'}$$

and was based on the requirement that the subtraction remove all quadratic divergences and that at one-loop level the resulting amplitudes amount to finite renormalizations of the result in dimensional regularization, so that the subtraction terms cancel out in physical observables, like the shifts of the vector boson masses. The prescription proposed in Ref. [90,91] instead manifestly preserves the generic Ward identities constraining the VPF [90]. In the following section, as an application of the calculation of section 4.1, I generalize the approach of Kniehl and Sirlin to the case of arbitrary quark masses, and verify the consistency of their prescription up to two-loop level [60]: it is clear that for the prescription to be viable, the result of perturbative computations of physical observables in dimensional regularization and via dispersive integration must be exactly the same.

---

[1]The on-shell mass is the parameter that governs the start of the cut on the complex $s$-plane in S-matrix theory, and is naturally associated to the application of Cutkosky rules.

In addition to provide an alternative method for the computation of amplitudes within perturbation theory, the dispersive approach can be used to implement additional contributions to the amplitudes, whenever perturbation theory is either not applicable or not reliable. For instance, in the case of the light-quark contribution to the running of $\alpha$, where the photon VPF has to be evaluated at zero momentum transfer and the effective scale is of the order of the masses of the quarks, perturbation theory cannot be used, and the standard procedure is to employ experimental data for the cross section $\sigma(e^+e^-) \rightarrow$ hadrons and obtain through the optical theorem the imaginary part of the photon VPF as a function of $s$. The photon self-energy is then calculated by evaluating numerically the dispersion integral [9, 12].

Another potentially relevant case where perturbation theory fails to give an acceptable description is associated with a heavy quark threshold. Perturbation theory up to $\mathcal{O}(\alpha\alpha_s)$ predicts a discontinuous step-like threshold behavior, which is a rather unrealistic approximation. On the other hand, one expects a dense spectrum of resonances just below the threshold, and only for a heavy top, above $\approx 130 \, \mathrm{GeV}$, the decay width becomes large enough to smear out the resonances, as the lifetime becomes shorter than the revolution period of a $t\bar{t}$ bound state [92][2]. It is also well-known that the Coulombic effects near threshold cannot be accommodated when the perturbation series in $\alpha_s$ is truncated at a finite order, as they correspond to soft multi-gluon exchange.[3]

An acceptable description of $\sigma(e^+e^- \rightarrow t\bar{t})$ therefore requires contributions to the absorptive part of the photon and Z self-energies besides the usual perturbative ones. As the $t - b$ doublet gives the leading electroweak corrections to most precision observables, the possibility arises that additional effects connected to the $t\bar{t}$ threshold be important in precision physics.

The question of whether these additional "non-perturbative" contributions to the imaginary parts of VPF's affect the real parts, and consequently the radiative corrections to electroweak precision observables, has been the subject of considerable attention in the last three years (see Refs. [90,91,36,51,92] and references therein). Several methods have been proposed to take into account the $t\bar{t}$ threshold contributions, based on the use of phenomenological potentials like Richardson's in the context of a non-relativistic approximation: a thorough discussion can be found in Ref. [91,36]. The resulting effect on the electroweak parameters seems to be generally sizable, even though with large theoretical uncertainties: in Ref. [36] it has been estimated that the threshold effects would enhance the usual QCD perturbative corrections to the leading top contribution of Eq.(4.33) by 25-40%.

---

[2]Compared to $t\bar{t}$ resonances, the contribution of $t\bar{b}$ bound states is small, as threshold effects are proportional to the squared reduced mass of the quarks, and can be safely neglected [93].

[3]The resummation of these long-distance effects amounts to multiplying the bare spectral function around threshold by the factor $\frac{4}{3}\pi\frac{\alpha_s}{v}\left(1 - e^{-4\pi\alpha_s/(3v)}\right)^{-1}$, for which the expansion parameter is $\alpha_s/v$ rather than $\alpha_s$, with $v = \sqrt{1 - 4m_t^2/s}$ the relative velocity of the quark pair in the center of mass system.

The suggestion has been made, however, that when one works at a specified order in perturbation theory, the effect of the potentially large threshold contributions on the real part of the VPF should cancel out in the dispersive integration over the real axis, if all perturbative and non-perturbative contributions are taken into account [94, 96], as a result of the analyticity of the S-matrix. This is in contrast to what happens in many other cases where the absorptive part of the VPF is *not* integrated over the cut in the complex plane (for instance, top production, $e^+e^- \to \tau^+\tau^-$ [95]). In other words, although individual hadronic bound states are governed by long-distance dynamics, and are therefore inherently non-perturbative, the dispersion integrals over these states are determined *uniquely* by the short-distance dynamics, and can be safely calculated in the framework of perturbation theory [97]. This point of view is supported by arguments by Takeuchi and collaborators [99], and Sirlin [51], that I briefly summarize in section 4.4.2. A significant progress in clarifying the problem of QCD corrections to electroweak observables has however come from a recent three-loop calculation of the QCD corrections to $\Delta\rho$ [103]. I will discuss some of the consequences of this new result in Sec. 4.5.

As the subject of the role of non-perturbative QCD corrections to precision observables is not settled to date, I will not dwell further on it. What is of relevance here is that dispersion relations might possibly be of use to implement non-perturbative effects, and that it is important to know the precise relationship between them and dimensional regularization. Indeed, as will be illustrated in Sec. 4.4.2, the implementation of non-perturbative effects depends sensitively on the subtraction prescription that is used. In order to test the consistency of the procedure, it is then crucial to verify whether the results for physical observables given by the dispersive method in perturbation theory coincide with the ones obtained in dimensional regularization. Finally, once the prescription has been chosen, it may be useful to implement threshold effects in the $\overline{MS}$ scheme, and the correspondence between the two methods must be elucidated. These are the points that I will investigate in the next section, using the results obtained in the first part of this chapter.

### 4.4.1 Connection with dimensional regularization

First of all, let us recall the basic elements of the procedure proposed by Kniehl and Sirlin (for a more detailed discussion see Refs. [90, 91, 36]). Using Eq.(4.1), we can write the Ward identity

$$q_\nu \Delta(s) \equiv \int d^4x\, e^{iq\cdot x} < 0 | \mathrm{T}^* \left[ \partial^\mu J_\mu(x) J_\nu^\dagger(0) \right] | 0 > = q^\mu \Pi_{\mu\nu}(s) \qquad (4.61)$$

where I dropped the indices $i, j$ of the vacuum polarization tensor for simplicity. Contracting the previous equation with $q^\nu$, one obtains $\Delta(s) = \Pi_L(s)$; then introducing the linear combination $\Phi(s) = (\Pi_L(s) - \Pi_T(s))/s$, the cofactor of $q^\mu q^\nu$ in Eq.(4.1), we can rewrite the transverse part in terms of $\Delta(s)$ and $\Phi(s)$,

$$\Pi_T(s) = -s\Phi(s) + \Pi_L(s) = -s\Phi(s) + \Delta(s), \qquad (4.62)$$

The advantage of expressing $\Pi_T$ in this way is that $\Phi(s)$ and $\Delta(s)$ are only logarithmically divergent for $s \to \infty$, unlike $\Pi_T(s)$ which is quadratically divergent as discussed above. For $\Phi(s)$ this is obvious for dimensional reasons (two powers of the external momentum have been extracted), while for $\Delta(s)$ this is because the soft breaking of the current $J_\mu$ by mass terms implies that its divergence involves at most operators of canonical dimension three, and one power of the momentum has been extracted. It is therefore possible to write unsubtracted dispersion relations for $\Phi(s)$ and $\Pi_L(s) = \Delta(s)$. The use of Eq.(4.62) will then produce a dispersion relation for the transverse part of the VPF that automatically enforces the relevant Ward identities. Re-expressing $\mathcal{I}m\Pi_L$ in terms of $\mathcal{I}m\Phi$ and $\mathcal{I}m\Pi_T$, we obtain

$$\Pi_T(s) = \frac{1}{\pi} \int_{(m_1+m_2)^2}^{\Lambda^2} ds' \left[ \frac{\mathcal{I}m\Pi_T(s')}{s'-s-i\epsilon} + \mathcal{I}m\Phi(s') \right]$$
$$+ \frac{1}{2\pi i} \oint_{|s'|=\Lambda^2} \frac{ds'}{s'} \left[ \Pi_L(s') - s\Phi(s') \right] \tag{4.63}$$

where the quadratic divergence is now removed by the contribution of $\mathcal{I}m\Phi(s')$, and the logarithmic divergence is regulated by the cutoff $\Lambda$. The integral over the large circle, where the $s$ in the denominator has been neglected compared to $s'$, comes from the fact that $\Pi_L(s)$ and $\Phi(s)$ do not vanish asymptotically but behave like constants modulo logarithms, so that they must be considered in the application of Cauchy theorem. Note that for conserved currents $\partial_\mu J^\mu(x) = 0$, and $\Pi_L(s)$ vanishes reducing then the Ward identity to the QED form $\Pi_T(s) = -s\Phi(s)$.

We can now analyze the relation between the two calculational approaches at the one and two-loop level, and use the one and two-loop results in dimensional regularization of section 4.1 to write the real parts of $\Pi_T^{V,A}(s)$ as

$$\Pi_T^{V,A}(s) = s\tilde{X} + (m_a \mp m_b)^2 \tilde{Y} + (m_a^2 - m_b^2) \ln\frac{m_a^2}{m_b^2} \tilde{Z} + F^{V,A}(s), \tag{4.64}$$

where the divergent constants $\tilde{X}$, $\tilde{Y}$ and $\tilde{Z}$ involve only the poles in $\epsilon$ and the logarithms of the scale $\mu$ of the expressions of section 4.1, the function $F^{V,A}$ behaves like a constant as $s \to \infty$, and the $-(+)$ sign refers to vector (axial-vector) current. The one and two-loop components of the functions $F^{V,A}$ and of the divergent constants can separated by writing

$$F^{V,A} = F_1^{V,A} + \frac{\alpha_s}{\pi} F_2^{V,A}, \tag{4.65}$$
$$\tilde{X} = \tilde{X}_1 + \frac{\alpha_s}{\pi}\tilde{X}_2, \qquad \tilde{Y} = \tilde{Y}_1 + \frac{\alpha_s}{\pi}\tilde{Y}_2, \qquad \tilde{Z} = \tilde{Z}_1 + \frac{\alpha_s}{\pi}\tilde{Z}_2.$$

Their expressions at the one-loop level are given by (see Eq.(4.6))[4]

$$\tilde{X}_1 = \frac{1}{\epsilon} - \ln\frac{m_a m_b}{\mu^2}, \qquad \tilde{Y}_1 = -\frac{3}{2\epsilon} + \frac{3}{2}\ln\frac{m_a m_b}{\mu^2}, \tag{4.66}$$

---

[4]The definition of $F_{1,2}^{V,A}(s)$ used here differs from the one used in Ref. [91], as we normalize the divergent constants to the dimensional regularization result Eq. (4.64) keeping only the ultraviolet divergences and the related logarithms; remember also that $\ln 4\pi$ and $\gamma$ constants have been absorbed in the normalization of the VPF.

with $\tilde{Z}_1 = 0$, while at the two-loop level, as can be easily read off from the asymptotic expression for $s \to \infty$ reported in App. B.1, they are given by

$$\tilde{X}_2 = \frac{1}{2\epsilon} - \ln \frac{m_a m_b}{\mu^2}$$

$$\tilde{Y}_2 = \frac{3}{2\epsilon^2} + \frac{1}{\epsilon}\left(\frac{11}{4} - 3\ln\frac{m_a m_b}{\mu^2}\right) - \frac{11}{2}\ln\frac{m_a m_b}{\mu^2} + 3\ln^2\frac{m_a m_b}{\mu^2}$$

$$\tilde{Z}_2 = -\frac{3}{2\epsilon} + 3\ln\frac{m_a m_b}{\mu^2}. \tag{4.67}$$

If one integrates the imaginary parts of the vector bosons VPF's at the one and two-loop level (given for example in App. B.2), the result of the dispersive integral Eq. (4.63) for arbitrary quark masses and for vector and axial-vector currents can be expressed as

$$\Pi_T^{V,A}(s) = s\, X + (m_a \mp m_b)^2 Y + (m_a^2 - m_b^2)\ln\frac{m_a^2}{m_b^2}\, Z + F^{V,A}(s)$$

$$+ \frac{1}{2\pi i}\oint_{|s'|=\Lambda^2}\frac{ds'}{s'}\left[\Pi_L^{V,A}(s') - s\Phi^{V,A}(s')\right], \tag{4.68}$$

where the upper sign again refers to vector current. Here the divergent constants $X, Y, Z$ involve logarithms of the cutoff scale $\Lambda$, and their one and two-loop parts can be separated as in Eq. (4.65); their explicit form is given at one-loop by

$$X_1 = \ln\frac{\Lambda^2}{m_a m_b} - \frac{5}{3}\,, \qquad Y_1 = -\frac{3}{2}\ln\frac{\Lambda^2}{m_a m_b} + 3, \tag{4.69}$$

with $Z_1 = 0$. The finite function $F^{V,A}(s)$ must be the same as the one in Eq. (4.64). Since the result of Eq. (4.63) should be consistent with the one obtained using dimensional regularization, the integral over the large circle must be absorbed in a redefinition of the divergent constants $X, Y, Z$. In particular, if $\Pi_L(s)$ and $\Phi(s)$ are evaluated in dimensional regularization, the effect of the second integral in Eq. (4.63) should be simply to replace $X_{1,2}$, $Y_{1,2}$, and $Z_{1,2}$ by $\tilde{X}_{1,2}$, $\tilde{Y}_{1,2}$, and $\tilde{Z}_{1,2}$. This is indeed true at one-loop level, and one has

$$\frac{1}{2\pi i}\oint_{|s'|=\Lambda^2} ds'\frac{\Phi^{V,A}(s')}{s'} = X_1 - \tilde{X}_1$$

$$\frac{1}{2\pi i}\oint_{|s'|=\Lambda^2} ds'\frac{\Pi_L^{V,A}(s')}{s'} = (m_a \mp m_b)^2(\tilde{Y}_1 - Y_1). \tag{4.70}$$

This point is crucial to ensure that the cancellation of divergences occurring in physical observables like the $\rho$ parameter and $\Delta r$ yields a total independence of these observables from the calculational approach used to obtain the vacuum polarization functions.

At the two-loop level the dispersive integral of Eq. (4.63) has been calculated in the cases $m_a = m_b$ and $m_b = 0$ [57], for which it has therefore been verified that the functions $F^{V,A}$ in Eq. (4.64, 4.68) coincide up to $\mathcal{O}(\alpha\alpha_s)$ once the integrals over the large circle are evaluated. In the case of arbitrary masses the integral Eq. (4.63) has not been calculated yet, but the divergent part (involving logarithms of $\Lambda$) can be easily obtained and is reported in Ref. [93,57].

It is therefore possible to use our asymptotic expressions for the $\mathcal{O}(\alpha\alpha_s)$ VPF's of App. B.1 to evaluate the integrals of the left-hand side of Eq.(4.68), obtain the two-loop constants $X_2, Y_2$ and $Z_2$, and compare at least the divergent part. Here again, since the result of Eq. (4.68) should be consistent with the one obtained in dimensional regularization, the integral over the large circle must be absorbable in a redefinition of the divergent constants $X_2, Y_2$ and $Z_2$. Thus, when we use dimensionally regularized expressions for $\Phi(s)$ and $\Pi_L(s)$, the integral involving $\Phi(s)$ in Eq. (4.68) will replace $X_2$ by $\tilde{X}_2$, and the one involving $\Pi_L(s)$ will replace $Y_2$ and $Z_2$ by respectively $\tilde{Y}_2$ and $\tilde{Z}_2$. One has therefore at two-loop

$$\frac{1}{2\pi i}\oint_{|s'|=\Lambda^2} ds' \frac{\Phi^{V,A}(s')}{s'} = X_2 - \tilde{X}_2,$$

$$\frac{1}{2\pi i}\oint_{|s'|=\Lambda^2} ds' \frac{\Pi_L^{V,A}(s')}{s'} = (m_a \mp m_b)^2(\tilde{Y}_2 - Y_2)$$
$$+ (m_a^2 - m_b^2)\ln\frac{m_a^2}{m_b^2}(\tilde{Z}_2 - Z_2). \tag{4.71}$$

Assuming that the procedure is consistent, we solve for $X_2, Y_2$, and $Z_2$, and obtain[5]

$$X_2 = \ln\frac{\Lambda^2}{m_a m_b} + 4\zeta(3) - \frac{55}{12}$$

$$Y_2 = \frac{3}{2}\ln^2\frac{\Lambda^2}{m_a m_b} - \frac{9}{2}\ln\frac{\Lambda^2}{m_a m_b} - \frac{3}{2}\ln^2\frac{m_a}{m_b} + \frac{3}{8} - 6\zeta(3) - \frac{3}{4}\pi^2$$

$$Z_2 = -\frac{3}{2}\ln\frac{\Lambda^2}{m_a m_b} + 5 \tag{4.72}$$

The divergent part of these constants agree with Ref. [57,93]. Inserting Eq.(4.71) into Eq.(4.63) finally leads to

$$\Pi_T(s) = \frac{1}{\pi}\int_{(m_1+m_2)^2}^{\Lambda^2} ds' \left[\frac{\mathcal{I}m\Pi_T(s')}{s'-s-i\epsilon} + \mathcal{I}m\Phi(s')\right] \tag{4.73}$$
$$+ s(\tilde{X} - X) + (m_a \mp m_b)^2(\tilde{Y} - Y) + (m_a^2 - m_b^2)\ln\frac{m_a^2}{m_b^2}(\tilde{Z} - Z)$$

Using Eq.(4.73), it is now possible in principle to implement in the general case of arbitrary quark masses non-perturbative effects [91, 36] also in the $\overline{\text{MS}}$ scheme for which a dimensional regularization calculation is required. This result [60] generalizes part of Ref. [91].

### 4.4.2   An application

Let us now briefly discuss the application of the dispersive method to physical observables, taking again the example of the fermionic contribution to the $\rho$ parameter. Using the fact that at zero-momentum transfer the transverse and longitudinal components of the $W$ and $Z$ self-energies are equal, and recalling that the vector component vanishes in this limit for equal mass quarks, we have written $\Delta\rho$ in terms of the the value at $s = 0$ of a function $f_{L,T}(s, m_a, m_b)$ defined in section 4.1.3. Since $\Delta\rho$ is an observable and finite physical quantity, $f_{L,T}(0, m_a, m_b)$ can be evaluated by means of an unsubtracted dispersion relation

$$f_{L,T}(s, m_a, m_b) = \frac{1}{\pi} \int_{(m_1 + m_2)^2}^{\Lambda^2} ds' \frac{\mathcal{I}m f_{L,T}(s', m_a, m_b)}{s' - s - i\epsilon} \qquad (4.74)$$

---

[5]Note that the sign of the $\zeta(3)$ term in $Y_2$ in Ref. [60] is incorrect due to a misprint, and that the expressions for the special cases considered in Ref. [91] were obtained using Ref. [54], and are therefore affected by the omission of the $\pi^2$ term in the mass renormalization mentioned in section 4.1.2.

The condition for Eq.(4.74) to be valid is that

$$\lim_{\Lambda^2 \to \infty} \oint_{|s'| = \Lambda^2} \frac{ds'}{s'} f_{L,T}(s', m_a, m_b) = 0 \qquad (4.75)$$

as a consequence of Cauchy theorem.

Using Eq.(B5), it is easy to verify at the two-loop level that $f_{L,T}(s, m_a, m_b)$ is free of ultraviolet divergences and tends to zero as $s \to \infty$ and therefore satisfies the condition of Eq.(4.74) in the case of arbitrary masses. Note that this happens for both transverse and longitudinal components of the VPF. This means that, as long as perturbative effects (up to $\mathcal{O}(\alpha\alpha_s)$) are concerned, a dispersion relation based on the longitudinal components as the one proposed in Ref. [90,91] [6] gives the same result as a dispersion relation based on the transverse components, as the one proposed in Ref. [56], despite the very different asymptotic behavior of the individual components. In fact, the two subtraction prescriptions of Refs. [56, 90] coincide at one-loop for arbitrary quark masses, but they differ by finite renormalizations to leading order in QCD (with the obvious exception of conserved vector current). The extra pieces cancel however in physical observables like $\Delta\rho$ and $\Delta r$ [7], giving the same results as in dimensional regularization. This is by no means a general feature: it has been shown in Ref. [98] that in the case of the leading order QCD corrections to the decay amplitude of the SM Higgs boson into leptons, only the prescription of Ref. [90] reproduces correctly the result obtained in dimensional regularization.

The observation, Eq.(4.75), that the two dispersive procedures are equivalent up to $\mathcal{O}(\alpha\alpha_s)$ motivated the work of Ref. [99], where the Operator Product Expansion (OPE) was used to show that non-perturbative QCD corrections to $\Delta\rho$ can be described equivalently using dispersion relations that involve longitudinal or transverse components of the VPF's. The basic idea of Takeuchi and collaborators is that the non-perturbative asymptotic behavior of the VPF's $\Pi_{T,L}(s)$ as $|s| \to \infty$ can be extracted from their OPE. The latter can be found in the appendix of Ref. [100], together with the relevant Wilson coefficients $C_i(q)$, which can also be extracted from the high-momentum expansion given here in App. B,

$$\Pi_L^{V,A}(q^2) = (\hat{m}_a(q) \mp \hat{m}_b(q))^2 C_{L,1}(q) + (\hat{m}_a(\mu) \mp \hat{m}_b(\mu))^2 C_{L,2}(q) + \dots, \qquad (4.76)$$

$$\Phi^{V,A}(q^2) = C_{\Phi,1}(q) + C_{\Phi,2}(q) \frac{(\hat{m}_a(q) \pm \hat{m}_b(q))^2}{q^2} + C_{\Phi,3}(q) \frac{(\hat{m}_a(q) \mp \hat{m}_b(q))^2}{q^2} + \dots$$

_________________

[6] Note that for $s = 0$ the integrand in the first line of Eq.(4.63) reduces to $\Pi_L(s')/s'$.

[7] I have verified this explicitly in both cases for arbitrary masses.

where the ellipses stand for terms suppressed by inverse powers of $q^2$. The long distance dynamics is embodied in the running masses $\hat{m}(\mu)$ in terms of which the OPE is built (products of masses are the only dimension two operators available), and in the vacuum expectation values of operators of dimension higher than two which are suppressed by inverse powers of $q^2$. The conclusion is that, since the Wilson coefficients depend only logarithmically on $q$, Eq.(4.75) is valid at any order in QCD and even including non-perturbative QCD corrections.

The immediate consequence of this result is that the incorporation of non-perturbative threshold effects through either of the two dispersion relations should give the same result.[8] The calculations of Refs. [57, 99] in the context of a non-relativistic approximation instead indicate that the two approaches give *opposite* results!

Eq. (4.75) and the argument by Takeuchi et al. can be recast in a different form, which is particularly simple. Sirlin has observed [51] that the difference of the two forms of Eq.(4.75) can be expressed as the integral over the large circle of a combination of functions $\Phi^{V,A}(s)$ (the denominator cancels). Applying Cauchy theorem again, the result can be alternatively phrased as the dispersive integral of a combination of spectral functions $\mathcal{I}m\Phi(s)$, a sum rule similar to the well-known Weinberg's [102]

$$\int^\infty ds \, \mathcal{I}m \left[ \sum_{i=a,b} \sum_{j=V,A} \frac{\Phi^j(s, m_i, m_i)}{2} - \Phi^V(s, m_a, m_b) - \Phi^A(s, m_a, m_b) \right] = 0 \tag{4.77}$$

With this sum rule in mind, Sirlin argues that, in the case of the top-bottom doublet, potentially large threshold effects can arise only from $\mathcal{I}m\Phi^{V,A}(m_t, m_t)$, because the other channels are suppressed by reduced mass considerations. Consequently possible large threshold contributions should cancel in the integral $\int^\infty ds \mathcal{I}m[\Phi^V(s, m_t, m_t) + \Phi^A(s, m_t, m_t)]$. The possibility that they cancel against each other seems inconsistent with the non-relativistic approximation developed in Ref. [91], where it is found that $\mathcal{I}m\Phi^V(s, m_t, m_t) \approx \mathcal{I}m\Phi^A(s, m_t, m_t)$ in the threshold region. The only reasonable conclusion is that large threshold contributions to $\mathcal{I}m\Phi^{V,A}$ cancel against other contributions when the integrals over all $s$ are considered. These integrals can be related to integrals in the asymptotic region, and can be expanded perturbatively, so that the cancellation must occur order by order in perturbation theory. This argument, without proving rigorously the cancellation of the threshold contributions, seems to point in the same direction as Ref. [94].

## 4.5 The choice of the scale of $\alpha_s$

Before concluding this chapter on QCD corrections with a overview of the incorporation of QCD corrections in the calculation of precision electroweak observables, let us turn our attention to a central issue for any QCD calculation: the choice of the scale at which the running coupling constant $\alpha_s$ must be evaluated. In the preceding sections I have purposely avoided this problem, but as soon as we consider a practical application of the calculations illustrated above, we are bound to specify the scale, and the question cannot be further postponed.

---

[8]In Ref. [101] the same authors of Ref. [99] apply the OPE method to the $H \to \ell\bar{\ell}$ case, and confirm to all orders in perturbation theory the result [98] that only the dispersion relation based on the Ward identity is consistent with dimensional regularization in that case.

In general, QCD predictions for a measurable quantity $A$ have the form

$$A = A_0 \left( 1 + C_1(Q) \frac{\alpha_s(Q)}{\pi} + C_2(Q) \frac{\alpha_s^2(Q)}{\pi^2} + \dots \right) \qquad (4.78)$$

The coefficients $C_i(Q)$ depend both on the exact definition of $\alpha_s$ (i.e. the scheme) and the choice of the scale $Q$. If a calculation could be performed at all orders in $\alpha_s(Q)$, the choice of scheme and scale would be immaterial: $A$ must be the same in any scheme, and independent of the scale chosen. If the perturbative series is truncated at a finite order, however, the prediction for $A$ is both scale and scheme dependent. For instance, a redefinition of coupling or a change of scale can, in principle, change the next-to-leading order coefficient to any desired value. Therefore, in order to set the scale appropriate to a particular process so that $C_1$, $C_2$, $\dots$ be meaningful, we need some sort of information on the higher order contributions of the perturbative expansion. There is no rigorous way to gain this information. It is possible, however, to make an educated guess.

Let us concentrate again on the contribution to $\Delta\rho$ of the (t-b) doublet. As I have mentioned above, there is a very recent result of a three loop calculation [103] in the limit of very heavy top that modifies Eq.(4.33) into

$$\Delta\rho_{top} = \frac{3 G_\mu m_t^2}{8\sqrt{2}\pi^2} \left( 1 - \frac{2}{9}(\pi^2 + 3) \frac{\alpha_s}{\pi} - 14.594 \frac{\alpha_s^2}{\pi^2} \right), \qquad (4.79)$$

where I have not specified the scale, and I have used the on-shell top mass $m_t$. As the only mass scale in the process is $m_t$, it seems reasonable to set $\alpha_s = \alpha_s(m_t)$. It turns out that this is not the case, and the fact that we know only the first two coefficients of the series implies that the error that we make by choosing the scale in Eq.(4.79) at $m_t$ is possibly relevant.

For this particular case, we have two different methods that give us very similar answers. The first method consists in considering some particular type of potentially large effects, like the gluon vacuum polarizations, and in trying to take them into account at higher orders. A heuristic procedure has been proposed for example by Brodsky, Mackenzie, and Lepage (BLM) [105]. An alternative procedure is based on an analysis of the scale dependence: the best choice for the scale is the one for which the scale dependence is minimal (Principle of Minimal Sensitivity, PMS) [106]. One can also use the Fastest Apparent Convergence (FAC) method, which consists in absorbing the whole QCD correction in a scale redefinition of the first term of the series. BLM, PMS and FAC applied to Eq.(4.79) give us somewhat similar answers [51], suggesting a smaller scale than $m_t$, and therefore a larger QCD correction. They also suggest large and increasing higher order coefficients for the series Eq.(4.79).

The second method consists in the use of an effective lagrangian approach, where the top is assumed to be very heavy, and integrated out. This approach is illustrated, for instance, in Ref. [104], and the result is that the scale appropriate for Eq.(4.79) is $m_t$, *provided* that Eq.(4.79) is expressed in terms of the $\overline{\text{MS}}$ top mass calculated at $m_t$. As shown in Ref. [51], the two approaches numerically give very similar answers. A very detailed discussion of all the points raised here can be found in [51], where it is also shown that a very good convergence pattern is obtained when the ratio between the pole and the $\overline{\text{MS}}$ mass $\hat{m}_t(m_t)$ is first optimized according to one of the methods (BLM or PMS) mentioned above, and then used in $\Delta\rho$. The result of Ref. [51] can be cast in a very convenient form for numerical calculations:

$$\Delta\rho_{top} = \frac{3 G_\mu m_t^2}{8\sqrt{2}\pi^2} \left( 1 - \frac{2}{9}(\pi^2 + 3)\frac{\alpha_s(\xi m_t)}{\pi} \right), \qquad (4.80)$$

where $\xi = 0.260^{+0.079}_{-0.056}$. Numerically, for $m_t = 180\,\text{GeV}$, this implies a QCD correction of $11.95 \pm 0.51\%$. The relative error on the QCD correction is $4.3\%$, and has a negligible impact on the electroweak observables. For instance, it induces a shift in the determination of $M_W$ of $\approx 3\,\text{MeV}$, and of $2 \times 10^{-5}$ in $\sin^2 \hat{\theta}_W(M_Z)$.

It is clear that large higher order corrections are linked in some way to the sensitivity of the process under consideration to infrared gluons. For example, it is easy to realize, and can be explicitly checked [94], that the diagrams of Fig. 4.1b are not sensitive to low momentum gluons, while the one of Fig. 4.2a, having the resonance of the top propagator exactly at the momentum transfer, is certainly very sensitive to the gluon infrared region. This is why, through the on-shell counterterm (diagram Fig. 4.2a evaluated at $p^2 = m_t^2$), the pole mass introduces in the perturbative series of $\Delta\rho$ large terms of $\mathcal{O}((n-1)!\beta_0^{n-1}\alpha_s^n)$, where $\beta_0$ is the first coefficient of the $\beta$ function, connected to the appearance of singularities in the Borel plane, the renormalons.

For what concerns the electroweak observables of interest, the VPF's have to be evaluated at $q^2$ equal to $M_Z^2$ or $M_W^2$, and the arguments used for $\Delta\rho$ cannot be applied. An effective field approach [104] again suggests to use $\alpha_s(m_t)$ for the leading quadratic contribution expressed in terms of $\hat{m}_t(m_t)$, while the logarithm $\ln\frac{m_t^2}{M_W^2}$ has to be shifted: $\ln\frac{m_t^2}{M_W^2} \rightarrow \ln\frac{m_t^2}{M_W^2} - 4/\beta_0 \ln\frac{\alpha_s(M_W)}{\alpha_s(m_t)}$.

## 4.6  QCD corrections to basic electroweak observables

In this section I consider the QCD corrections entering the determination of $M_W$ and $\sin^2 \hat{\theta}_W(M_Z)$. I remind that QCD corrections for the form factor $\hat{k}(M_Z^2)$ that connects $\sin^2 \theta_{eff}^{lept}$ and $\sin^2 \hat{\theta}_W(M_Z)$ have been considered already in Sec. 2.4. Unlike the case of processes involving external hadrons, like $e^+\, e^- \rightarrow hadrons$, which receive pure QCD radiative corrections, the tree level predictions for $M_W$ and $\sin^2 \hat{\theta}_W(M_Z)$ are affected only at $\mathcal{O}(\alpha\alpha_s)$, by gluonic corrections to quark loop insertions on vector boson propagators. Despite the large value of the QCD coupling constant, we can therefore expect this kind of effects to become relevant only when the one-loop correction is already sizable. This is the case of the large top contribution to the radiative corrections $\Delta r$ and $\Delta\hat{r}$.

The incorporation of perturbative QCD effects in the basic electroweak corrections $\Delta r$ and $\Delta \hat{r}$ has been considered in detail in Ref. [36]. Here I give just a summary of their results, in order to make the implementation of $\mathcal{O}(\alpha \alpha_s)$ effects possible, using the results of this chapter and of App. B. Normalizing the VPF's as I have done throughout this chapter, and extracting from $A_{ZZ}(q^2)$ and $A_{WW}(q^2)$ a factor $\frac{\alpha}{\pi}$ at one-loop level and $\frac{\alpha}{\pi}\frac{\alpha_s}{\pi}$ at two-loop, we find the contributions of a quark isodoublet $(u, d)$ to the $Z$ and $W$ self-energies as

$$A_{ZZ}^{doubl}(q^2) = - \sum_{i=u,d} \frac{1}{16\hat{s}^2\hat{c}^2} \left[ (1 - 4\hat{s}^2 C_{3i} Q_i)^2 \Pi^V(q^2, m_i, m_i) + \Pi^A(q^2, m_i, m_i) \right],$$

(4.81a)

and

$$A_{WW}^{doubl}(q^2) = -\frac{1}{8\hat{s}^2} \left[ \Pi^V(q^2, m_u, m_d) + \Pi^A(q^2, m_u, m_d) \right],$$

(4.81b)

where $Q_i$ and $C_{3i} = \pm 1$ are the electric charge and the isospin eigenvalue of the quark flavor, respectively. Including the decoupling of the top quark according to the Marciano-Rosner convention [38], and neglecting terms that contribute in $\mathcal{O}(\hat{\alpha}^2)$ but not $\mathcal{O}(\hat{\alpha}\alpha_s)$, we find [36]

$$\Delta\hat{r}^{doubl} = \left[ \text{Re}\left( \frac{A_{ZZ}(M_Z^2)}{M_Z^2} - \frac{A_{WW}(0)}{M_W^2} \right) + \sum_{i=u,d} Q_i^2 \frac{\partial}{\partial q^2} \Pi^V(q^2, m_i, m_i) \Big|_{q^2=0} + D \right]_{\overline{\text{MS}}}$$

(4.82)

where the subscript $\overline{\text{MS}}$ indicates that the poles in $\epsilon$ have been removed and that $\mu$ has been set to $M_Z$. $D$ is a contribution deriving from the use of the decoupling as described in section 2.3, and can be written as

$$D = \frac{\alpha}{\pi} \frac{2\hat{s}^2 - \frac{8}{3}\hat{s}^4 - 1}{6\hat{s}^2\hat{c}^2} \left[ \left(1 + \frac{\alpha_s}{\pi}\right) \ln \frac{m_t^2}{M_Z^2} - \frac{15}{4}\frac{\alpha_s}{\pi} \right].$$

(4.83)

Similarly, we obtain for $\Delta r$ in the on-shell scheme

$$\Delta r^{doubl} = \sum_{i=u,d} Q_i^2 \frac{\partial}{\partial q^2} \Pi^V(q^2, m_i, m_i) \Big|_{q^2=0} + \frac{1}{8s^2 M_W^2} \sum_{i=V,A} \Pi^i(0, m_u, m_d)$$

$$- \frac{1}{16s^4 M_Z^2} \text{Re} \sum_{i=u,d} \left[ (1 - 4s^2 C_{3i} Q_i)^2 \Pi^V(M_Z^2, m_i, m_i) + \Pi^A(M_Z^2, m_i, m_i) \right]$$

$$+ \frac{c^2 - s^2}{8s^4 M_W^2} \text{Re} \left[ \Pi^V(M_W^2, m_u, m_d) + \Pi^A(M_W^2, m_u, m_d) \right].$$

(4.84)

The expression for $\Delta r$ in the $\overline{\text{MS}}$ scheme differs from Eq.(4.84) by subleading terms of order $\mathcal{O}(\alpha^2)$, that I do not consider here. We see that for $m_u \gg M_Z, M_W$, $\Delta r^{doubl}$ reduces to $c^2/s^2\Delta\rho$. One can use for this leading term the result of the three loop calculation of Ref. [103], or Eq.(4.80). From the argument of the preceding section, it seems convenient to use $\overline{\text{MS}}$ masses for the quarks. To this end, Eqs.(4.16) and (4.17) have to be employed.

Now all the ingredients for a detailed evaluation of the hadronic contribution to $\Delta r$ and $\Delta \hat{r}$ have been given. Using the results of this chapter and of App. B, together with the estimate of higher order corrections of Ref. [51], a very accurate determination of the QCD effects in the prediction of $M_W$ and $\sin^2 \hat{\theta}_W(M_Z)$ should be possible, including the light quark mass effects. Of course, different options are possible (the use of Eq.(4.80) or (4.79), of on-shell quark masses or $\overline{\text{MS}}$, etc.) and they all contribute to define a theoretical error for the QCD corrections. This can be expected to be beyond the present and future experimental resolution.

# Chapter 5

# Higher order electroweak corrections

Potentially large two-loop contributions originated by electroweak interactions are the subject of this chapter. After a brief introduction, I will present the main features of a calculation of the $\rho$ parameter in the SM in the limit of heavy top mass [107–109]. Leading and next-to-leading terms in $m_t$ are kept up to $\mathcal{O}(G_\mu^2 m_t^2 M_Z^2)$. Although the result cannot be applied directly to the case of $M_W$ and $\sin^2 \theta_{eff}^{lept}$, an extrapolation indicates that next-to-leading $\mathcal{O}(G_\mu^2 m_t^2 M_Z^2)$ terms can contribute significantly to the theoretical prediction of these observables, and provides an estimate of the current theoretical error in the determination of high-energy observables.

## 5.1 Higher order effects of electroweak origin

In Sec. 2.5 I have anticipated that large masses, like $m_t$ and $M_H$, potentially induce large radiative corrections to precision observables. This can be regarded as a consequence of the violation of the decoupling theorem occurring in the SM. The investigation of the actual impact of the large masses on the high-energy phenomenology at the two-loop level is therefore of great importance. Part of this program has already been completed with the study of the leading dependence on $m_t$ and $M_H$ of all the precision observables at two-loop level.

### 5.1.1 Large Higgs mass corrections

An elegant effective Lagrangian approach has been used in Ref. [110] to investigate the leading two-loop effect of a large Higgs mass on the various precision observables in electroweak physics. Barbieri et al. have shown that the physical $M_H^2$ corrections can all be reduced to the corresponding effects in the two parameters $\epsilon_1$ and $\epsilon_3$ [112] (or in a different language $T$ and $S$ [113]), that are often used to parametrize the effect of new physics on the precision observables. Self-energy diagrams that contribute to $\mathcal{O}(g^4 M_H^4)$ are present, but cancel in all relevant physical amplitudes. The contribution to the isospin breaking parameter $\epsilon_1$ is the same as the one to the traditional $\Delta\rho$, and the actual calculation yields $\Delta\rho \simeq 5.03 \times 10^{-5} \left(M_H/\text{TeV}\right)^2$, when the one-loop corrections are expressed in terms of $G_\mu$, $\alpha$, $M_Z$, and the on-shell mass $M_H$. This result confirms the pioneer work of Ref. [111]. For a Higgs mass below $\approx$ 2TeV, the correction is extremely small when compared to the current accuracy. Since $M_H \gtrsim 1$TeV is unlikely to be consistent with a perturbative treatment of the SM in general, this means that the two-loop $M_H^2$ corrections to the $\rho$ parameter and to the theoretical determination of $M_W$ and $\sin^2 \hat{\theta}_W(M_Z)$ are practically irrelevant. As an illustration, keeping such a contribution in $\Delta r$ and $\Delta \hat{r}$ would shift $M_W$ by $\delta M_W \approx 3$MeV, and $\sin^2 \hat{\theta}_W(M_Z)$ by a mere $1.7 \times 10^{-5}$, for $M_H \simeq 1$TeV. A similar pattern is found for the parameter $\epsilon_3$, which enters the decay width of the $Z$ boson, $\Delta\epsilon_3 \simeq 3.16 \times 10^{-5} \left(M_H/\text{TeV}\right)^2$. Consequently, the effect on the $Z$ width, $\Gamma_Z$, is also tiny, $\Delta\Gamma_Z/\Gamma_Z \simeq 0.53 \times 10^{-4} \left(M_H/\text{TeV}\right)^2$.

### 5.1.2  Large top mass corrections

The leading two-loop corrections in the limit of a heavy quark doublet have been first calculated some time ago [114] in the approximation of a light Higgs mass, $M_H \ll m_t$, and found quite small. As the experimental bound on $M_H$ became higher, the approximation used in Ref. [114] turned out to be inadequate. This motivated a calculation [115] of the leading $G_\mu^2 m_t^4$ corrections in the case of a heavy Higgs boson whose mass can be comparable to $m_t$ (with $m_b = 0$). Rather unexpectedly the result, a function of the ratio $M_H/m_t$, was found to give significant contributions to the precision observables, of the order of the future experimental accuracy, for values of $M_H$ comparable to the top mass.

To study this kind of effects it has proven useful, both from a conceptual and a practical point of view, to think of the electroweak sector of the SM as a weak perturbation around an underlying pure Yukawa theory, obtained from the SM by turning off the gauge interactions [118]. Such a theory describes the ideal case of top and Higgs masses much larger than the gauge vector boson and light fermion masses, with gauge interactions becoming negligible compared to those associated with the top-scalar sector. In this scenario all relevant physical observables of present interest are evaluated at a momentum scale much lower than the top or Higgs masses and that can be set to zero, at least in first approximation. This makes it possible to relate physical quantities of interest, e.g. the ratio between $M_W$ and $M_Z$, to Green functions of the Yukawa theory. Indeed, in the spontaneously broken phase of the SM, a set of Ward identities relates vector boson Green functions at vanishing external momenta to appropriate Green functions of the corresponding Goldstone bosons. For instance the relevant self-energies entering into the radiative corrections to the $\rho$ parameter can be expressed, neglecting subleading contributions in the heavy top, as (see Sec. 5.3)

$$\frac{A_{WW}(0)}{M_W^2} = -\left.\frac{\partial}{\partial q^2}\Pi_{\Phi\Phi}(q^2)\right|_{q^2=0} \tag{5.1a}$$

$$\frac{A_{ZZ}(0)}{M_Z^2} = -\left.\frac{\partial}{\partial q^2}\Pi_{\Phi_2\Phi_2}(q^2)\right|_{q^2=0}. \tag{5.1b}$$

where the $A(0)$'s are the transverse part of the vector boson self-energies at zero momentum transfer and the $\Pi$'s are the two-point functions associated with the unphysical counterparts. Using this correspondence one can, in practice, derive the leading contribution to the radiative corrections working entirely in the framework of the Yukawa model without reference to the original theory, this framework allowing a considerable simplification in the evaluation of physical effects at higher orders. Along these lines, the leading $\mathcal{O}(G_\mu^2 m_t^4)$ two-loop terms have been evaluated for both the $\rho$ parameter and the $Z \to b\bar{b}$ partial width, for an arbitrary ratio of the top and Higgs masses [115–117, 107]. These corrections represent effects proportional to $g_t^4$, that depend only upon the two parameters of the Yukawa Lagrangian $g_t$ and $\lambda$, the top Yukawa coupling and the quartic self-interaction of the scalars, respectively. $\Delta\rho$ is the result of interest here, because it is proportional to the leading top contributions to $\Delta r$ and $\Delta\hat{r}$, the relevant radiative corrections for the determination of $M_W$ and $\sin^2\hat{\theta}_W(M_Z)$. It can be written as

$$\rho = \frac{1}{1 - \Delta\rho}, \tag{5.2a}$$

with

$$\Delta\rho = x_t + \frac{x_t^2}{3} \, R\left(\frac{M_H}{m_t}\right), \tag{5.2b}$$

where $R(x)$ is a function that reaches a minimum of $-11.8$ for $x \simeq 5.7$, and $x_t = 3\frac{G_\mu m_t^2}{8\pi^2\sqrt{2}}$. The effect of such a contribution can be estimated using Eqs. (2.15) and (2.21). When the function $R$ reaches its minimum, and for $m_t = 180\text{GeV}$, $\delta M_W \approx 23\text{MeV}$ and $\delta \sin^2\theta_{eff}^{lept} \approx 1.4 \times 10^{-4}$.

While the above picture is theoretically clean and very useful for practical computations, up to now its limitations have not been fully clarified. Concerning the Higgs, it is still possible that this particle is much heavier than all the remaining spectrum. Instead, in the case of the top, we have seen that there are strong indications [1, 2] that its mass is only moderately larger than the vector boson masses, $m_t \approx 2M_Z$. This raises the question of the corrections to the asymptotic regime here referred to as the Yukawa limit. Compared to the leading order results, these corrections are of order $(M_Z/m_t)^2$, and could be sizable, as can be verified in explicit one-loop examples. Consider for instance the function $f(m_t^2/M_W^2)$ appearing in the evaluation of the one-loop box diagram for $B^0 - \bar{B}^0$ oscillation:

$$f(z) = \frac{1}{4} + \frac{9}{4(1-z)} - \frac{3}{2(1-z)^2} + \frac{3}{2}\frac{z^2 \ln z}{(z-1)^3}. \tag{5.3}$$

The asymptotic value $f(\infty) = 1/4$ is rather far from the realistic value $f(5) \simeq 0.54$, and in this case the subleading contributions amount to a 100% correction.

Coming to two-loop effects, it is important to establish whether the present estimates of the purely electroweak top contribution, based on a calculation in the Yukawa limit, are realistic or can be influenced by subleading effects that cannot be neglected in view of the future experimental precision. The first step in this direction seems to be an investigation of the two-loop next-to-leading top corrections, namely contributions that scale as $g_t^2 g^2$ where $g$ is the gauge coupling. To keep the computation as simple as possible I will focus on neutrino scattering on a leptonic target. For this process I present a calculation [108, 109] of the electroweak corrections of $\mathcal{O}(G_\mu^2 m_t^2 M_Z^2)$ to the $\rho$ parameter, defined as the ratio of neutral-to-charged current amplitudes, at zero momentum transfer. There are two important advantages in considering $\rho$ instead of other observables: first, the natural relation $\rho_0 = 1$ (cf. Eq.(2.4)) of the bare Lagrangian ensures that only one-loop renormalization is needed[1]; secondly, the irreducible two-point functions to be considered are evaluated at zero momentum transfer, which greatly simplifies the calculation. The computation can be regarded as an attempt to check the validity of the expansion in powers of the gauge coupling constant around the Yukawa limit, until the full two-loop results will be available. At the same time we will be able to provide an estimate of the error associated with the two-loop electroweak effects. For the processes under examination, we find large subleading corrections, of the same sign and of about the same magnitude as the leading one.

## 5.2 The example of $\Delta\rho$: a complete calculation at $\mathcal{O}(G_\mu^2 m_t^2 M_W^2)$

In this section I outline the computation of the electroweak corrections of $\mathcal{O}(G_\mu^2 m_t^2 M_Z^2)$ to the $\rho$ parameter. Specifically, I identify $\rho$ with the cofactor of the $J_Z J_Z$ interaction in neutral current amplitudes, expressed in units of $G_\mu$. As well known, radiative effects lead also to a modification of the mixing angle, described by a form factor usually called $\kappa(q^2)$. These effects will not be discussed here.

We begin by writing the relation between the $\mu$-decay constant and the charged current amplitude expressed in terms of bare quantities. At the two-loop level, neglecting terms that do not contribute at $\mathcal{O}(G_\mu^2 m_t^2 M_Z^2)$, we have

$$\frac{G_\mu}{\sqrt{2}} = \frac{g_0^2}{8 M_{W_0}^2} \left\{ 1 - \frac{A_{WW}}{M_{W_0}^2} + V_W + M_W^2 B_W + \frac{A_{WW}^2}{M_W^4} - \frac{A_{WW} V_W}{M_W^2} \right\} \quad , \tag{5.4}$$

where $g_0$ and $M_{W_0}$ are the bare $SU(2)_L$ coupling and $W$ mass, respectively, $A_{WW} \equiv A_{WW}(0)$ is the transverse part of the $W$ self-energy at zero momentum transfer, and the quantities $V_W$ and $B_W$ represent the relevant vertex and box corrections. At the bare level, using the fact that $M_{Z_0}^2 c_0^2 = M_{W_0}^2$, where $c_0 \equiv \cos\theta_{W_0}$, the $\rho$ parameter can be written as:

$$\rho = \frac{\left( 1 - \dfrac{A_{ZZ}}{M_{Z_0}^2} + V_Z + M_{Z_0}^2 c_0^2 B_Z + \dfrac{A_{ZZ}^2}{M_Z^4} - \dfrac{A_{ZZ} V_Z}{M_Z^2} \right)}{\left( 1 - \dfrac{A_{WW}}{M_{W_0}^2} + V_W + M_{W_0}^2 B_W + \dfrac{A_{WW}^2}{M_W^4} - \dfrac{A_{WW} V_W}{M_W^2} \right)} \quad , \tag{5.5}$$

where the self-energies are evaluated at $q^2 = 0$, and $V_Z$ and $B_Z$ are the corresponding vertex and box contribution in neutral current amplitude. To the order we are interested in,

---

[1] A natural relation is a constraint between coupling constants and masses that results from the symmetry structure of the Lagrangian before spontaneous symmetry breaking. The counterterms needed to renormalize the parameters of the broken theory possess the symmetry of the unbroken Lagrangian, and, consequently, not all of them are independent. As a result, the radiative corrections to a natural relation are finite at one-loop.

remembering that at one-loop only self-energies give rise to $m_t^2$ terms, Eq.(5.5) reduces to:

$$\rho = 1 + \left(\frac{A_{WW}}{M_{W_0}^2} - \frac{A_{ZZ}}{M_{Z_0}^2}\right) + (V_Z - V_W) + (M_{W_0}^2 + A_{WW})(B_Z - B_W)$$
$$+ \left(\frac{A_{WW}}{M_W^2} - \frac{A_{ZZ}}{M_Z^2}\right)\left(-\frac{A_{ZZ}}{M_Z^2} + (V_Z - V_W) - M_W^2 B_W\right) \quad . \tag{5.6}$$

We proceed by separating the self-energies into one-loop and two-loop contributions:

$$A_{ZZ} = A_{ZZ}^{(1)} + A_{ZZ}^{(2)}; \qquad A_{WW} = A_{WW}^{(1)} + A_{WW}^{(2)} \tag{5.7}$$

with the understanding that the one-loop term is still expressed in terms of bare parameters. The one-loop part can be further decomposed into pure bosonic ($b$) and fermionic ($f$) terms:

$$A_{ZZ}^{(1)} = A_{ZZ}^{b(1)} + A_{ZZ}^{f(1)}; \qquad A_{WW}^{(1)} = A_{WW}^{b(1)} + A_{WW}^{f(1)}, \tag{5.8}$$

and the one-loop fermionic contribution to the $\rho$ parameter, assuming a vanishing bottom mass, can be expressed as follows:

$$X_d^0 = \left(\frac{A_{WW}^f}{M_{W_0}^2} - \frac{A_{ZZ}^f}{M_{Z_0}^2}\right)^{(1)} = \frac{g_0^2}{8 M_{W_0}^2} f(m_{t_0}^2, \epsilon) \tag{5.9a}$$

$$f(m_t^2, \epsilon) \equiv \frac{3}{2\pi^2} \frac{1}{(4 - 2\epsilon)} m_t^2 \epsilon \Gamma(\epsilon) \left(\frac{4\pi\mu^2}{m_t^2}\right)^\epsilon , \tag{5.9b}$$

where $\epsilon$ is related to the dimension $d$ of the space-time by $\epsilon = (4 - d)/2$ and $\mu$ is the 't-Hooft mass scale.

As we want to express the final result in terms of the physical $Z$ mass, we perform the shift $M_{Z_0}^2 = M_Z^2 - \text{Re } A_{ZZ}(M_Z^2) \simeq M_Z^2 - A_{ZZ}$. Using the decompositions given in Eqs. (5.7) and (5.8), recalling that one-loop vertices and boxes do not involve the top quark, and keeping only terms up to $\mathcal{O}(G_\mu^2 m_t^2 M_Z^2)$, we obtain after simple algebra:

$$\rho = 1 + X_d^0 + X_d\left(-\frac{A_{WW}}{M_W^2} + V_W + M_W^2 B_W\right)$$
$$+ \left(\frac{A_{WW}^b/c_0^2 - A_{ZZ}^b}{M_Z^2}\right)^{(1)} + \left(\frac{A_{WW}}{M_W^2} - \frac{A_{ZZ}}{M_Z^2}\right)^{(2)}$$
$$+ (V_Z - V_W) + M_Z^2 c_0^2(B_Z - B_W) - X_d(V_W + 2 M_W^2 B_W)$$
$$+ X_d\left[\left(\frac{A_{WW}}{M_W^2} - \frac{A_{ZZ}}{M_Z^2}\right) + (V_Z - V_W) + M_W^2(B_Z - B_W)\right], \tag{5.10}$$

where $X_d$ is the quantity introduced in Eq. (5.9), now expressed in terms of renormalized parameters.

We observe that the Eq.(5.10) further simplifies if we express the one-loop fermionic contribution in terms of the Fermi constant $G_\mu$. Indeed, as can be seen from Eq. (5.4), the first line of Eq. (5.10) reproduces the effective coupling in the charged sector:

$$X_d^0 \left(1 - \frac{A_{WW}}{M_W^2} + V_W + M_W^2 B_W \right) = \frac{g_0^2}{8 M_{W_0}^2} \left(1 - \frac{A_{WW}}{M_W^2} + V_W + M_W^2 B_W \right) f(m_{t_0}^2, \epsilon)$$
$$\simeq \frac{G_\mu}{\sqrt{2}} \, f(m_{t_0}^2, \epsilon) \quad . \tag{5.11}$$

Up to now, apart from the use of the physical $Z$ mass, we have not specified any particular renormalization condition. In order to simplify the structure of the counterterms, we have found it convenient to perform the calculation using the $\overline{\text{MS}}$ parameter $\sin^2 \hat{\theta}_W (M_Z)$ (henceforth abbreviated as $\hat{s}^2$). Indeed, while in the on-shell (OS) scheme the counterterm associated to the quantity $s^2 = 1 - M_W^2/M_Z^2$ contains terms proportional to $m_t^2$ and gives rise to $\mathcal{O}(G_\mu^2 m_t^2 M_Z^2)$ contributions to $\rho$, the counterterm related to $\hat{s}^2$ does not exhibit any $m_t^2$ dependence and this greatly simplifies our task (cf. Sec. 2.4)[2]. Therefore, to the order we are interested in, we can directly replace $c_0^2$ with $\hat{c}^2$ in Eq.(5.10) ($\hat{c}^2 \equiv 1 - \hat{s}^2$). It will always be possible to recover the result in the pure OS scheme, by appropriately shifting $\hat{s}^2$ in the one-loop expression for $\rho$.

We now notice that the one-loop contribution is still written in terms of bare quantities. To put $\rho$ in its final form, we split it into the usual $\mathcal{O}(\alpha)$ result, $\Delta \rho^{(1)}$, plus the counterterm part, $\Delta \rho_c$, namely

$$\frac{G_\mu}{\sqrt{2}} \, f(m_{t_0}^2, \epsilon) + \left( \frac{A_{WW}^b/\hat{c}^2 - A_{ZZ}^b}{M_Z^2} \right)^{(1)} + (V_Z - V_W)^{(1)} + M_Z^2 \hat{c}^2 (B_Z - B_W)^{(1)} \equiv$$
$$\equiv \Delta \rho^{(1)} + \Delta \rho_c, \tag{5.12}$$

with

$$\Delta \rho^{(1)} = \Delta \rho^{f(1)} + \Delta \rho^{b(1)}, \tag{5.13a}$$

$$\Delta \rho^{f(1)} = N_c x_t \equiv N_c \frac{G_\mu m_t^2}{8 \pi^2 \sqrt{2}}, \tag{5.13b}$$

$$\Delta \rho^{b(1)} = \frac{\hat{\alpha}}{4 \pi \hat{s}^2} \left[ \frac{3}{4 \hat{s}^2} \ln \hat{c}^2 - \frac{7}{4} + \frac{2 \, c_Z}{\hat{c}^2} + \hat{s}^2 \, G(\xi, \hat{c}^2) \right], \tag{5.13c}$$

where $N_c$ is the color factor, and $\hat{\alpha}$ is the $\overline{\text{MS}}$ coupling defined in Sec. 2.4 (the decoupling of the top quark is irrelevant at the order we are working). In Eq.(5.13c)

$$c_Z = \frac{\hat{c}^2}{4} (5 - 3 I_3) - 3 \left( \frac{I_3}{8} - \frac{\hat{s}^2}{2} Q + \hat{s}^4 \, I_3 \, Q^2 \right), \tag{5.14a}$$

where $I_3$ and $Q$ are the isospin and electric charge of the target ($I_3 = -1$ for electrons), and

$$G(\xi, \hat{c}^2) = \frac{3}{4} \frac{\xi}{\hat{s}^2} \left[ \frac{\ln \hat{c}^2 - \ln \xi}{\hat{c}^2 - \xi} + \frac{1}{\hat{c}^2} \frac{\ln \xi}{1 - \xi} \right], \tag{5.14b}$$

---

[2]The mass counterterm for the unphysical scalars is related to the tadpole counterterm and therefore determined from the tadpole diagrams [126].

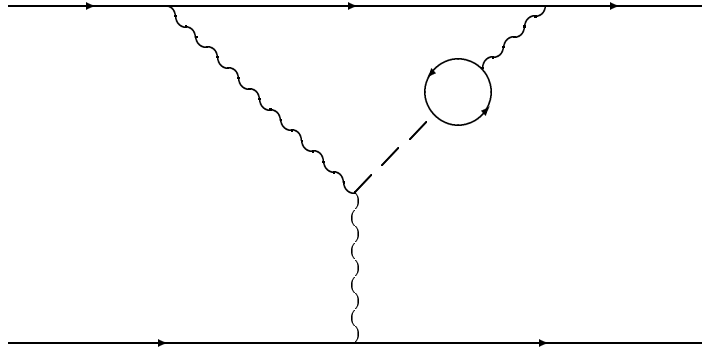

Figure 5.1: Vertex diagram contributing to $\mathcal{O}(G_\mu^2 m_t^2 M_Z^2)$.

with $\xi \equiv M_H^2/M_Z^2$. Using Eqs. (5.10), (5.11), and (5.12), we can express $\rho$ as follows:

$$\rho = 1 + \Delta\rho^{(1)} + N_c x_t \Delta\rho^{(1)} + \Delta\rho^{(2)}, \tag{5.15}$$

where the previous relation defines the two-loop contribution, $\Delta\rho^{(2)}$, as:

$$\Delta\rho^{(2)} = \Delta\rho_C + \left(\frac{A_{WW}}{M_W^2} - \frac{A_{ZZ}}{M_Z^2}\right)^{(2)} + (V_Z - V_W)^{(2)} + M_Z^2 \hat{c}^2 (B_Z - B_W)^{(2)} \\ - X_d \left(V_W + 2\,M_W^2 B_W\right). \tag{5.16}$$

Eq.(5.15) suggests that a possible way to take into account higher order effects is to write $\rho$ as

$$\rho = \frac{1}{(1 - \Delta\rho^{f(1)})}(1 + \Delta\rho^{b(1)} + \Delta\rho^{(2)}), \tag{5.17}$$

where the resummation of $\Delta\rho^{f(1)}$ can be theoretically justified on the basis of $1/N_c$ expansion arguments [119].

From Eq.(5.16) it is immediate to see that, unlike the case of the leading $\mathcal{O}(G_\mu^2 m_t^4)$ contribution, the calculation of the self-energies is not sufficient to compute the $\mathcal{O}(G_\mu^2 m_t^2 M_Z^2)$ corrections. Indeed vertex and box diagrams do contribute to the order we are interested in. At the two-loop level the only box graphs that can give rise to $m_t^2$ terms are those obtained from a one-loop box diagram either by inserting a fermionic self-energy in a vector boson propagator line, or by expanding a bare coupling. The former can be taken into account, to the order we are working at, by replacing in the corresponding one-loop diagram the bare propagator mass by the physical one[3]. The latter do not give any contribution in the $\overline{\text{MS}}$ framework, because of the structure of the counterterms. For what concerns the vertex graphs, besides those with fermionic insertions in the one-loop diagrams as discussed above for boxes, one has to take into account the mixing between unphysical scalars and vector bosons in an internal line (see Fig. 5.1).

---

[3]In fact, the insertion of the on-shell mass counterterm cancels all $m_t^2$ terms coming from this kind of diagrams.

The calculation of the vacuum polarizations is performed following the lines of Ref. [107]. In that paper two-loop self-energy diagrams were evaluated employing techniques used in the Current Algebra formulation of radiative corrections [44]. This framework provides an efficient way to enforce the relevant Ward identities while discussing at the same time several Feynman diagrams. As soon as subleading terms are taken into account, the relevant Ward identities do not have any more the simple form of Eq.(5.1) in a general gauge, and complicate considerably[4]. In the next section, as an illustration of the method, I will show how the Ward identities concerning the leading top contributions (Yukawa limit) can be derived using Current Algebra. The two-loop self-energy integrals are computed using the a Taylor expansion method outlined in Ref. [130, 131], and described in App. A.2. As the transfer momentum is zero, the two-loop vertex integrals of the diagrams of Fig. 5.1 are equivalent to self-energy integrals, and can be solved accordingly. The calculation has been performed using an anticommuting $\gamma_5$; we have checked explicitly that we never encounter traces involving $\gamma_5$ and more than two other gamma matrices or the antisymmetric tensor, so the inconsistency of this definition has no way to manifest in the actual calculation. The advantage is not only computational, as alternative definitions of $\gamma_5$ break the Ward identities of the theory.

We define the renormalized top mass as the pole mass (see Sec. 4.1.1), using the counterterm (in the 't Hooft-Feynman gauge[5]), $\delta m \equiv m - m_0$,

$$\delta m_t \equiv \Sigma(m_t) = \frac{\hat{g}^2}{16\pi^2} \frac{m_t^3}{M_W^2} \left(\frac{4\pi\mu^2}{e^\gamma m_t^2}\right)^\epsilon \left\{ -\frac{3}{8\epsilon} \left[1 + zt \left(1 - \frac{14}{9}\hat{s}^2\right)\right] \right.$$
$$\left. + \frac{1}{4}\mathcal{A} + zt \left(\frac{25\hat{s}^2 - 18}{36}\right) + \mathcal{O}(zt^{\frac{3}{2}}) \right\}, \qquad (5.18a)$$

with

$$\mathcal{A} = wt \ln wt - 4 + \frac{zt}{2}(1 + \ln zt) + \frac{ht}{2} + \frac{ht}{4}(6 - ht)\ln ht - \frac{ht - 4}{4}\sqrt{ht}\, g(ht). \qquad (5.18b)$$

I have used $ht \equiv (M_H/m_t)^2$, $zt \equiv (M_Z/m_t)^2$, $wt \equiv (M_W/m_t)^2$, and

$$g(x) = \begin{cases} \sqrt{4-x}\,\left(\pi - 2\arcsin\sqrt{x/4}\right) & 0 < x \leq 4 \\[2mm] 2\sqrt{x/4 - 1}\,\ln\left(\frac{1-\sqrt{1-4/x}}{1+\sqrt{1-4/x}}\right) & x > 4\,. \end{cases} \qquad (5.19)$$

Eq.(5.18) complements Eq.(4.12), valid for strong interactions, and it is obtained from diagrams like the one in Fig. 4.2a, where the gluon is replaced by scalars and vector bosons.

---

[4] The calculation described here was performed in the 't Hooft–Feynman gauge; the Ward identities in App. C are valid in that particular gauge. In the framework of the Background Field Method it has been verified [120] that the Ward identities maintain the form of Eq.(5.1) up to $\mathcal{O}(G_\mu^2 m_t^2 M_z^2)$.

[5] Unless tadpoles are included, the pole mass counterterm is gauge dependent, see also Ref. [63].

Explicitly, we find for $\Delta \rho^{(2)}$, in units $N_c \left( \hat{\alpha}/(16\pi \hat{s}^2 \hat{c}^2) \, m_t^2/M_Z^2 \right)^2 \simeq N_c x_t^2$:

$$\Delta \rho^{(2)} = 25 - 4\,ht + \frac{(-4+ht)\,\sqrt{ht}\,g(ht)}{2} + \left(-6 - 6\,ht + \frac{ht^2}{2}\right)\ln ht$$

$$+ \left(\frac{1}{2} - \frac{1}{ht}\right)\pi^2 + \left(\frac{6}{ht} - 15 + 12\,ht - 3\,ht^2\right)Li_2(1-ht) + \left(9\,ht - 15 - \frac{3\,ht^2}{2}\right)\phi\left(\frac{ht}{4}\right)$$

$$+ zt\left[\frac{25}{2} + \frac{4}{ht} - 10\,\hat{c}^2 + \frac{3}{\hat{s}^2} + \frac{277\,\hat{s}^2}{9} - \frac{4\,\hat{s}^2}{ht} + \left(9 + \frac{3}{\hat{s}^4} - \frac{6}{\hat{s}^2} - 6\,\hat{s}^2\right)\ln \hat{c}^2\right.$$

$$+ 3\left(5 - 6\,\hat{s}^2\right)\ln zt + 6\,I_3\,\hat{c}^2 + \left(2 - \frac{4}{ht} - 8\,\hat{s}^2 + \frac{28\,\hat{s}^2}{ht}\right)\ln ht$$

$$+ \pi^2\left(-\frac{7}{3} - \frac{2}{3\,ht^2} + \frac{1}{ht} - \frac{56\,\hat{s}^2}{27} + \frac{2\,\hat{s}^2}{3\,ht^2} - \frac{\hat{s}^2}{ht}\right)$$

$$+ \frac{12\,(-4+ht)\,\hat{s}^2}{ht}\Lambda\left(-1 + \frac{4}{ht}\right) + \left(2\,ht\,\hat{c}^2 - \frac{2\,(-2+3\,ht)\,\hat{c}^2}{ht^2}\right)Li_2(1-ht)$$

$$+ \left. \left(-2 - \frac{8}{ht} + 5\,\hat{s}^2 + \frac{24\,\hat{s}^2}{ht^2} - \frac{10\,\hat{s}^2}{ht} + ht\,\hat{c}^2\right)\phi\left(\frac{ht}{4}\right)\right] \tag{5.20a}$$

for $M_H \gg M_Z$, while in the region $M_H \ll M_Z$,

$$\Delta \rho^{(2)} = 19 - 2\,\pi^2 - 4\,\pi\,\sqrt{ht} + ht\,\left(-\frac{27}{2} + 2\,\pi^2 - 6\ln ht - 5\ln \hat{c}^2 + 3\ln zt\right)$$

$$+ zt\left[-\frac{11}{2} + \frac{3}{\hat{s}^2} + \frac{319\,\hat{s}^2}{9} + 6\,I_3\,\hat{c}^2 + \pi^2\left(-\frac{7}{3} - \frac{56\,\hat{s}^2}{27}\right)\right.$$

$$+ \left.\left(7 + \frac{3}{\hat{s}^4} - \frac{6}{\hat{s}^2} - 4\,\hat{s}^2\right)\ln \hat{c}^2 + \left(21 - 16\,\hat{s}^2\right)\ln zt\right]. \tag{5.20b}$$

In Eqs. (5.20)

$$\Lambda\left(-1 + \frac{4}{x}\right) = \begin{cases} -\frac{1}{2\sqrt{x}}\,g(x) + \frac{\pi}{2}\,\sqrt{4/x - 1} & 0 < x \leq 4 \\[2mm] -\frac{1}{2\sqrt{x}}\,g(x) & x > 4\,, \end{cases} \tag{5.21a}$$

$$Li_2(x) = -\int_0^x dt \frac{\ln(1-t)}{t}\,, \tag{5.21b}$$

and

$$\phi(z) = \begin{cases} 4\sqrt{\frac{z}{1-z}}Cl_2(2\arcsin\sqrt{z}) & 0 < z \leq 1 \\[2mm] \frac{1}{\lambda}\left[-4Li_2(\frac{1-\lambda}{2}) + 2\ln^2(\frac{1-\lambda}{2}) - \ln^2(4z) + \pi^2/3\right] & z > 1\,, \end{cases} \tag{5.21c}$$

where $Cl_2(x) = \text{Im}\,Li_2(e^{ix})$ is the Clausen function with

$$\lambda = \sqrt{1 - \frac{1}{z}}\,. \tag{5.21d}$$

The first two lines of Eq. (5.20a) represent the leading $\mathcal{O}(G_\mu^2 m_t^4)$ result [115, 116], which is completely independent of the gauge sector of the theory. The rest of Eq. (5.20a) is proportional to $zt = M_Z^2/m_t^2$, and represents the first correction to the Yukawa limit. To completely identify what we call the $\mathcal{O}(G_\mu^2 m_t^2 M_Z^2)$ contribution, we also have to provide a hierarchy in the couplings. Assuming $g_t \gg g, g'$, where $g'$ is the $U(1)$ coupling constant, we have the two possibilities $\lambda \gg g$ (Eq.(5.20a)), or $\lambda \ll g$ (Eq.(5.20b)).

Eqs.(5.20) exhibit a process-dependent contribution, i.e. $6\,zt\,I_3\,\hat{c}^2$, that comes from $B_Z{}^{(2)}$. This reflects the fact that, already at one-loop, the box diagrams in neutral current depend on the process under consideration [37] (cf. Eq.(5.14a)). Because we choose to write our result in terms of the physical $Z$ mass and $\hat{c}$, we obtain a process-dependent term by re-expressing $M_W$, in the one-loop $WW$ box diagram in neutral current, in favor of $M_Z$ using Eq.(2.19b), which at leading order reads

$$M_W^2 = M_Z^2 \hat{c}^2 \hat{\rho} \simeq M_Z^2 \hat{c}^2 \left(1 + N_c \frac{\hat{\alpha}}{4\pi \hat{s}^2 \hat{c}^2} \frac{m_t^2}{4M_Z^2}\right) \ . \tag{5.22}$$

## 5.3 Ward identities by Current Algebra

The local gauge symmetry of the SM Lagrangian induces relations between different Green functions, that are usually called Slavnov-Taylor identities [121], and generalize the Ward-Takahashi identities which derive from global symmetries. The use of the Becchi-Rouet-Stora (BRS) invariance of the SM Lagrangian makes their derivation very simple (see for ex. [122]). In particular, in a renormalizable gauge parametrized by $\xi$ (the 't Hooft-Feynman gauge corresponds to $\xi = 1$), we find for the longitudinal parts of the unrenormalized gauge field propagators

$$\begin{aligned}
&\langle \partial_\mu A^\mu(x) \partial_\nu A^\nu(0) \rangle = -i\xi \delta^4(x), \\
&\langle \partial_\mu A^\mu(x) \partial_\nu Z^\nu(0) \rangle + \xi M_Z \partial_\mu A^\mu(x) \Phi_2(0) \rangle = 0, \\
&\langle \partial_\mu Z^\mu(x) \partial_\nu Z^\nu(0) \rangle + \xi M_Z \langle \partial_\mu Z^\mu(x) \Phi_2(0) \rangle \\
&+ \xi M_Z \langle \Phi_2(x) \partial_\nu Z^\nu(0) \rangle + \xi^2 M_Z^2 \langle \Phi_2(x) \Phi_2(0) \rangle = i\xi \delta^4(x),
\end{aligned} \tag{5.23}$$

where $A^\mu$, $Z^\mu$, and $\Phi_2$ are the photon, $Z$, and neutral Goldstone boson fields, and I have used the short-hand notation $\langle ... \rangle$ for the vacuum expectation value (v.e.v.) of the covariant time-ordered product $\langle 0 \,|\, T^* ... |\, 0 \rangle$. The case of $W$ is completely analogous. The identities Eqs. (5.23) make manifest the connection between the longitudinal degrees of freedom of the gauge bosons and their unphysical counterparts. As we are interested in vacuum polarization functions, we can extract identities for the self-energies of the corresponding fields at an arbitrary $q^2$. At one-loop level, the result is essentially the one given in Sec. 4.3. At two-loop, however, the identities complicate considerably, and involve products of one-loop self-energies as well as irreducible VPF's, as can be seen, for instance, from Eq.(7.2) of Ref. [123]. Of course, at leading order in $m_t$, they reduce to Eqs.(5.1), but they are not of practical use in the case at hand, where the next-to-leading contributions need to be retained. An alternative is represented by the use of Current Algebra.

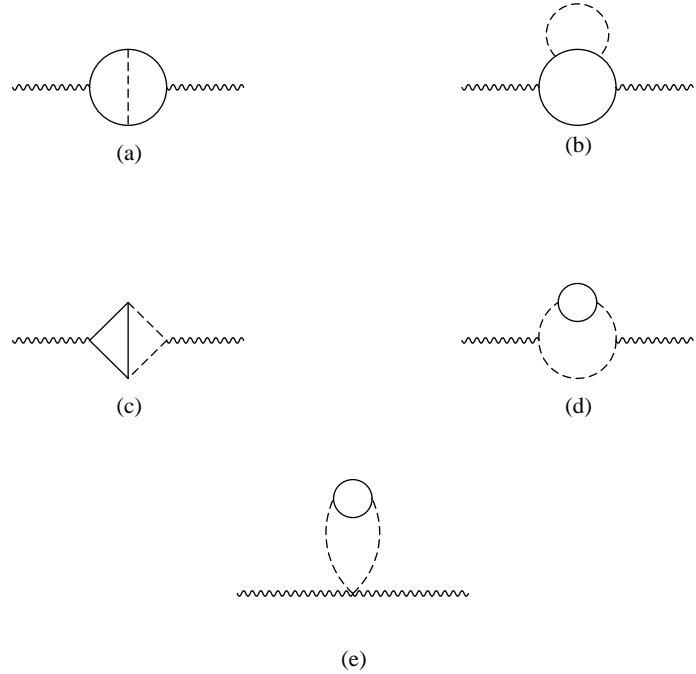

Figure 5.2: Two-loop diagrams contributing to the vector boson self-energies at $\mathcal{O}(G_\mu^2 m_t^4)$.

In this section I present a Current Algebra derivation of Ward identities at $q^2 = 0$ for the subset of diagrams contributing to leading $\mathcal{O}(G_\mu^2 m_t^4)$. The same method can be applied to the remaining subleading contributions, and yields the identities reported in App. C. Just by power-counting inspection it is easy to realize that the only diagrams which can contribute to the leading $m_t^4$ term are those containing, besides top and bottom, the physical and unphysical scalars. A suitable choice of the tadpole counterterm allows us to neglect the tadpole diagrams in the calculation [126]. Consequently, the only topologies that contribute at leading order in $m_t$ are those depicted in Fig. 5.2. In the figure, wavy lines represent vector bosons, dashed lines physical or unphysical scalars, while solid lines are fermions.

In order to fix the notation, I write the part of the SM Lagrangian density that describes the interaction of the $W$, $Z$ and scalars with fermions as

$$\mathcal{L}_{int} = -\frac{g}{\sqrt{2}}(W_\mu^\dagger J_W^\mu + \text{h.c.}) - \frac{g}{c} Z_\mu J_Z^\mu - \frac{g}{2\,M_W}\left[\Phi_1 S_1 + \Phi_2 S_2 + \sqrt{2}\,(\Phi^\dagger S + \text{h.c.})\right], \tag{5.24}$$

where $J_Z^\mu$ and $J_W^\mu$ are the fermionic currents coupled to $Z$ and $W$ respectively, $W^\dagger$ is the field that creates a $W^+$ meson, $\Phi_1$ is the physical Higgs boson, $\Phi_2$ and $\Phi$ the unphysical counterparts associated with the $Z$ and $W$.

$$S_1 = \bar{\psi}\,m^0\,\psi, \tag{5.25a}$$

$$S_2 = 2\,\partial_\mu J_Z^\mu = -i\bar{\psi}\,m^0\,C_3\gamma_5\psi, \tag{5.25b}$$

$$S = - i\, \partial_\mu J_W^\mu = \bar{\psi} \Gamma \psi, \tag{5.25c}$$

are the hadronic currents coupled to the scalar fields. In Eqs. (5.25), $\psi$ represents the column vector $\psi \equiv (t, b)^T$, $m^0$, $C_3$ and $\Gamma$ are the $2 \times 2$ matrices

$$m^0 = \begin{pmatrix} m_t^0 & 0 \\ 0 & 0 \end{pmatrix} \tag{5.26a}$$

$$C_3 = \begin{pmatrix} 1 & 0 \\ 0 & -1 \end{pmatrix} \tag{5.26b}$$

$$\Gamma = \begin{pmatrix} 0 & 0 \\ -m_t^0\, a_+ & 0 \end{pmatrix}, \tag{5.26c}$$

$a_+ \equiv \frac{1+\gamma_5}{2}$, and the superscript $0$ on $m_t$ refers to the bare mass. It is evident from Eqs. (5.26a) and (5.26c) that we consider only the third generation, and take the bottom quark as massless. We begin by studying $A_{ZZ}(0)$. Using current correlation functions, and working in $n$ dimension, we can combine the amplitudes of Figs. 5.2a,b, where the continuous line represents a top and the dashed one a Higgs or $\Phi_2$, into the expression

$$\Pi_{(a,b)}^{\mu\nu}(q^2) = \sum_{j=1}^{2} \frac{g^4}{4\, c^2 M_W^2} \frac{1}{2} \int_k \frac{1}{k^2 - m_j^2} \times$$

$$\times \int_y e^{-iq\cdot y} \int_{x_1} e^{ik\cdot x_1} \int_{x_2} e^{-ik\cdot x_2} \langle 0\, |\, T^* \left[ J_Z^\mu(y)\, J_Z^\nu(0)\, S_j(x_1)\, S_j(x_2) \right] |\, 0 \rangle, \tag{5.27}$$

with the short-hand notation $\int_k = \int d^n k / ((2\pi)^n \mu^{n-4})$ and $\int_{x_i} = \int d^n x$, where $\mu$ is the 't Hooft mass scale, $T^*$ is the covariant time-ordered product and $m_1 \equiv M_H$ and $m_2 \equiv M_Z$. In the case of the unphysical scalar, Eq. (5.27) is valid in the 't Hooft-Feynman gauge.

In order to trigger Ward identities we contract $\Pi_{(a,b)}^{\mu\nu}(q^2)$ with $q^\mu q^\nu$. Contraction of a current operator with its four-momentum gives rise to a term involving the divergence of a current plus an equal time commutator that reduces the number of operators inside the time-ordered product by one unit. Noticing that

$$\frac{\partial}{\partial q^2} \left\{ q_\mu q_\nu \Pi_{ZZ}^{\mu\nu}(q^2) \right\}_{q^2=0} = A_{ZZ}(0), \tag{5.28}$$

we obtain

$$A_{(a,b)}(0) = \frac{g^4}{8\, c^2 M_W^2} \frac{\partial}{\partial q^2} \int_k \int_y e^{-iq\cdot y} \times$$

$$\times \left\{ \sum_{j=1}^{2} \int_{x_1} e^{ik\cdot x_1} \int_{x_2} e^{-ik\cdot x_2} \frac{\langle S_2(y)\, S_2(0)\, S_j(x_1)\, S_j(x_2) \rangle}{4(k^2 - m_j^2)} \right.$$

$$\left. + i \int_x e^{ik\cdot x} \langle S_2(y)\, S_1(x)\, S_2(0) \rangle \left[ \frac{1}{k^2 - m_1^2} - \frac{1}{(k-q)^2 - m_2^2} \right] \right\}_{q^2=0}$$

$$+ \frac{g^4}{16\, c^2 M_W^2} \int_k \left\{ \left( 1 - \frac{4}{n} \frac{k^2}{k^2 - m_1^2} \right) \int_x e^{ik\cdot x} \frac{\langle S_2(x)\, S_2(0) \rangle}{(k^2 - m_1^2)^2} + (1 \leftrightarrow 2) \right\}. \tag{5.29}$$

In Eq.(5.29) the notation $(1 \leftrightarrow 2)$ represents a term obtained by the previous one inside the curly bracket by the substitution $1 \leftrightarrow 2$.

We now examine the contributions that are described by the topology shown in Fig. 5.2c. Let us consider the case when one dashed line represents a $\Phi_1$ and the other one a $\Phi_2$. We write

$$\Pi^{\alpha\nu}_{(c)}(q^2) = -\frac{g^4}{4\,c^2 M_W^2} \int_k \int_y e^{-iq\cdot y} \int_x e^{ik\cdot x} \frac{(2k-q)^\nu \,\langle J_z^\alpha(y)\, S_1(x)\, S_2(0)\rangle}{[k^2 - m_1^2][(k-q)^2 - m_2^2]}. \tag{5.30}$$

Contraction with $q^\alpha$ gives

$$q_\alpha \Pi^{\alpha\nu}_{(c)}(q^2) = \frac{ig^4}{8\,c^2 M_W^2} \int_k \frac{(2k-q)^\nu}{[k^2 - m_1^2][(k-q)^2 - m_2^2]} \times$$
$$\times \left\{ \int_y e^{-iq\cdot y} \int_x e^{ik\cdot x} \langle S_2(y)\, S_1(x)\, S_2(0)\rangle + i \int_x \left[ e^{i(k-q)\cdot x} \langle S_2(x)\, S_2(0)\rangle \right.\right.$$
$$\left.\left. - e^{ik\cdot x}\langle S_1(x)\, S_1(0)\rangle \right] \right\}. \tag{5.31}$$

Recalling that

$$\frac{\partial}{\partial q_\mu}\left(q_\alpha \Pi^{\alpha\nu}(q^2)\right) = \Pi^{\mu\nu}(q^2) + q_\alpha \frac{\partial}{\partial q_\mu}\Pi^{\alpha\nu}(q^2) \tag{5.32}$$

it follows that the contribution of Fig. 5.2c to $\Pi^{\mu\nu}_{zz}(0)$ is obtained by differentiating Eq.(5.31) with respect to $q^\mu$ and then setting $q^2 = 0$. Consider now the first term in Eq.(5.31),

$$T^\nu = \int_k \frac{(2k-q)^\nu}{[k^2 - m_1^2][(k-q)^2 - m_2^2]} \left\{ \, \cdots \, \right\} \tag{5.33}$$

where $\left\{ \cdots \right\}$ represents the three-point correlation function times the appropriate constants. By Lorentz invariance, $T^\nu$ should have the form $T^\nu = \alpha(q^2)\, q^\nu$, and therefore

$$\frac{\partial}{\partial q_\mu}T^\mu = 2\,\frac{\partial\alpha}{\partial q^2}(q^2)\, q^\mu q^\nu + \alpha(q^2)\, g^{\mu\nu}. \tag{5.34}$$

The contribution to $A(q^2)$ is then given by $\alpha(q^2)$. It is easy to show that

$$\alpha(0) = \frac{\partial}{\partial q^2}(q_\nu T^\nu)\bigg|_{q^2=0}. \tag{5.35}$$

Using Eq.(5.35), we can write for the transverse part of $\Pi^{\mu\nu}_{(c)}$:

$$A_{(c)}(0) = \frac{ig^4}{8\,c^2 M_W^2} \frac{\partial}{\partial q^2} \left\{ \int_y e^{-iq\cdot y} \int_x e^{ik\cdot x} \langle S_2(y)\, S_1(x)\, S_2(0)\rangle \times \right.$$
$$\times \int_k \left[ \frac{1}{(k-q)^2 - m_2^2} - \frac{1}{k^2 - m_1^2} + \frac{m_1^2 - m_2^2}{(k^2 - m_1^2)[(k-q)^2 - m_2^2]} \right] \Bigg\}_{q^2=0}$$
$$- \frac{g^4}{8\,c^2 M_W^2} \int_k \left\{ \left( 1 - \frac{4}{n}\frac{k^2}{k^2 - m_1^2} \right) \int_x e^{ik\cdot x} \frac{\langle S_2(x)\, S_2(0)\rangle}{(k^2 - m_1^2)(k^2 - m_2^2)} + (1 \leftrightarrow 2) \right\}. \tag{5.36}$$

The contribution of $\Phi_1$ and $\Phi_2$ to Figs. 5.2d,e can be similarly computed:

$$A_{d,e}(0) = \frac{g^4}{16\,c^2 M_W^2} \int_k \left\{ \left( 1 - \frac{4}{n}\frac{k^2}{k^2 - m_2^2} \right) \int_x e^{ik\cdot x} \frac{\langle S_1(x)\,S_1(0)\rangle}{(k^2 - m_1^2)^2} + (1 \leftrightarrow 2) \right\}. \tag{5.37}$$

The diagrams involving the charged scalar $\Phi$ can be considered along the same lines, and give

$$A_{(a-e)}(0) = \frac{g^4}{8\,c^2 M_W^2} \frac{\partial}{\partial q^2} \left\{ \int_k \frac{1}{k^2 - M_W^2} \right.$$
$$\left. \times \int_y e^{-iq\cdot y} \int_{x_1} e^{ik\cdot x_1} \int_{x_2} e^{-ik\cdot x_2} \langle S_2(y)\,S_2(0)\,S^\dagger(x_1)\,S(x_2)\rangle \right\}_{q^2=0}. \tag{5.38}$$

Finally, summing Eqs. (5.29), and (5.37)–(5.38), we find that the subset of diagrams defined above (see Fig. 5.2) contributes the transverse part of the $Z$ self-energy at $q^2 = 0$ a factor

$$\frac{A_{ZZ}^{\;lead}(0)}{M_Z^2} = \frac{g^4}{8 M_W^4} \frac{\partial}{\partial q^2} \int_k \left\{ \left\{ \int_y e^{-iq\cdot y} \int_{x_1} e^{ik\cdot x_1} \int_{x_2} e^{-ik\cdot x_2} \right. \right.$$
$$\times \left[ \frac{\langle S_2(y)\,S_2(0)\,S^\dagger(x_1)\,S(x_2)\rangle}{k^2 - M_W^2} + \sum_{j=1}^{2} \frac{\langle S_2(y)\,S_2(0)\,S_j(x_1)\,S_j(x_2)\rangle}{4(k^2 - m_j^2)} \right]$$
$$+ i(M_H^2 - M_Z^2) \int_y e^{-iq\cdot y} \int_x e^{ik\cdot x} \frac{\langle S_2(y)\,S_1(x)\,S_2(0)\rangle}{(k^2 - M_H^2)((k-q)^2 - M_Z^2)} \right\}_{q^2=0}$$
$$+ \frac{g^4}{16 M_W^4} \int_k \frac{(M_H^2 - M_Z^2)^2}{(k^2 - M_H^2)^2 (k^2 - M_Z^2)^2} \times$$
$$\left. \times \left\{ \left( 1 - \frac{4}{n}\frac{k^2}{k^2 - m_2^2} \right) \int_x e^{ik\cdot x} \langle S_1(x)\,S_1(0)\rangle + (1 \leftrightarrow 2) \right\} \right\}. \tag{5.39a}$$

The discussion of "leading" diagrams of the $W$ self-energy can be performed on the same footing as in the $Z$ case. Here I present only the result:

$$\frac{A_{WW}^{\;lead}(0)}{M_W^2} = \frac{g^4}{8 M_W^4} \frac{\partial}{\partial q^2} \int_k \left\{ \left\{ \int_y e^{-iq\cdot y} \int_{x_1} e^{ik\cdot x_1} \int_{x_2} e^{-ik\cdot x_2} \right. \right.$$
$$\times \left[ \frac{\langle S^\dagger(y)S(0)\left( S^\dagger(x_1)S(x_2) + \text{h.c.} \right)\rangle}{(k^2 - M_W^2)} + \sum_{j=1}^{2} \frac{\langle S^\dagger(y)\,S(0)\,S_j(x_1)\,S_j(x_2)\rangle}{2(k^2 - m_j^2)} \right]$$
$$+ \int_y e^{-iq\cdot y} \int_x e^{ik\cdot x} \left[ i(M_H^2 - M_W^2) \frac{\langle \left( S^\dagger(y)S(0) + \text{h.c.} \right) S_1(x)\rangle}{(k^2 - M_H^2)((k-q)^2 - M_W^2)} \right.$$
$$\left. \left. + (M_Z^2 - M_W^2) \frac{\langle \left( S^\dagger(y)S(0) - \text{h.c.} \right) S_2(x)\rangle}{(k^2 - M_Z^2)((k-q)^2 - M_W^2)} \right] \right\}_{q^2=0}$$
$$+ \frac{g^4}{16 M_W^4} \int_k \left\{ \frac{(m_1^2 - M_W^2)^2}{(k^2 - m_1^2)^2 (k^2 - M_W^2)^2} \left[ \left( 1 - \frac{4}{n}\frac{k^2}{k^2 - M_W^2} \right) \int_x e^{ik\cdot x} \langle S_1(x)\,S_1(0)\rangle \right. \right.$$

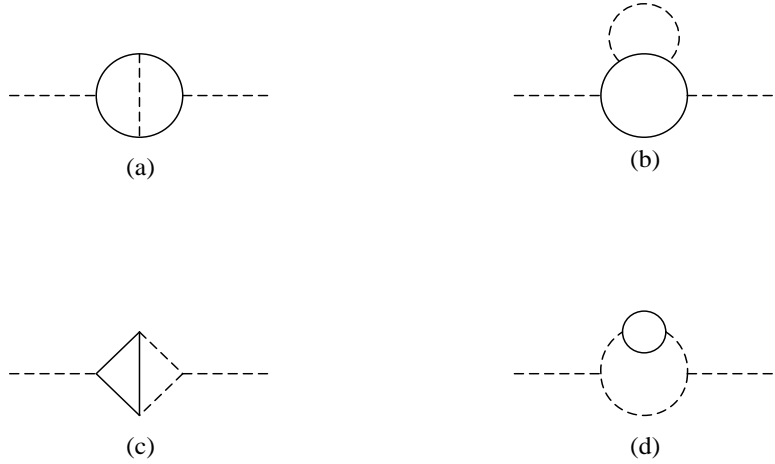

Figure 5.3: Two-loop diagrams contributing to $\Pi_{\Phi\Phi}$ and to $\Pi_{\Phi_2\Phi_2}$ in a Yukawa theory.

$$+\ 2\left(1-\frac{4}{n}\frac{k^2}{k^2-m_1^2}\right)\int_x e^{ik\cdot x}\langle S^\dagger(x)\,S(0)\rangle\bigg]+(1\leftrightarrow 2)\bigg\}.\qquad(5.39b)$$

In order to understand the connection between Eqs. (5.39) and the r.h.s. of Eqs. (5.1) consider for example the four-point correlation functions in Eq.(5.39a). In the limit $g$, $g' \to 0$ or $M_W$, $M_Z \to 0$, their contribution to $\Delta\rho$ is proportional to $g_t^4$. In fact each $S$ operator contains an $m_t = g_t v$ factor, where $v$ is the v.e.v., while the coefficient in front is proportional to $1/v^4$. In this limit the terms involving four-point functions represent exactly the contribution to $\frac{\partial}{\partial q^2}\Pi_{\Phi_2\Phi_2}(q^2)\big|_{q^2=0}$ of the diagrams shown in Figs. 5.3a and 5.3b. A similar connection can be established between the three- and two-point correlation functions in Eq.(5.39a) and the diagrams 5.3c and 5.3d respectively. We notice that Figs. 5.3c and 5.3d involve trilinear scalar couplings proportional to $M_H^2$. Although no such coupling is present in the diagrams of Fig. 5.2, we recover these terms using Eq.(5.35) and simple algebraic manipulations. Equation (5.39b) shows an analogous connection between its various contributions and $\frac{\partial}{\partial q^2}\Pi_{\Phi\Phi}(q^2)\big|_{q^2=0}$ computed in the Yukawa theory.

We have explicitly verified the Ward identities Eqs. (5.39), which represent an important check of the calculation. It might be worth emphasizing that Eqs. (5.39) establish a correspondence between a whole class of graphs evaluated at $q^2 = 0$, and no additional approximation has been applied. They are valid for the first term of the heavy top mass expansion (Eq.(5.1)), as well as for the next terms of the expansion; their physical content is therefore different from Eq.(5.1), which they generalize. Obviously, there are other 1PI diagrams that can contribute to subleading order. Ward identities dealing with them are considered in App. C.

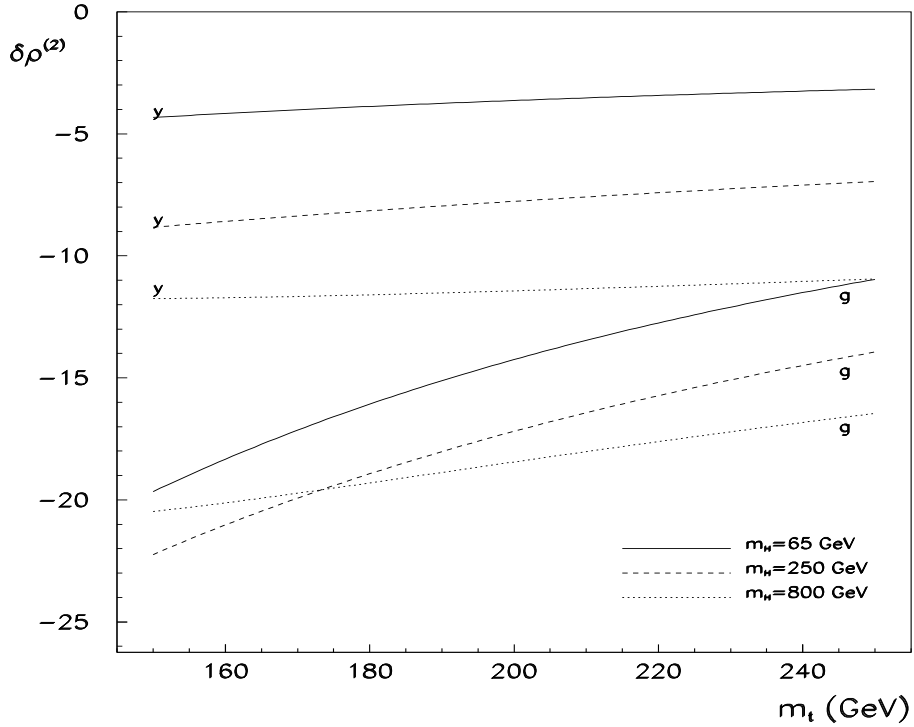

Figure 5.4: $\Delta\rho^{(2)}$ for $\nu_\mu e$ scattering, in units $N_c x_t^2$ as function of $m_t$ for few values of $M_H$: including only the $\mathcal{O}(G_\mu^2 m_t^4)$ contribution ($y$), and with both the $\mathcal{O}(G_\mu^2 m_t^4)$ and $\mathcal{O}(G_\mu^2 m_t^2 M_Z^2)$ terms ($g$).

## 5.4 Scheme dependence and numerical results

In Sec. 5.2, the calculation of $\Delta\rho$ was performed in the $\overline{MS}$ scheme, expressing the one-loop contribution in terms of $\hat{\alpha}$, $M_Z$, and $\hat{s}^2$ (with the exception of the top one-loop contribution, which is conveniently expressed in terms of $G_\mu$, as we have seen). I present now the corresponding expression for $\Delta\rho^{(2)}$ in terms of $\alpha$, $G_\mu$, $M_Z$ and the on-shell (OS) parameter $c^2 \equiv M_W^2/M_Z^2$. The relations [11]

$$\frac{\hat{\alpha}}{4\pi\hat{s}^2} = \frac{G_\mu M_W^2}{2\sqrt{2}\pi^2} \frac{1 - \Delta\hat{r}_W}{1 + (\frac{2\delta\epsilon}{e})_{\overline{MS}}} \simeq \frac{G_\mu M_Z^2 c^2}{2\sqrt{2}\pi^2} \tag{5.40a}$$

$$\hat{c}^2 = c^2(1 - Y_{\overline{MS}}) \simeq c^2(1 - N_c x_t) \tag{5.40b}$$

allow us to replace the $\overline{MS}$ quantities $\hat{\alpha}$, $\hat{s}^2$ by $G_\mu$ and $c^2$. Furthermore, Eq.(5.40b) creates additional contributions to $\Delta\rho^{(2)}$. The one-loop result is then given by Eqs. (5.13) with the substitutions $\hat{\alpha}/(4\pi\hat{s}^2) \to (G_\mu M_Z^2 c^2)/(2\sqrt{2}\pi^2)$, and $\hat{s}^2$, $\hat{c}^2 \to s^2$, $c^2$, while for the two-loop contribution we have

$$\Delta\rho_{OS}^{(2)} = \Delta\rho^{(2)}(\hat{s}^2, \hat{c}^2 \to s^2, c^2) \tag{5.41a}$$

$$+ N_c x_t^2 zt \left[ -\frac{3c^4}{s^4} \ln c^2 - \frac{3c^2}{s^2} - 3I_3 + 12Q - 24s^2(1+c^2)I_3Q^2 + 4c^2G'(\xi, c^2) \right],$$

where

$$G'(\xi, c^2) = \frac{3}{4}\xi \left[ c^2 \frac{\ln(c^2/\xi)}{(c^2 - \xi)^2} - \frac{1}{c^2 - \xi} + \frac{1}{c^2}\frac{\ln \xi}{1 - \xi} \right]. \tag{5.41b}$$

In Eq.(5.41a) $\Delta\rho^{(2)}(\hat{s}^2, \hat{c}^2 \to s^2, c^2)$ represents a term obtained from Eqs. (5.20) applying the same substitutions as in the one-loop case.

Fig. 5.4 shows $\Delta\rho^{(2)}$ (Eqs. (5.20)) as a function of $m_t$ for different values of $M_H$. As a comparison, we also show the values obtained including only the $\mathcal{O}(G_\mu^2 m_t^4)$ contribution. The process under consideration is $\nu_\mu e$ scattering. From the figure it is evident that the inclusion of corrections suppressed by a factor $M_Z^2/m_t^2$ with respect to the leading term has a quite substantial effect. For $\nu - \nu$ scattering the effect is even more pronounced. To have a better understanding of the size of these effects and the difference induced by the choice of the renormalization scheme, I present in Table 5.1 the values of $\Delta\rho^{(2)}$ and $\Delta\rho_{OS}^{(2)}$ for $zt = 0$, and for $zt = 0.2, 0.3$, corresponding to $m_t = \approx 200, 170\,\mathrm{GeV}$, as a function of $r = M_H/m_t$. In preparing the table we matched the values coming from (5.20a) and (5.20b) where they meet. We see that in the region of light Higgs the $\mathcal{O}(G_\mu^2 m_t^2 M_Z^2)$ corrections are much larger than the $m_t^4$ term, that is actually suppressed by accidental cancellations, while for large Higgs mass, in the TeV region, their contribution is still 50% of the leading part. It is intriguing to note that the numbers shown in Table 5.1 are very close to the corresponding ones obtained in Ref. [107] in the case of a model with $SU(2)$ symmetry, where $g'$ has been set to zero.

From Eq.(5.41a) we see that the process-dependence is more pronounced in the OS framework. This is easily understood by noticing that the expansion of the bare couplings in the one-loop box diagrams gives rise, unlike the $\overline{\mathrm{MS}}$ case, to $m_t^2$ contributions. It is worth mentioning that in the $\overline{\mathrm{MS}}$ framework the process-dependence can be confined entirely in the one-loop contribution. This can be achieved by expressing the one-loop vertices and boxes using $\overline{\mathrm{MS}}$ couplings and the physical $W$ and $Z$ masses whenever they appear in the propagators. Such a procedure is frequently used in one-loop calculations [11]. A numerical investigation shows that this procedure minimizes $\Delta\rho^{(2)}$, for any value of $M_H, m_t$. I present the values of $\Delta\rho^{(2)}$ for this particular $\overline{\mathrm{MS}}$ prescription in Table 5.2 in the case $m_t = 180\,\mathrm{GeV}$, and I will use them for the estimates of the next section. Table 5.1, instead, has been obtained following Sec. 5.2. In Fig. 5.5 I plot $\Delta\rho^{(2)}$ for $\nu_\mu e$ scattering as a function of the ratio between $M_H$ and $m_t$, expressing again the one-loop result in terms of $\overline{\mathrm{MS}}$ couplings and physical vector boson masses. For a comparison, the upper curve shows only the leading $\mathcal{O}(G_\mu^2 m_t^4)$ term. It is interesting to note the very large difference in the light Higgs regime.

Our investigation of the two-loop next-to-leading top corrections to the $\rho$ parameter reveals several interesting features. i) The $\mathcal{O}(G_\mu^2 m_t^2 M_Z^2)$ corrections are not universal but depend upon the physical process under consideration. Indeed such terms can originate from self-energies as well as from vertex and box diagrams, which are process dependent. ii) The asymptotic result, obtained in a Yukawa theory, does not seem to be a realistic approximation for values of the top mass compatible with the experimental constraints. In this range, the first order gauge corrections are numerically comparable to the $\mathcal{O}(G_\mu^2 m_t^4)$ contribution. The values of $m_t$ for which $\Delta\rho^{(2)}$ is well approximated by the $\mathcal{O}(G_\mu^2 m_t^4)$ contribution are very large. Typically, $\Delta\rho^{(2)}$ starts to be within 10% of the leading $m_t^4$ value for $m_t \approx 800\,\text{GeV}$. iii) The $\mathcal{O}(G_\mu^2 m_t^2 M_z^2)$ contribution has the same sign of the leading term and enhances its effect.

| $r = \dfrac{M_H}{m_t}$ | $zt = 0$ | $\overline{\text{MS}}$ | | $OS$ | |
|:---:|:---:|:---:|:---:|:---:|:---:|
| | | $zt = 0.2$ | $zt = 0.3$ | $zt = 0.2$ | $zt = 0.3$ |
| 0.1 | $-$ 1.8 | $-12.6$ | $-15.8$ | $-12.7$ | $-16.0$ |
| 0.2 | $-$ 2.7 | $-13.3$ | $-16.5$ | $-13.5$ | $-16.8$ |
| 0.3 | $-$ 3.5 | $-13.9$ | $-17.0$ | $-14.2$ | $-17.4$ |
| 0.4 | $-$ 4.1 | $-14.5$ | $-17.6$ | $-14.9$ | $-18.1$ |
| 0.5 | $-$ 4.7 | $-15.2$ | $-18.3$ | $-15.7$ | $-18.9$ |
| 0.6 | $-$ 5.2 | $-15.8$ | $-18.9$ | $-16.6$ | $-19.7$ |
| 0.7 | $-$ 5.7 | $-16.2$ | $-19.8$ | $-16.9$ | $-20.9$ |
| 0.8 | $-$ 6.2 | $-16.4$ | $-20.1$ | $-17.1$ | $-21.0$ |
| 0.9 | $-$ 6.6 | $-16.5$ | $-20.1$ | $-17.4$ | $-21.2$ |
| 1.0 | $-$ 6.9 | $-16.6$ | $-20.1$ | $-17.6$ | $-21.3$ |
| 1.1 | $-$ 7.3 | $-16.8$ | $-20.2$ | $-17.8$ | $-21.4$ |
| 1.2 | $-$ 7.6 | $-16.9$ | $-20.2$ | $-18.0$ | $-21.6$ |
| 1.3 | $-$ 7.9 | $-17.0$ | $-20.2$ | $-18.2$ | $-21.7$ |
| 1.4 | $-$ 8.2 | $-17.2$ | $-20.3$ | $-18.4$ | $-21.9$ |
| 1.5 | $-$ 8.4 | $-17.3$ | $-20.3$ | $-18.6$ | $-22.0$ |
| 1.6 | $-$ 8.7 | $-17.4$ | $-20.4$ | $-18.7$ | $-22.1$ |
| 1.7 | $-$ 8.9 | $-17.5$ | $-20.5$ | $-18.9$ | $-22.3$ |
| 1.8 | $-$ 9.1 | $-17.6$ | $-20.5$ | $-19.1$ | $-22.4$ |
| 1.9 | $-$ 9.3 | $-17.7$ | $-20.6$ | $-19.2$ | $-22.6$ |
| 2.0 | $-$ 9.5 | $-17.8$ | $-20.6$ | $-19.4$ | $-22.7$ |
| 2.5 | $-10.2$ | $-18.2$ | $-20.9$ | $-20.0$ | $-23.3$ |
| 3.0 | $-10.8$ | $-18.4$ | $-20.8$ | $-20.4$ | $-23.5$ |
| 3.5 | $-11.2$ | $-18.3$ | $-20.6$ | $-20.6$ | $-23.6$ |
| 4.0 | $-11.4$ | $-18.3$ | $-20.4$ | $-20.6$ | $-23.5$ |
| 4.5 | $-11.6$ | $-18.2$ | $-20.1$ | $-20.6$ | $-23.4$ |
| 5.0 | $-11.7$ | $-18.0$ | $-19.8$ | $-20.5$ | $-23.3$ |
| 5.5 | $-11.8$ | $-17.8$ | $-19.4$ | $-20.4$ | $-23.1$ |
| 6.0 | $-11.8$ | $-17.5$ | $-19.0$ | $-20.3$ | $-22.9$ |

Table 5.1: $\Delta\rho^{(2)}$ ($\overline{\text{MS}}$) and $\Delta\rho_{OS}^{(2)}$ ($OS$) relevant to $\nu_\mu\, e$ scattering for $zt \equiv M_z^2/m_t^2 = 0.2, 0.3$, in units $N_c x_t^2$ as a function of $r = M_H/m_t$. The column $zt = 0$ is the result of the Yukawa theory.

| $M_H/m_t$ | $\Delta\rho^{(2)}{}_{\overline{\text{MS}}}$ | $M_H/m_t$ | $\Delta\rho^{(2)}{}_{\overline{\text{MS}}}$ |
|-----------|------------|-----------|------------|
| 0.1 | $-11.3$ | 1.5 | $-15.8$ |
| 0.2 | $-11.9$ | 1.6 | $-15.9$ |
| 0.3 | $-12.5$ | 1.7 | $-16.0$ |
| 0.4 | $-13.0$ | 1.8 | $-16.1$ |
| 0.5 | $-13.6$ | 1.9 | $-16.1$ |
| 0.6 | $-14.4$ | 2.0 | $-16.2$ |
| 0.7 | $-15.2$ | 2.5 | $-16.6$ |
| 0.8 | $-15.3$ | 3.0 | $-16.6$ |
| 0.9 | $-15.3$ | 3.5 | $-16.5$ |
| 1.0 | $-15.4$ | 4.0 | $-16.3$ |
| 1.1 | $-15.5$ | 4.5 | $-16.1$ |
| 1.2 | $-15.6$ | 5.0 | $-15.8$ |
| 1.3 | $-15.7$ | 5.5 | $-15.5$ |
| 1.4 | $-15.7$ | 6.0 | $-15.2$ |

Table 5.2: $\Delta\rho^{(2)}$ $(\overline{\text{MS}})$ relevant to $\nu_\mu e$ scattering for $zt \equiv M_z^2/m_t^2 = 0.2, 0.3$, in units $N_c x_t^2$ as a function of $r = M_H/m_t$.

## 5.5 Extrapolation to $M_W$ and $\sin^2\theta_{eff}^{lept}$

The result illustrated in the previous section, obtained at $q^2 = 0$, cannot be directly applied to LEP physics. We have seen that there are terms $\mathcal{O}(G_\mu^2 m_t^2 M_z^2)$ that depend on the specific process under consideration, and we know that $\Delta r$ and $\Delta\hat{r}$ involve two-loop vacuum polarization functions evaluated at $q^2 = M_Z^2$ or $M_W^2$. They differ from the self-energies at $q^2 = 0$ by terms of $\mathcal{O}(G_\mu^2 m_t^2 M_z^2)$. However, the size of the correction that we have found, relative to the leading $\mathcal{O}(G_\mu^2 m_t^4)$ term, suggests to attempt a quantitative estimate of the theoretical error that we can make when we include the leading corrections, but neglect the ones of $\mathcal{O}(G_\mu^2 m_t^2 M_z^2)$. A complete calculation of this kind of effects in $\Delta r$ and $\Delta\hat{r}$, the relevant radiative corrections in the prediction of $M_W$ and $\sin^2\theta_{eff}^{lept}$, is under way. Meanwhile, we can first notice that for $m_t$ in the current range the result of the previous section is not a complete surprise. Contributions of next-to-leading order are expected to be roughly $(\hat{\alpha}/\pi\hat{s}^2)^2 m_t^2/M_z^2 \ln(m_t^2/M_z^2) \approx 6 \times 10^{-4}$ for $m_t = 180\,\text{GeV}$, of the order of magnitude of the uncertainty due to the hadronic contribution to $\Delta\alpha$.

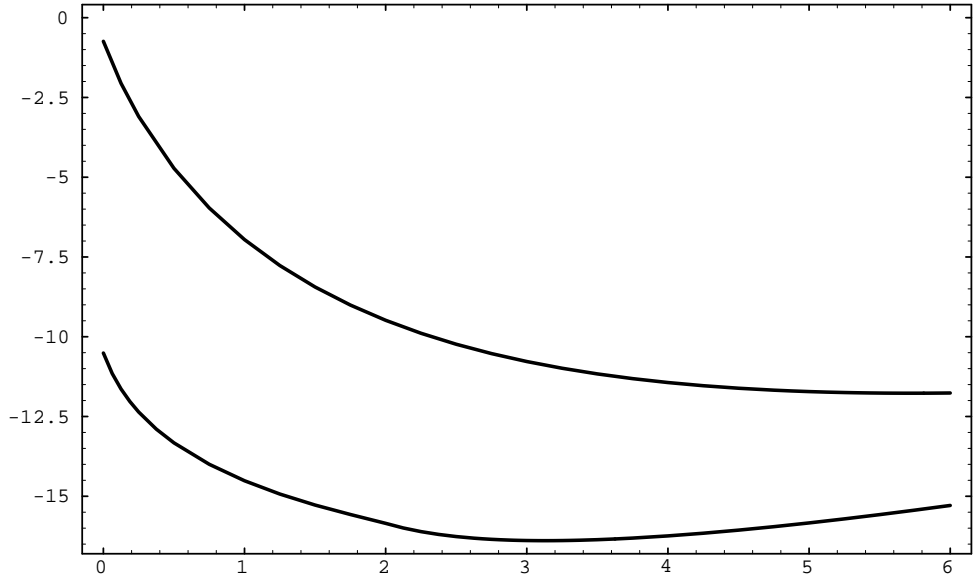

**Figure 5.5:** $\Delta\rho^{(2)}$ for $\nu_\mu e$ scattering, in units $N_c x_t^2$ for $m_t = 180\text{GeV}$ as a function of $M_H/m_t$: including only the $\mathcal{O}(G_\mu^2 m_t^4)$ contribution (upper curve), and with both the $\mathcal{O}(G_\mu^2 m_t^4)$ and $\mathcal{O}(G_\mu^2 m_t^2 M_Z^2)$ terms (lower curve). The one-loop bosonic contribution is expressed in terms of $\overline{\text{MS}}$ couplings and physical vector boson masses.

More specifically, we can assume that the ratio between the $\mathcal{O}(G_\mu^2 m_t^2 M_Z^2)$ and the $\mathcal{O}(G_\mu^2 m_t^4)$ contributions in $\Delta\rho^{(2)}$ be representative of the unknown two-loop top effects in $\Delta r$ and $\Delta\hat{r}$. We can then use this ratio to estimate the additional contributions to $\Delta r$ and $\Delta\hat{r}$ simply multiplying it by the known $\mathcal{O}(G_\mu^2 m_t^4)$ terms of these quantities. The shifts in the W mass and the $\overline{\text{MS}}$ mixing angle, $\sin^2\hat{\theta}_W(M_Z)$ (and consequently in $\sin^2\theta_{eff}^{lept}$), due to these additional contributions can be estimated using Eqs. (2.15) and (2.21) once more. In Table 5.5 we show, for a few values of $m_t$ and $M_H$, the effect of our estimate of the unknown top contributions on the W mass and $\sin^2\theta_{eff}^{lept}$. We have assumed that Eq.(2.27) is not changed by effects of $\mathcal{O}(G_\mu^2 m_t^2 M_Z^2)$, which we can decouple from the definition of $\sin^2\hat{\theta}_W(M_Z)$, and that the implementation of the decoupling does not give relevant $\mathcal{O}(G_\mu^2 m_t^2 M_Z^2)$ effects in $\Delta\hat{r}$. The ratio of subleading and leading terms in $\Delta\rho^{(2)}$ has been computed using expressions slightly different from Eqs. (5.20). In fact, as mentioned in the previous section, we decided to minimize the two-loop corrections of our $\overline{\text{MS}}$ calculation by writing the one-loop contribution in terms of the $\overline{\text{MS}}$ couplings and of the physical masses of W and Z. Such a procedure [11] has the further advantage of eliminating the process-dependent terms. From the third column, it can immediately be seen that, for a fixed value of the top mass, the effect is more pronounced for a light Higgs boson. This is not surprising, bearing in mind the fact that the $\mathcal{O}(G_\mu^2 m_t^4)$ term is a monotonically increasing (in modulus) function of $M_H$.

It is important to stress that the numbers presented in Table 5.5, more than a definite estimate of the shifts in $M_W$ and $\sin^2\theta_{eff}^{lept}$, should be taken as an indication that subleading two-loop $m_t$ effects could be larger than what is naïvely expected. Their size is probably comparable to, and possibly larger than, the theoretical uncertainty due to the hadronic contribution to the photonic self-energy. As we have seen, the latter amounts to 13MeV and $2.5 \times 10^{-4}$ in $M_W$ and $\sin^2 \theta_{eff}^{lept}$, respectively.

| $m_t$ (GeV) | $M_H$ (GeV) | $R$ % | $\delta M_W$ (MeV) | $\delta \sin^2 \theta_{eff}^{lept}$ $(10^{-4})$ |
|---|---|---|---|---|
| 150 | 65 | 247 | −10 | 0.6 |
|  | 250 | 100 | −8 | 0.5 |
|  | 800 | 35 | −4 | 0.2 |
| 175 | 65 | 234 | −16 | 0.9 |
|  | 250 | 94 | −14 | 0.8 |
|  | 800 | 38 | −8 | 0.5 |
| 200 | 65 | 221 | −23 | 1.4 |
|  | 250 | 88 | −20 | 1.2 |
|  | 800 | 38 | −13 | 0.7 |

Table 5.3: Calculated ratio ($R$) of the $\mathcal{O}(G_\mu^2 m_t^2 M_Z^2)$ and the $\mathcal{O}(G_\mu^2 m_t^4)$ contributions in $\Delta \rho^{(2)}$, for a few values of $m_t$ and $M_H$. The corresponding estimate of the shifts in the W mass and $\sin^2 \theta_{eff}^{lept}$ are also presented (see text).

In conclusion, the calculation of $\Delta \rho^{(2)}$ presented here shows that, at least in the specific case of the $\rho$ parameter for neutrino-lepton scattering, two-loop electroweak top contributions are not well approximated by the $\mathcal{O}(G_\mu^2 m_t^4)$ term. It seems reasonable to expect a similar behavior in the more relevant cases of high-energy precision observables like $M_W$ and $\sin^2 \theta_{eff}^{lept}$. Although at present this kind of effects are well beyond the experimental resolution, the precision foreseen in the next decade (see Table 2.3) seems to justify a theoretical effort to investigate the reliability of the current analyses. In other words, a complete two-loop calculation for the most important precision observables may be worthwhile.

# Chapter 6

# Conclusions

In the preceding chapters I have focussed on the theoretical determination of the two precision observables $M_W$ and $\sin^2 \theta_{eff}^{lept}$, that appear to be most promising for a significant improvement of the experimental precision in the next decade. I have analyzed the major sources of uncertainty in these determinations, and tried to estimate their impact, in relation to the expected experimental accuracy. In doing so, I have also presented original calculations that, as in the case of QCD corrections, find applications in other areas of precision physics. A few points can now summarize the results of the investigations presented in the various chapters.

- The whole set of one-loop radiative corrections is needed to describe present data. Current experiments are beginning to probe radiative corrections beyond the leading fermionic contributions, and to become sensitive to the symmetry breaking sector of the theory. Even if this may not be reflected in a high sensitivity to the mass of the Higgs boson, the comparison with different models (supersymmetric SM, different symmetry breaking mechanism, etc.) will certainly be extremely valuable.

- At the moment, the uncertainty on the hadronic contribution to the running of $\alpha$ is one of the major obstacles to an improvement of the theoretical error. New low-energy data can help to solve this problem, but a clarification of diverging analyses [13, 22, 23] is also needed.

- For what concerns the QCD corrections entering the prediction of $M_W$ and $\sin^2 \theta_{eff}^{lept}$, we can safely argue that they are well under control. I have provided a set of explicit formulae that allow the computation of the universal QCD corrections to a number of processes, involving vector and scalar bosons, for any value of the transfer momentum and masses. The leading term proportional to $m_t^2$ is known up to $\mathcal{O}(\alpha_s^2)$ and the theoretical error can be estimated, for instance, according to Ref. [51].

- As shown in the specific case of the $\rho$ parameter at low-energy, next-to-leading $\mathcal{O}(G_\mu^2 m_t^2 M_Z^2)$ corrections to electroweak observables can be relatively sizable, and the leading $\mathcal{O}(G_\mu^2 m_t^4)$ term does not provide a good approximation of the two-loop electroweak corrections.

An extrapolation to high-energy observables indicates the need for an explicit calculation of $\mathcal{O}(G_\mu^2 m_t^2 M_Z^2)$ effects on the theoretical determination of $M_W$ and $\sin^2 \theta_{eff}^{lept}$. For the moment, it seems appropriate to consider this extrapolation as an estimate of the theoretical error associated with higher order electroweak contributions.

- A very recent work [137] confirms this point of view: a comparison between five different computer programs for the evaluation of radiative corrections, based on very different theoretical approaches, indicates discrepancies up to $\approx 25\text{MeV}$ and $\approx 2 \times 10^{-4}$ for $M_W$ and $\sin^2 \theta_{eff}^{lept}$, respectively, in the current $m_t$ range. These uncertainties or scheme ambiguities are of the same order of magnitude of the next-to-leading contributions I have discussed, and may be interpreted as indications of sizable higher order effects.

- Among the higher order contributions that I have considered, it appears that only the QCD corrections to the one-loop quadratic top contribution may be relevant at the present level of precision. However, if the performance of the experiments already planned will be at the level of expectations, a meaningful theoretical interpretation will require a complete study of two-loop effects for the most relevant precision observables. After all, it should be kept in mind that the accuracy of the $Z$-resonance data at the end of the LEP program has abundantly surpassed all the expectations of a few years ago.

# Appendix A

# Two-loop integrals

## A.1 Scalar integrals for $O(\alpha\alpha_s)$ calculations

In this appendix I provide the expressions of the one and two-loop scalar integrals which are needed in the evaluation of the two-loop vacuum polarization functions discussed in Ch. 4.

It has been shown in Ref. [123] that it is possible to express any two-loop VPF's exclusively in terms of one and two-loop integrals involving only products of scalar propagators. In other words, the numerator of two-loop integrals involved in the calculation of electroweak self-energies can always be cancelled against appropriate combinations of the factors $(p_i^2 - m_i^2)$ appearing in the denominators. This is not apparent from the two loop integrals reported here, some of which do involve integration momenta in the numerator, but can be easily understood. In fact, any numerator that cannot be cancelled against the corresponding denominator corresponds to a vector or tensor sub-integral in one of the two variables of integration. These one-loop sub-integrals can be further reduced to one-loop scalar sub-integrals according to a well-known procedure (see for ex. Ref. [28]), reducing the whole integral to a sum of two-loop scalar integrals. This is the case, for instance, of the integral

$$\int \int d^n k \ d^n p \frac{k \cdot p}{((k-q)^2 - m_1^2)^a((p-q)^2 - m_2^2)^b}.  \tag{A1}$$

This general decomposition is not possible, however, in the case of two-loop vertices and boxes, for which the tensor structure is more involved.

### A.1.1 Relevant one-loop integrals at $\mathcal{O}(\epsilon)$

In two-loop calculations one-loop integrals are needed at $\mathcal{O}(\epsilon)$. Products of two disconnected one-loop integrals can be found as a result of the above mentioned decomposition in scalar integrals, and of the insertion of counterterms in the first order diagrams. Following the notation previously set ($n = 4 - 2\epsilon$, $\alpha = -m_a^2/q^2$, etc.), we have:

$$D_A = -\frac{i}{q^2}(-q^2 e^\gamma)^\epsilon \int \frac{d^n k}{\pi^{n/2}} \frac{1}{k^2 - m_a^2} = \alpha^{1-\epsilon} \ e^{\epsilon\gamma} \ \Gamma(\epsilon - 1)$$

$$= -\alpha \left[ \frac{1}{\epsilon} + 1 - \log \alpha + \epsilon \left( \frac{1}{2} \log^2 \alpha - \log \alpha + 1 + \frac{\pi^2}{12} \right) + \mathcal{O}(\epsilon^2) \right], \qquad \text{(A2)}$$

$$\begin{aligned} K &= -i(-q^2 e^\gamma)^\epsilon \int \frac{d^n k}{\pi^{n/2}} \frac{1}{[k^2 - m_a^2][(k-q)^2 - m_b^2]} \\ &= \frac{1}{2} \left[ \frac{1}{\epsilon} + 2 + (\alpha - \beta + \lambda^{\frac{1}{2}}) \log x_a - \log \alpha \right] + \frac{1}{2} [\alpha \leftrightarrow \beta] + \mathcal{O}(\epsilon). \end{aligned} \qquad \text{(A3)}$$

The $\mathcal{O}(\epsilon)$ part of $K$ can be found in Ref. [128]. Note that in the case of the calculations $\mathcal{O}(\alpha \alpha_s)$ discussed in Ch. 4 the $\mathcal{O}(\epsilon)$ term of $K$ is not needed: its total contribution, via terms like $K^2$, $K D_A$, $K \partial K / \partial \alpha$, $\cdots$, vanishes in the final result. The derivatives of $K$ and $D_A$ with respect to $\alpha$ enter sometimes in the expressions of two-loop integrals and are given by

$$\frac{\partial D_A}{\partial \alpha} = -\frac{1}{\epsilon} \left[ 1 - \epsilon \log \alpha + \epsilon^2 \left( \frac{1}{2} \log^2 \alpha + \frac{\pi^2}{12} \right) \right] + \mathcal{O}(\epsilon^2) \qquad \text{(A4)}$$

$$\frac{\partial K}{\partial \alpha} = \frac{1}{2\lambda^{\frac{1}{2}}} \left[ (1 + \alpha - \beta + \lambda^{\frac{1}{2}}) \log x_a + (1 + \alpha - \beta - \lambda^{\frac{1}{2}}) \log x_b \right] + \mathcal{O}(\epsilon)$$

The following relations are useful for numerical evaluation of the VPFs; in the equal mass case, for $D = -1 - 4\alpha > 0$:

$$\frac{1}{2} \sqrt{1 + 4\alpha} \ln x = -D^{1/2} \arctan D^{-1/2} \qquad \text{(A5)}$$

where $x = \frac{4\alpha}{(1 + \sqrt{1 + 4\alpha})^2}$. In the arbitrary mass case:

$$\text{Re} \left[ \lambda^{\frac{1}{2}} (\ln x_a + \ln x_b) \right] = -4\Omega(\alpha, \beta) \qquad \text{(A6)}$$

where

$$\Omega(\alpha, \beta) = \begin{cases} \sqrt{\frac{-\lambda}{4}} \arccos \left[ -\frac{1 + \alpha + \beta}{2\sqrt{\alpha\beta}} \right] & \lambda < 0, \\ \\ \frac{\lambda^{\frac{1}{2}}}{4} \ln \left| \frac{1 + \alpha + \beta + \lambda^{\frac{1}{2}}}{1 + \alpha + \beta - \lambda^{\frac{1}{2}}} \right| & \lambda > 0. \end{cases} \qquad \text{(A7)}$$

## A.1.2  Scalar two-loop integrals.

All the two-loop integrals needed for the calculations described in Ch. 4 can be expressed in a compact form in terms of polylogarithms of algebraic functions of the masses and external momentum. This is far from being a general feature of two-loop scalar integrals, and it has been noted [67] that this is the case only if all three particle cuts of the diagram satisfy the condition

$$m_1^2 m_2^2 m_3^2 \, p^2 \, \Pi_\pm \left( p^2 - (m_1 \pm m_2 \pm m_3)^2 \right) = 0, \qquad \text{(A8)}$$

where $m_i$ are the three masses involved, $p$ the total momentum going across the cut, and the product is over all the combinations of signs. As the gluon is massless, all the two-loop integrals used in the calculations of Ch. 4 satisfy this condition. To simplify the notation, I use for the definition of the two-loop integrals the following abbreviations:

$$< X > = -(-q^2 e^{\gamma})^{2\epsilon} \int \frac{d^n k_1}{\pi^{n/2}} \int \frac{d^n k_2}{\pi^{n/2}} \; X \qquad (A9)$$

and

$$K_{1,2} \equiv k_{1,2}^2 - m_a^2 \; , \qquad K_0 \equiv (k_1 - k_2)^2 \; , \qquad Q_{1,2} \equiv (k_{1,2} - q)^2 - m_b^2. \qquad (A10)$$

With the help of the variables defined in Eqs. (2.9)–(2.10), the two-loop integrals are

$$P \equiv < \frac{K_0}{q^2 K_1 K_2 Q_1 Q_2} > = (1 - \alpha + \beta) K D_A + (1 + \alpha - \beta) K D_B$$
$$- \frac{1}{2}(D_A - D_B)^2 - \frac{1}{2}\lambda K^2;$$

$$W_A \equiv \frac{1}{(q^2)} < \frac{1}{K_1^2 K_2} > = -D_A \frac{\partial D_A}{\partial \alpha};$$

$$T_A \equiv \frac{1}{(q^2)^2} < \frac{Q_1}{K_1^2 K_2} > = D_A^2 - D_A(1 - \alpha + \beta)\frac{\partial D_A}{\partial \alpha};$$

$$U_A \equiv < \frac{1}{K_1^2 K_2 Q_1} > = -D_A \frac{\partial K}{\partial \alpha};$$

$$M_A \equiv \frac{1}{q^2} < \frac{1}{K_1 K_2 K_0} > = -\frac{1}{\alpha}\frac{1 - \epsilon}{1 - 2\epsilon} D_A^2;$$

$$V_A \equiv < \frac{1}{K_1^2 K_2 K_0} > = -\frac{1}{2\alpha^2}\frac{1 - \epsilon}{1 - 2\epsilon} D_A \left( D_A - 2\alpha \frac{\partial D_A}{\partial \alpha} \right);$$

$$R_A \equiv \frac{1}{q^2} < \frac{Q_1}{K_1^2 K_2 K_0} > = M_A + (1 - \alpha + \beta)V_A;$$

$$N_A \equiv \frac{1}{q^2} < \frac{(k_1 - 2k_2) \cdot q}{K_1 K_2 K_0 Q_1} > = \alpha \left[ 1 + \frac{\log x_a}{1 - x_a} \right] \left[ 1 + \frac{x_b \log x_b}{1 - x_b} \right];$$

$$J_A \equiv < \frac{1}{K_1 K_2 K_0 Q_1} >$$
$$= \frac{1}{2}(1 + \epsilon)K^2 + \frac{1}{2} \left[ 3 - \frac{x_a \log^2 x_a}{(1 - x_a)^2} - \frac{x_b \log^2 x_b}{(1 - x_b)^2} - G(x_a) + G(x_b) \right];$$

$$L \equiv < \frac{1}{q^2 K_1 Q_2 K_0} >$$
$$= \frac{1}{2} \left[ \frac{7}{8} + \frac{1}{4}(2 + \epsilon)K^2 + M_A + N_A - (1 - \alpha + \beta)J_A \right] + \frac{1}{2}[\alpha \leftrightarrow \beta];$$

$$Q_A \equiv \frac{1}{q^2} < \frac{Q_1}{K_1 K_2 Q_2 K_0} > = \frac{1}{2}\left[ (1 - \alpha + \beta)J_A - L + M_A \right] + N_A;$$

$$S_A \equiv < \frac{q^2}{K_1^2 K_2 Q_1 K_0} >$$
$$= \frac{1}{\lambda^{\frac{1}{2}}} \left[ \frac{\log x_a}{1 - x_a} \left( 1 + \frac{\log x_a}{1 - x_a} \right) - \frac{x_b \log x_b}{1 - x_b} \left( 1 + \frac{x_b \log x_b}{1 - x_b} \right) \right] - (1 + \epsilon)K \frac{\partial K}{\partial \alpha}$$
$$+ \frac{1}{2\alpha} \left[ \lambda^{\frac{1}{2}} G(x_a x_b) - \frac{1}{2}(1 - \alpha + \beta + \lambda^{\frac{1}{2}})G(x_a) + \frac{1}{2}(1 - \alpha + \beta - \lambda^{\frac{1}{2}})G(x_b) \right];$$

$$I \equiv \; < \frac{q^2}{K_1 K_2 Q_1 Q_2 K_0} >= F(1) + F(x_a x_b) - F(x_a) - F(x_b). \tag{A11}$$

The functions $F(x)$ and $G(x)$ are given in Eq.(4.22). Some of the preceding expressions can be obtained from one-loop integrals and the "master" integral $I$ by differentiation with respect to the masses, and exploiting symmetries in the shift and interchange of the integration variables $k_1$ and $k_2$, as well as of the masses $m_a$ and $m_b$. Others can be found in Ref. [134,55]. All the calculations described in Ch. 4 can be performed on the basis of this (non-minimal) set of integrals, using again symmetries and differentiation with respect to the masses. The master integrals $I$ has been calculated by Broadhurst in two different ways in Ref. [133].

## A.2    Small momentum expansion of VPF's

The two-loop self-energy integrals necessary for the calculation of Sec. 5.2 are evaluated at $q^2 = 0$; the use of Ward identities sometimes introduces derivatives of two-loop integrals at $q^2 = 0$, and they all involve many different masses. In the approximation that the bottom is massless, there are at least four different scales entering that calculation: $M_Z$, $M_W$, $M_H$, and $m_t$. The number of relevant Feynman diagrams is also quite large. It is therefore useful to develop an algorithm that can automatically compute the Taylor expansion of two-loop integrals in the squared momentum transfer $q^2$ [129–131]. For the case at hand we can restrict to the first two terms of such expansion. The coefficients of the momentum expansion correspond to two-loop vacuum integrals and can be computed exactly, for any value of the masses involved, and expressed in terms of polylogarithmic functions.

Formally, we can write the Taylor expansion around $q^2 = 0$ as

$$A(q^2) = \sum_{j=0}^{\infty} \frac{1}{j!(n/2)_j} \left( \frac{q^2}{4} \right)^j \left( \Box_q^j A(q^2) \right) \Big|_{q=0}, \tag{A12}$$

where $\Box_q \equiv \partial^2/\partial q_\alpha \partial q^\alpha$, and $(a)_j = \Gamma(a+j)/\Gamma(a)$. In practice this is equivalent to expanding each denominator containing the external momentum according to

$$\frac{1}{(k-q)^2 - m^2} \to \frac{1}{k^2 - m^2} \left( 1 + \frac{2k \cdot q - q^2}{k^2 - m^2} + \frac{4(k \cdot q)^2}{(k^2 - m^2)^2} + \mathcal{O}(q^3) \right). \tag{A13}$$

A rigorous theory of mass and momentum expansions of Feynman integral can be found in Ref. [132].

As mentioned above, the coefficients of the momentum expansion are vacuum integrals, i.e. integrals with zero external momentum. They can always be algebraically reduced to the form

$$\int d^n k \, d^n p \frac{1}{(k^2 - m_1^2)^a (p^2 - m_2^2)^b ((k-p)^2 - m_3^2)^c}, \tag{A14}$$

where $a, b, c$ are integers. Analytic expressions for any value of $\{a, b, c\}$ can be obtained [130] in terms of Appell's hypergeometric functions of two variables, but a more convenient representation in terms of simple polylogarithms follows from the realization that any integral with a given set of $\{a, b, c\}$ can be further reduced to a combination of integrals with $a, b, c \leq 1$. Integrals with all exponents equal to one can be expressed for any $m_{1,2,3}$ in terms of dilogarithms or Clausen functions [130]. Integrals with one of the exponents equal to zero correspond to the product of two disconnected one-loop integrals and can be trivially obtained from Eq.(A2). This reduction can be achieved equivalently by the use of recurrence relations based on integration-by-parts technique [130], or by differentiation of the basic $a = b = c = 1$ integral with respect to the masses. We have checked that the two approaches give the same result. Note also that vertex diagrams at $q^2 = 0$ can be treated by the same method, and effectively correspond to the same kind of integrals.

# Appendix B

# QCD corrections: useful formulae

## B.1  Asymptotic expressions

The expressions obtained in Ch. 4 for the VPF's of electroweak vector bosons and scalars are as general as possible, but in many applications it is sufficient to know some asymptotic behavior of the general formulae. As the asymptotic expansions are usually not elementary, in this Appendix I give a list of useful expressions for the most relevant limiting cases: equal quark masses, one massless quark, and high momentum transfer. The expressions for vanishing external momentum in the case of vector boson self-energies have been discussed in Sec. 4.1.3. The notation will follow Ch. 4.

### B.1.1  Vector boson self-energies

From the one-loop VPF of Eq. (4.6), it is easy to obtain for $m_b = m_a$

$$\Pi_T^V(s) = \frac{s}{3} \left[ \frac{1}{\epsilon} - \rho_a + \frac{5}{3} - 4\alpha + (1 + 4\alpha)^{1/2}(1 - 2\alpha) \ln \frac{4\alpha}{(1 + \sqrt{1 + 4\alpha})^2} \right],$$

$$\Pi_T^A(s) = \frac{s}{3} \left[ \left( \frac{1}{\epsilon} - \rho_a \right)(1 + 6\alpha) + \frac{5}{3} + 8\alpha + (1 + 4\alpha)^{3/2} \ln \frac{4\alpha}{(1 + \sqrt{1 + 4\alpha})^2} \right],$$

$$\Pi_L^A(s) = 2s\alpha \left[ \frac{1}{\epsilon} - \rho_a + 2 + (1 + 4\alpha)^{1/2} \ln \frac{4\alpha}{(1 + \sqrt{1 + 4\alpha})^2} \right], \tag{B1}$$

and of course in this case $\Pi_L^V = 0$. For $m_b = 0$ one has

$$\Pi_T^{V,A}(s) = \frac{s}{6} \left[ (2 + 3\alpha) \left( \frac{1}{\epsilon} - \rho_a \right) + \frac{10}{3} + 2\alpha - \alpha^2 + (2 + 3\alpha - \alpha^3) \ln \frac{\alpha}{1 + \alpha} \right],$$

$$\Pi_L^{V,A}(s) = \frac{s}{2} \alpha \left[ \frac{1}{\epsilon} - \rho_a + 2 + \alpha + (1 + \alpha)^2 \ln \frac{\alpha}{1 + \alpha} \right]. \tag{B2}$$

At the **two-loop** level, with all masses defined on-shell, the asymptotic expressions allow several checks of the general formulae of Eq. (4.20), as the VPF's have been known for some time in particular cases. Note that here the color factor $N_c = 3$ is included, unlike in the one-loop expressions.

**Case $m_b = m_a$**

The real parts are given by

$$\Pi_T^V(s) = s \left\{ \frac{1}{2\epsilon} - \rho + \frac{55}{12} - \frac{26}{3}\alpha + \sqrt{1+4\alpha}(1-6\alpha)\ln x - \frac{2}{3}\alpha(4+\alpha)\ln^2 x \right.$$
$$+ \frac{2}{3}(4\alpha^2 - 1)\left[F(1) + F(x^2) - 2F(x)\right]$$
$$\left. - \frac{4}{3}(1-2\alpha)\sqrt{1+4\alpha}\left[G(x^2) - G(x)\right] \right\},$$

$$\Pi_T^A(s) = s \left\{ -\frac{6\alpha}{\epsilon^2} + (1 + 24\alpha\rho - 22\alpha)\frac{1}{2\epsilon} - (1 + 12\alpha\rho - 22\alpha)\rho + \frac{55}{12} - \frac{19}{6}\alpha \right.$$
$$+ 4\alpha^2 + (1 + 12\alpha + 4\alpha^2)\sqrt{1+4\alpha}\ln x + \frac{2}{3}\alpha(5 + 11\alpha + 6\alpha^2)\ln^2 x$$
$$- \alpha\pi^2 - \frac{2}{3}(1+2\alpha)(1+4\alpha)\left[F(1) + F(x^2) - 2F(x)\right]$$
$$\left. - \frac{4}{3}(1+4\alpha)^{3/2}\left[G(x^2) - G(x)\right] \right\}, \tag{B3}$$

$$\Pi_L^A(s) = 2s\alpha \left\{ -\frac{3}{\epsilon^2} + \left(6\rho - \frac{11}{2}\right)\frac{1}{\epsilon} + \frac{3}{4} - 6\alpha - \frac{\pi^2}{2} + 3(3-2\alpha)\sqrt{1+4\alpha}\ln x \right.$$
$$+ (11 - 6\rho)\rho + (3 + 4\alpha - 6\alpha^2)\ln^2 x$$
$$\left. - 2(1+2\alpha)\left[F(1) + F(x^2) - 2F(x)\right] - 4\sqrt{1+4\alpha}\left[G(x^2) - G(x)\right] \right\},$$

with $x = 4\alpha/(1 + \sqrt{1+4\alpha})^2$, and the functions $F(x)$ and $G(x)$ have been introduced in Chap. 4. The vector part of the longitudinal component vanishes in this case: this is expected to occur as a consequence of a QED-like Ward identity.

In the equal mass case, $\Pi_T^V(s)$ coincides, up to a color factor, with the irreducible part of the photon self-energy in QED which has been calculated in the fifties by the pioneers of Ref. [58]. $\Pi_T^{V,A}(s)$ has also been derived using a dispersive approach in [56], and more recently in [57]. In fact, we have expanded the expressions of the previous sections around $m_b = m_a$ and $m_b = 0$ retaining terms up to order $(m_b - m_a)^2/s$ and $m_b^2/s$, respectively. The rather lengthy expressions [136] will not be given here.

**Case $m_b = 0$**

For one massless quark, the coefficients of $\Pi_{L,T}^-(s)$ vanish, so that $\Pi_{L,T}^V(s) = \Pi_{L,T}^A(s) = s\,\Pi_{L,T}^+(s)$, and one obtains for the real parts

$$\Pi_T^{V,A}(s) = s \left\{ -\frac{3\alpha}{2\epsilon^2} + \left(1 - \frac{11}{2}\alpha + 6\alpha\rho\right)\frac{1}{2\epsilon} - \left(1 - \frac{11}{2}\alpha + 3\alpha\rho\right)\rho + \frac{55}{12} - \frac{71}{24}\alpha \right.$$
$$- \frac{\pi^2}{4}\alpha + (1+\alpha)\left(1 + \frac{3}{2}\alpha - \frac{5}{6}\alpha^2\right)\ln x + \frac{\alpha}{6}(1-3\alpha)(1+\alpha)\ln^2 x$$
$$\left. - \frac{5}{6}\alpha^2 + \frac{1}{3}(2+\alpha)(\alpha-1)G(x) + \frac{1}{3}(\alpha-2)(1+\alpha)^2\left[F(1) - F(x)\right] \right\},$$

$$\Pi_L^{V,A}(s) = s\alpha \left\{ -\frac{3}{2\epsilon^2} + \left( -\frac{11}{4} + 3\rho \right)\frac{1}{\epsilon} + \frac{3}{8} + \frac{7}{2}\alpha + \frac{11}{2}\rho - 3\rho^2 - \frac{\pi^2}{4} \right.$$

$$+ \frac{9}{2}(1+\alpha)^2 \ln x + (3+2\alpha)(1+\alpha)\frac{\ln^2 x}{2} + (1+\alpha)G(x)$$

$$\left. - (1+\alpha)^2 \left[ F(1) - F(x) \right] \right\}, \tag{B4}$$

with $x = \alpha/(1+\alpha)$. The expression of $\Pi_T^{V,A}(s)$ in this special case has been obtained in Ref. [57] by means of a dispersion integration; the connection with Kniehl's result is discussed in Sec. 4.4.

**Case $s \gg m_{a,b}^2$**

From Eqs. (4.20), in the limit $|s| \gg m_{a,b}^2$, one obtains for $\Pi_{T,L}^{V,A}(s)$

$$\Pi_T^{V,A}(|s| \to \infty) = s \left[ \frac{1}{2\epsilon} - \ln\frac{-s}{\mu^2} + \frac{55}{12} - 4\zeta(3) \right]$$

$$+ (m_a \mp m_b)^2 \left[ \frac{3}{2\epsilon^2} + \frac{1}{2\epsilon}\left( \frac{11}{2} - 3\rho_a - 3\rho_b \right) - \frac{11}{4}(\rho_a + \rho_b) - \frac{11}{8} \right.$$

$$\left. + \frac{3}{4}(\rho_a + \rho_b)^2 - \frac{9}{4}(\ln\alpha + \ln\beta) - \frac{3}{2}\ln\alpha\ln\beta + 6\zeta(3) + \frac{\pi^2}{4} \right]$$

$$+ (m_a^2 - m_b^2)\ln\frac{m_a^2}{m_b^2}\left[ -\frac{3}{2\epsilon} + \frac{3}{2}(\rho_a + \rho_b) - \frac{3}{4}(\ln\alpha + \ln\beta) + \frac{3}{4} \right]$$

$$+ 3(m_a^2 + m_b^2)(\ln\alpha + \ln\beta), \tag{B5}$$

$$\Pi_L^{V,A}(|s| \to \infty) = (m_a \mp m_b)^2 \left[ \frac{3}{2\epsilon^2} + \frac{1}{2\epsilon}\left( \frac{11}{2} - 3\rho_a - 3\rho_b \right) - \frac{11}{4}(\rho_a + \rho_b) - \frac{3}{8} \right.$$

$$\left. + \frac{3}{4}(\rho_a + \rho_b)^2 - \frac{9}{4}(\ln\alpha + \ln\beta) - \frac{3}{2}\ln\alpha\ln\beta + 6\zeta(3) + \frac{\pi^2}{4} \right] \tag{B6}$$

$$+ (m_a^2 - m_b^2)\ln\frac{m_a^2}{m_b^2}\left[ -\frac{3}{2\epsilon} + \frac{3}{2}(\rho_a + \rho_b) - \frac{3}{4}(\ln\alpha + \ln\beta) - 5 \right],$$

with the upper (lower) sign corresponding to vector (axial) current, and $\zeta(3) = 1.20206\ldots$ As expected, only the transverse part of the vacuum polarization function is quadratically divergent for $|q| \to \infty$. This divergent term, the expression of which is in agreement with the one obtained in Ref. [73], is the same for axial and axial-vector currents as expected from chiral symmetry. Moreover, as required by the Kinoshita-Lee-Nauenberg theorem [61,62], this term does not contain any mass singularity as $m_{a,b}$ tend to zero. As the vector part of the longitudinal component should vanish for $m_a = m_b$, it must be proportional to $(m_a - m_b)$ or $(m_a^2 - m_b^2)$, and since the axial-vector component can be obtained by changing the sign of one of the two masses, it must be proportional $(m_a + m_b)$ or $(m_a^2 - m_b^2)$ ($\ln m_a^2/m_b^2$ alone would have introduced mass singularities). This behavior is exhibited in Eq.(B6).

### B.1.2   Scalar self-energies

Referring to the results of Sec. 4.2, I begin giving the one-loop scalar VPF in the limit of transfer momentum much larger or much smaller than the quark masses; in this case the self-energies read

$$s\Pi^+(s \ll m_{u,d}^2) = -(m_u^2 + m_d^2)\left(\frac{1}{\epsilon} - \ln\frac{m_u m_d}{\mu^2} + 1\right) + \frac{1}{2}\frac{m_u^4 + m_d^4}{m_u^2 - m_d^2}\ln\frac{m_u^2}{m_d^2},$$

$$\Pi^-(s \ll m_{u,d}^2) = -\frac{1}{\epsilon} + \ln\frac{m_u m_d}{\mu^2} - 1 + \frac{1}{2}\frac{m_u^2 + m_d^2}{m_u^2 - m_d^2}\ln\frac{m_u^2}{m_d^2}, \tag{B7}$$

$$s\Pi^+(s \gg m_{u,d}^2) = \frac{1}{2\epsilon} - \frac{1}{2}\ln\frac{-s}{\mu^2} + 1. \tag{B8}$$

We can now consider the two-loop case, listing some limiting cases of the result of Eqs. (4.50).

**Equal masses**

In analogy with the one-loop case illustrated in Sec. 4.2.1, we can derive from Eqs. (4.20) the expressions of the self-energies of neutral scalar and pseudoscalar Higgs bosons, setting the quark masses to $m_q$, and using the appropriate couplings. Using $x = 4\alpha/(1 + \sqrt{1 + 4\alpha})^2$ with $\alpha = -m_q^2/s$,

$$\Pi_q^{S,A}(s) = \frac{G_F}{2\sqrt{2}\pi^2}\frac{\alpha_s}{\pi}s\,h_q^2\,(S_q, A_q) \tag{B9}$$

with $S_q = \Pi_q^+(s) - \alpha\Pi^-(s)$ and $A_q = \Pi_q^+(s) + \alpha\Pi^-(s)$ given by[1]

$$\begin{aligned}
S_q =& -\frac{3}{2\epsilon^2}(1 + 12\alpha) - \frac{1}{\epsilon}\left(\frac{11}{4} - 3\rho_a + 42\alpha - 36\alpha\rho_a\right) + \frac{11}{2}\rho_a - 3\rho_a^2 + 84\alpha\rho_a \\
& - 36\alpha\rho_a^2 + \frac{3}{8} - 73\alpha - \frac{\pi^2}{4}(1 + 12\alpha) + \frac{3}{2}(1 + 4\alpha)^{\frac{1}{2}}(14\alpha + 3)\ln x \\
& + (\frac{3}{2} + 14\alpha + 29\alpha^2)\ln^2 x - 2(1 + 4\alpha)^{\frac{3}{2}}\left[G(x^2) - G(x)\right] \\
& - (1 + 2\alpha)(1 + 4\alpha)\left[F(1) + F(x^2) - 2F(x)\right]; \\
A_q =& -\frac{3}{2\epsilon^2}(1 + 4\alpha) - \frac{1}{\epsilon}\left(\frac{11}{4} - 3\rho_a + 14\alpha - 12\alpha\rho_a\right) + \frac{11}{2}\rho_a - 3\rho_a^2 + 28\alpha\rho_a \\
& - 12\alpha\rho_a^2 + \frac{3}{8} - 33\alpha - \frac{\pi^2}{4}(1 + 4\alpha) + \frac{3}{2}(1 + 4\alpha)^{\frac{1}{2}}(3 - 2\alpha)\ln x \\
& + (\frac{3}{2} + 2\alpha - 3\alpha^2)\ln^2 x - 2(1 + 4\alpha)^{\frac{1}{2}}\left[G(x^2) - G(x)\right] \\
& - (1 + 2\alpha)\left[F(1) + F(x^2) - 2F(x)\right]. \tag{B10}
\end{aligned}$$

---

[1] The expression of $\Pi_q^s$ recently derived in Ref. [84] in the case of the SM Higgs boson is in agreement with our result.

**One massless quark**

In the limit where one of the quarks is massless, $m_d = 0$, the coefficients of $\Pi_d^+$ and $\Pi^-$ vanish while $\Pi_u^+$ takes the simpler form (with $x = \alpha/(1 + \alpha)$)

$$
\begin{aligned}
\Pi_u^+(s) = &-\frac{3}{2\epsilon^2}(1 + 4\alpha) - \frac{1}{\epsilon}\left[\frac{11}{4} + 14\alpha - 3\rho_a - 12\alpha\rho_a\right] + \frac{1}{2}\rho_a(11 + 56\alpha) \\
&- 3\rho_a^2(1 + 4\alpha) + \frac{3}{8} - \frac{53}{2}\alpha - \frac{\pi^2}{4}(1 + 4\alpha) + \frac{9}{2}(1 + \alpha)^2\ln x \\
&+ \frac{1}{2}(1 + \alpha)^2(3 + 2\alpha)\ln^2 x + (1 + \alpha)^2[F(x) - F(1)] + (1 + \alpha)G(x)
\end{aligned}
\tag{B11}
$$

As expected, in this limit the expression of $\Pi_u^+(s)$ is free of mass singularities.

**Case $s \gg m_{u,d}^2$**

When the masses are very small compared to the momentum transfer, the coefficient of $\Pi^-$ vanishes and $\Pi_q^+$ reads

$$
\begin{aligned}
s\Pi_q^+ = &-\frac{3}{2\epsilon^2} + \frac{1}{\epsilon}\left(3\ln\frac{m_q^2}{\mu^2} - \frac{11}{4}\right) + \frac{3}{8} - \frac{\pi^2}{4} - 6\zeta(3) \\
&+ 10\ln\frac{m_q^2}{\mu^2} - 3\ln^2\frac{m_q^2}{\mu^2} - \frac{9}{2}\ln\frac{-s}{\mu^2} + \frac{3}{2}\ln^2\frac{m_q^2}{-s}
\end{aligned}
\tag{B12}
$$

**Case $s \ll m_{u,d}^2$**

Finally, in the limit where the momentum squared is much smaller than the internal quark masses squared, the components $\Pi_q^+$ and $\Pi^-$ read

$$
\begin{aligned}
s\Pi_q^+ = &\frac{6}{\epsilon^2}m_+ + \frac{1}{\epsilon}\left(14m_+ - 3m_+\rho_- - 3m_-\rho_- - 6m_+\rho_+\right) - \frac{3}{4}\frac{m_+^3}{m_-^2}\rho_-^2 \\
&+ \frac{3}{4}\frac{m_+^2}{m_-}\rho_-^2 + m_+\left(3\rho_+^2 - 14\rho_+ - 7\rho_- + 3\rho_+\rho_- + \frac{9}{4}\rho_-^2 + 30\right) \\
&+ \pi^2 m_+ + m_-\left(\frac{3}{4}\rho_-^2 - 7\rho_- + 3\rho_+\rho_-\right); \\
\Pi^- = &\frac{6}{\epsilon^2} + \frac{1}{\epsilon}(14 - 6\rho_+) - \frac{3}{2}\frac{m_+^2}{m_-^2}\rho_-^2 + 30 - 14\rho_+ + 3\rho_+^2 + \pi^2 + \frac{3}{2}\rho_-^2,
\end{aligned}
\tag{B13}
$$

with $\rho_\pm = \ln\frac{m_u^2}{\mu^2} \pm \ln\frac{m_d^2}{\mu^2}$ and $m_\pm = m_u^2 \pm m_d^2$.

## B.2   Imaginary parts and partial widths

In all the expressions for the VPF's the momentum transfer has been defined to be in the space-like region, $q^2 < 0$. When continued to the physical region above the threshold for the production of two fermions, $s \geq (m_a + m_b)^2$, the self-energies acquire imaginary parts, related to the decay widths of the bosons into fermions, at this order in perturbation theory, by

$$\mathcal{I}m\Pi(q^2) = M \; \Gamma(q^2) \qquad (B14)$$

where $M$ is the mass of the boson[2].

Adding a small imaginary part $-i\epsilon$ to the fermion masses squared, the analytical continuation is consistently defined. From the general expressions of Eqs. (4.6,4.20), the imaginary parts can be straightforwardly obtained using

$$\mathcal{I}m \ln x_{a,b} = \pi \qquad , \qquad \mathcal{I}m \ln^2 x_{a,b} = 2\pi \ln|x_{a,b}|.$$

In particular, for some of the integrals present in the two-loop VPF, we obtain[3]

$$\begin{aligned}
\mathcal{J} = \frac{1}{\pi}\mathcal{I}m\mathcal{I} = &-2\left[4\text{Li}_2(x_a x_b) - 2\text{Li}_2(x_a) - 2\text{Li}_2(x_b) + 2\ln|x_a x_b|\right. \\
&\times \left.\ln(1 - x_a x_b) - \ln|x_a|\ln(1 - x_a) - \ln|x_b|\ln(1 - x_b)\right],
\end{aligned} \qquad (B15)$$

$$\begin{aligned}
\mathcal{J}' = \frac{1}{\pi}\mathcal{I}m\mathcal{I}' = &\ln|x_a x_b|\left(\frac{3}{2} + \alpha + \beta - \frac{3}{2}\lambda^{\frac{1}{2}}\right) + 4\lambda^{\frac{1}{2}}\ln(1 - x_a x_b) \\
&+ \frac{\beta - \alpha}{2}\ln\frac{\alpha}{\beta} - (\beta - \alpha + \lambda^{\frac{1}{2}})\ln(1 - x_a) - (\alpha - \beta + \lambda^{\frac{1}{2}})\ln(1 - x_b).
\end{aligned} \qquad (B16)$$

### B.2.1 Vector boson self-energies

From the complete one-loop expressions of Eqs. (4.6) we have the following imaginary parts

$$\begin{aligned}
\mathcal{I}m\Pi_T^+(s) &= \frac{\pi}{2}\lambda^{\frac{1}{2}}\left[(1 + \alpha + \beta) - \frac{1}{3}\lambda\right], & \mathcal{I}m\Pi_T^-(s) &= \pi\lambda^{\frac{1}{2}} \\
\mathcal{I}m\Pi_L^+(s) &= \frac{\pi}{2}\lambda^{\frac{1}{2}}\left[\lambda - 1 - \alpha - \beta\right], & \mathcal{I}m\Pi_L^-(s) &= \pi\lambda^{\frac{1}{2}}
\end{aligned} \qquad (B17)$$

or, equivalently,

$$\begin{aligned}
\mathcal{I}m\Pi_T^{V,A}(s) &= \frac{\pi}{2}s\lambda^{\frac{1}{2}}\left[(1 + \alpha + \beta) - \frac{1}{3}\lambda \; \pm \; 2\frac{m_a m_b}{s}\right], \\
\mathcal{I}m\Pi_L^{V,A}(s) &= \frac{\pi}{2}s\lambda^{\frac{1}{2}}\left[\lambda - 1 - \alpha - \beta \; \pm \; 2\frac{m_a m_b}{s}\right].
\end{aligned} \qquad (B18)$$

From the complete two-loop expressions of Eqs. (4.20) we obtain for $\mathcal{I}m\Pi_{T,L}^{\pm}$

$$\frac{1}{\pi}\mathcal{I}m\Pi_T^+ = \frac{1}{6}(11 + 19\alpha + 19\beta + 12\alpha\beta - 5\lambda)\lambda^{\frac{1}{2}} + \frac{4}{3}(\alpha - \beta)\ln\frac{(1 - x_b)}{(1 - x_a)}$$

---

[2]At higher orders in perturbation theory, and for $q^2 = M^2$, one should take into account the wave function renormalization of the decaying particle, including a factor $1 - \Pi'(M^2)$ in the denominator of the r.h.s. of Eq.(B14)

[3]There are small differences with the formulae of Ref. [59] and [74] due to trivial algebraic simplification, and to the correction of a couple of misprints.

$$+ \frac{\ln |x_a x_b|}{6} \left[ 3(1 + \alpha + \beta)(1 + \alpha + \beta + 2\alpha\beta - \lambda) + 4\alpha + 4\beta + 26\alpha\beta \right]$$

$$+ \left( \frac{2}{3} - \frac{1 + \alpha + \beta}{2} \lambda^{\frac{1}{2}} \right) (\alpha - \beta) \ln \frac{\alpha}{\beta}$$

$$+ \frac{1}{3} \left[ (\alpha + \beta - 2)\lambda - 12\alpha\beta \right] \mathcal{J} - \frac{2}{3} \left[ 3(1 + \alpha + \beta) - \lambda \right] \mathcal{J}';$$

$$\frac{1}{\pi} \mathcal{I}m \Pi_T^- = (9 + \alpha + \beta)\lambda^{\frac{1}{2}} + \left[ 4(1 + \alpha + \beta) + 2\alpha\beta - \lambda \right] \ln |x_a x_b|$$

$$+ (\alpha - \beta)\lambda^{\frac{1}{2}} \ln \frac{\beta}{\alpha} - 2(1 + \alpha + \beta)\mathcal{J} - 4\mathcal{J}';$$

$$\frac{1}{\pi} \mathcal{I}m \Pi_L^+ = \frac{1}{2}(9\lambda - 9 - 9\alpha - 9\beta - 12\alpha\beta)\lambda^{\frac{1}{2}} - \left[ (\alpha + \beta)\lambda - 4\alpha\beta \right] \mathcal{J}$$

$$+ \frac{1}{2} \left[ (2\lambda - 2 - \alpha - \beta)\lambda - \alpha\beta(6\alpha + 6\beta + 16\alpha\beta) \right] \ln |x_a x_b|$$

$$+ \frac{1}{2}(2\lambda + 1 + \alpha + \beta)\lambda^{\frac{1}{2}}(\beta - \alpha) \ln \frac{\beta}{\alpha} + 2[1 + \alpha + \beta - \lambda]\mathcal{J}';$$

$$\frac{1}{\pi} \mathcal{I}m \Pi_L^- = 3(3 - \alpha - \beta)\lambda^{\frac{1}{2}} + \left[ 4(1 + \alpha + \beta) - 6\alpha\beta - \lambda \right] \ln |x_a x_b|$$

$$+ (\alpha - \beta)\lambda^{\frac{1}{2}} \ln \frac{\beta}{\alpha} - 2(1 + \alpha + \beta)\mathcal{J} - 4\mathcal{J}'. \tag{B19}$$

**Equal masses**

$$\mathcal{I}m \Pi_T^V(s) = \frac{2\pi}{3} s \left\{ \frac{3}{2} \lambda^{\frac{1}{2}}(1 - 6\alpha) - 2\alpha(4 + \alpha) \ln |x| - (1 - 4\alpha^2)J \right.$$

$$\left. - 2(1 - 2\alpha)J' \right\},$$

$$\mathcal{I}m \Pi_T^A(s) = \frac{2\pi}{3} s \left\{ \frac{3}{2} \lambda^{\frac{1}{2}}(1 + 12\alpha + 4\alpha^2) + 2\alpha(5 + 11\alpha + 6\alpha^2) \ln |x| \right.$$

$$\left. - (1 + 2\alpha)\lambda J - 2\lambda J' \right\}, \tag{B20}$$

$$\mathcal{I}m \Pi_L^A(s) = 4\pi s \alpha \left\{ \frac{3}{2} \lambda^{\frac{1}{2}}(3 - 2\alpha) + (3 + 4\alpha - 6\alpha^2) \ln |x| - (1 + 2\alpha)J - 2J' \right\},$$

where

$$J = -4 \left[ 2\text{Li}_2(x^2) - 2\text{Li}_2(x) + 2 \ln |x| \ln(1 - x^2) - \ln |x| \ln(1 - x) \right],$$

$$J' = (3 + 4\alpha - 3\lambda^{\frac{1}{2}}) \ln |x| + 4\lambda^{\frac{1}{2}} \ln(1 - x^2) - 2\lambda^{\frac{1}{2}} \ln(1 - x), \tag{B21}$$

with $\lambda^{\frac{1}{2}} = \sqrt{1 + 4\alpha}$.

**One massless quark**

$$\mathcal{I}m \Pi_T^{V,A}(s) = \pi s \left\{ 1 + \frac{5}{2}\alpha + \frac{2}{3}\alpha^2 - \frac{5}{6}\alpha^3 - \frac{1}{3}(1 + \alpha)^2(4 - 5\alpha) \ln(1 + \alpha) \right.$$

$$\left. - \frac{\alpha}{3}(5\alpha^2 + 4\alpha - 5) \ln(-\alpha) \right.$$

$$- \frac{2}{3}(2-\alpha)(1+\alpha)^2 \left[ 2\mathrm{Li}_2 \left( \frac{\alpha}{\alpha+1} \right) - \ln(1+\alpha) \ln \frac{-\alpha}{1+\alpha} \right] \Big\},$$

$$\mathcal{I}m\Pi_L^{V,A}(s) = -\pi s\alpha(1+\alpha) \Big\{ (1+\alpha) \left[ -\frac{9}{2} + (5+2\alpha)\ln(1+\alpha) \right]$$
$$- (3+7\alpha+2\alpha^2)\ln(-\alpha)$$
$$+ 2(1+\alpha) \left[ 2\mathrm{Li}_2 \left( \frac{\alpha}{\alpha+1} \right) - \ln(1+\alpha) \ln \frac{-\alpha}{1+\alpha} \right] \Big\}. \tag{B22}$$

Close forms for $\mathcal{I}m\Pi_{T,L}^{V,A}(s)$ in the general case $m_a \neq m_b \neq 0$ have been also derived in the past by a number of authors [56, 68, 69] (in the first reference only the transverse part is given) by directly calculating the QCD corrections to the flavor changing decay of a vector boson. The results that we obtain here using a completely different method agree with those of Ref. [56, 68] and also with Ref. [69] once some obvious mistakes in the integrals of their Appendix ($J_1$ and $J_2$) are corrected; see also Ref. [70].

### B.2.2   Scalar self-energies

We now consider the partial decay widths of the various Higgs bosons into quark-antiquark pairs. At $\mathcal{O}(G_\mu \alpha_s)$, the partial decay width of a charged Higgs boson into $u\bar{d}$ pairs is given by ($s = M_{H^+}^2$)

$$\Gamma(H^{+-} \to u\bar{d}) = \frac{G_\mu \alpha_s M_{H^+}}{2\sqrt{2}\pi^3} \left[ h_u^2 \mathcal{I}m\Pi_u^+(s) + h_d^2 \mathcal{I}m\Pi_D^+(s) + 2h_u h_d \frac{m_u m_d}{s} \mathcal{I}m\Pi^-(s) \right].$$

The imaginary parts of $\Pi_{u,d}^+$ and $\Pi^-$ can be derived along the same lines as discussed previously, and one obtains for $\mathcal{I}m\Pi^\pm$ in the case of arbitrary masses

$$\frac{1}{\pi}\mathcal{I}m\Pi_u^+(s) = \left[ \frac{3}{2}(1+\alpha+\beta)\lambda^{\frac{1}{2}} + \left( \frac{3}{2}+\alpha+\beta \right)\lambda + 5\alpha\beta \right] (\ln|x_a x_b|$$
$$+ (1+\alpha+\beta)(\alpha-\beta)\lambda^{\frac{1}{2}} + \ln\frac{\alpha}{\beta}$$
$$+ \frac{9}{2}(1+\alpha+\beta)\lambda^{\frac{1}{2}} - (1+\alpha+\beta)^2\mathcal{J} - 2(1+\alpha+\beta)\mathcal{J}';$$

$$\mathcal{I}m\Pi_d^+(s) = \mathcal{I}m\Pi_u^+(s) \ [\ m_u \leftrightarrow m_d\ ]; \tag{B23}$$

$$\frac{1}{\pi}\mathcal{I}m\Pi^-(s) = -2 \left[ \lambda + 2(1+\alpha+\beta) \right] \ln|x_a x_b|$$
$$- 12\lambda^{\frac{1}{2}} - 2(\alpha-\beta)\lambda^{\frac{1}{2}} \ln\frac{\alpha}{\beta} + 2(1+\alpha+\beta)\mathcal{J} + 4\mathcal{J}'.$$

From this formulae one can derive again the expressions of the hadronic decay widths in the previous special situations of physical relevance. In the limit where one of the quark is massless, $m_d \to 0$, one has for $\mathcal{I}m\Pi_u^+(s)$

$$C_q = \frac{1}{\pi}\mathcal{I}m\Pi_u^+(s) = \frac{9}{2}(1+\alpha)^2 + (1+\alpha)(3+7\alpha+2\alpha^2)\ln\frac{-\alpha}{1+\alpha} - 2(1+\alpha)^2$$
$$\times \left[ \frac{\ln(1+\alpha)}{1+\alpha} + 2\mathrm{Li}_2 \left( \frac{\alpha}{\alpha+1} \right) - \ln(1+\alpha)\ln\frac{-\alpha}{1+\alpha} \right] \tag{B24}$$

In the case of scalar and pseudoscalar neutral Higgs bosons, the partial decay widths $\Gamma(S, A \to q\bar{q})$ are given by

$$\Gamma[(S, A) \to q\bar{q}] = \frac{G_F}{2\sqrt{2}\pi^2} \frac{\alpha_s}{\pi} h_q^2 M_{S,A} \, \mathcal{I}m(S_q, A_q) \tag{B25}$$

where $\mathcal{I}m(S_q, A_q) = \mathcal{I}m\Pi_q^+(s, \beta = \alpha) \mp \alpha \mathcal{I}m\Pi^-(s, \beta = \alpha)$ are given by

$$\frac{1}{\pi}\mathcal{I}mS_q = \frac{3}{2}(1 + 4\alpha)^{\frac{1}{2}}(14\alpha + 3) + (58\alpha^2 + 28\alpha + 3)\ln|x| - 4(1 + 4\alpha)\mathcal{L}$$
$$\frac{1}{\pi}\mathcal{I}mA_q = \frac{3}{2}(1 + 4\alpha)^{\frac{1}{2}}(3 - 2\alpha) + (3 + 4\alpha - 6\alpha^2)\ln|x| - 4\mathcal{L} \tag{B26}$$

where

$$\mathcal{L} = \frac{1}{2}J' + \frac{1}{4}(1 + 2\alpha)J. \tag{B27}$$

Finally, we note that in the limit where the quark masses are much smaller than the Higgs boson masses, the QCD corrections to the decay widths will exhibit the well known logarithmic behavior [78, 79] which, because of chiral symmetry, is the same for the scalar, pseudoscalar and charged Higgs boson

$$\frac{1}{\pi}\mathcal{I}m\Pi_q \to \frac{9}{2} + 3\ln\frac{m_q^2}{M_\Phi^2}. \tag{B28}$$

The large logarithm is responsible, for example, of the very large QCD corrections to the decay width of the SM Higgs boson into $b\bar{b}$, which are of the order of 35% for a Higgs mass of only $\approx 100\,\text{GeV}$. However, it is possible to resum the large logarithmic terms: this is equivalent to replacing the on-shell quark masses by running masses defined at $M_\Phi^2$ in the first order amplitude [79]. We can see that explicitly, in the case the $\overline{\text{MS}}$ mass $\hat{m}_q(M_\Phi)$ is used. The relation between on-shell and $\overline{\text{MS}}$ mass is shown in Eq.(4.13), and we can write the partial width of the scalar $\Phi$ in the limit of vanishing quark masses as

$$\Gamma = \Gamma_0 \left[1 + \frac{2}{3} \frac{\alpha_s}{\pi} \left(\frac{9}{2} + 3\ln\frac{m_q^2}{M_\Phi^2}\right)\right], \tag{B29a}$$

with

$$\Gamma_0 = \frac{3G_\mu M_\Phi}{4\sqrt{2}\pi} h_q^2. \tag{B29b}$$

As $h_q$ is proportional to $m_q$, shifting the mass according to Eq.(4.13) cancels the logarithm in the r.h.s. of Eq.(B29a).

# Appendix C

# Ward identities for electroweak VPF's

In this appendix I report some Ward identities for the two-loop self-energy diagrams that contribute to $\mathcal{O}(G_\mu^2 m_t^2 M_Z^2)$. They were derived by the method described in Sec. 5.3 and have been used in the calculation of Sec. 5.2. We can write the $W$ and $Z$ self-energies at $q^2 = 0$ as the sum of different contributions:

$$A(0) = A^{lead}(0) + A^H(0) + A^{noH}(0) \tag{C1}$$

where the $A^{lead}(0)$'s are given in Eqs. (5.39), and represent the set of diagrams of Fig. 5.2, i.e. diagrams contributing to $\mathcal{O}(G_\mu^2 m_t^4)$. $A^H$ and $A^{noH}$ represent all the other contributions of Feynman diagrams involving the top quark. Some of the graphs contributing to them can be obtained from Fig. 5.2 replacing the scalar lines with vector boson lines, the others are displayed in Fig. C.1. The diagrams involving fermionic insertions on a scalar propagator (Fig. C.1b) formally contribute to leading $\mathcal{O}(G_\mu^2 m_t^4)$. However, when the appropriate mass counterterms of the physical and unphysical scalars are included, they are effectively of $\mathcal{O}(G_\mu^2 m_t^2 M_Z^2)$.

## C.1   Subleading with Higgs

I first consider only the irreducible two-point diagrams that contain the Higgs boson and the top quark, but do not contribute to $\mathcal{O}(G_\mu^2 m_t^4)$. They are depicted in Fig. C.1, where the dashed line represents a Higgs boson. The short-hand notation $\langle \ldots \rangle$ stands for $\langle 0|T^* \ldots |0\rangle$, while $\int_k \equiv \mu^{4-n} \int \frac{d^n k}{(2\pi)^n}$ and $\int_x \equiv \int d^n x$. We have explicitly checked these Ward identities as part of the calculation in Ref. [109].

$$A_{ZZ}^H(0) =$$

$$\frac{1}{n} \frac{\hat{g}^4}{2\hat{c}^4} \frac{\partial}{\partial q^\alpha} \left\{ \int_k \int_x \int_y \frac{e^{ik\cdot x - iq\cdot y}}{(k^2 - M_Z^2)[(k-q)^2 - M_H^2]} \langle\, S_2(y) J_Z^\alpha(x) S_1(0) \,\rangle \right\}_{q^2=0}$$

$$+ \frac{\hat{g}^4}{4\hat{c}^4} \int_k \int_x \frac{e^{ik\cdot x}}{(k^2 - M_Z^2)(k^2 - M_H^2)^2} \left\langle S_1(x) S_1(0) \right\rangle$$

$$- \frac{\hat{g}^4 M_W^2}{n\,\hat{c}^6} \int_k \int_x e^{ik\cdot x} \frac{\left\langle J_Z^\alpha(x) J_{Z\,\alpha}(0) \right\rangle}{\left(k^2 - M_Z^2\right)^2 \left(k^2 - M_H^2\right)}$$

$$- (M_H^2 - M_Z^2)\,\frac{1}{n}\frac{\hat{g}^4}{2\hat{c}^4} \int_k \frac{\int_x e^{ik\cdot x}\langle S_2(x) S_2(0) \rangle - i\,\langle S_1(0) \rangle}{\left(k^2 - M_Z^2\right)^2 \left(k^2 - M_H^2\right)^2} \tag{C2}$$

$$A_{WW}^H(0) =$$

$$- \frac{i}{n}\frac{\hat{g}^4}{4}\frac{\partial}{\partial q^\alpha} \int_k \int_x \int_y e^{-iq\cdot y + ik\cdot x} \frac{\left\langle \left( S^\dagger(y) J_W^\alpha(x) - \text{h.c.} \right) S_1(0) \right\rangle}{(k^2 - M_W^2)\left[(k-q)^2 - M_H^2\right]} \Bigg|_{q^2=0}$$

$$+ \frac{\hat{g}^4}{4} \int_k \int_x \frac{e^{ik\cdot x}\langle S_1(x) S_1(0)\rangle}{(k^2 - M_W^2)(k^2 - M_H^2)} - \frac{\hat{g}^4 M_W^2}{2\,n} \int_k \int_x \frac{e^{ik\cdot x}\langle J_W^\alpha(x) J_W^{\dagger\alpha}(0)\rangle}{(k^2 - M_W^2)^2 (k^2 - M_H^2)}$$

$$- \frac{\hat{g}^4}{n}(M_H^2 - M_W^2) \int_k \frac{\int_x e^{ik\cdot x}\langle S^\dagger(x) S(0)\rangle - \frac{i}{2}\langle S_1(0)\rangle}{\left(k^2 - M_W^2\right)^2 (k^2 - M_H^2)^2}. \tag{C3}$$

## C.2 Subleading without Higgs

Here I consider the two-loop diagrams involving the top quark but not the Higgs boson that contribute to $\mathcal{O}(G_\mu^2 m_t^2 M_Z^2)$. Some infrared divergences (that cancel in the sum) have been regulated by a photon mass $\lambda$. Note that the last two lines of Eq.(C5), corresponding to the diagrams in Fig.C.1e and to some residual terms coming from the four- and three-point functions, do not contribute to $\mathcal{O}(G_\mu^2 m_t^2 M_Z^2)$, as can be easily verified.

$$A_{ZZ}^{noH}(0) = -\frac{\hat{g}^4}{8\hat{c}^2}\frac{\partial}{\partial q^2} \int_k \int_{x_1} \int_{x_2} \int_y e^{iqy + ik(x_1 - x_2)} \times$$

$$\times \left\langle S_2(y) S_2(0) \left[ \frac{J_W^{\dagger\lambda}(x_1) J_{W\lambda}(x_2)}{k^2 - M_W^2} + \frac{J_Z^\lambda(x_1) J_{Z\lambda}(x_2)}{\hat{c}^2(k^2 - M_Z^2)} + \frac{\hat{s}^2 J_\gamma^\lambda(x_1) J_{\gamma\lambda}(x_2)}{k^2} \right] \right\rangle \Bigg|_{q^2=0}$$

$$+ \frac{\hat{g}^4}{2n\,\hat{c}^2} \frac{\partial}{\partial q_\alpha} \int_k \int_{x,y} e^{iqy - ikx} \frac{\left\langle S_2(y) \left( S^\dagger(x) J_W^\alpha(0) + \text{h.c.} \right) \right\rangle}{(k^2 - M_W^2)((k-q)^2 - M_W^2)} \Bigg|_{q^2=0}$$

$$+ M_W^2 \hat{g}^4 \frac{\hat{c}^2 - \hat{s}^2}{n\,\hat{c}^2} \int_k \int_x \frac{e^{ikx}}{(k^2 - M_W^2)^3}\langle J_W^{\dagger\alpha}(x) J_{W\alpha}(0)\rangle + \frac{4}{n}\hat{g}^4 \int_{k,x} \frac{\langle J_Z^\alpha(x) J_{\alpha Z}(0)\rangle}{(k^2 - M_W^2)^2}$$

$$+ \frac{\hat{g}^4}{\hat{c}^2} \int_{k,x} \frac{e^{ikx}\langle S^\dagger(x) S(0)\rangle}{(k^2 - M_W^2)^3} - \frac{i}{2}(\hat{c}^2 + 2\hat{s}^2) \int_k \frac{\langle S_1(0)\rangle}{(k^2 - M_W^2)^3}. \tag{C4}$$

$$A_{WW}^{noH}(0) = -\frac{\hat{g}^4}{4}\frac{\partial}{\partial q^2} \int_k \int_{x_1} \int_{x_2} \int_y e^{iqy + ik(x_1 - x_2)} \times$$

$$\times \left\langle S(y) S^\dagger(0) \left[ \frac{J_W^{\dagger\lambda}(x_1) J_{W\lambda}(x_2)}{k^2 - M_W^2} + \frac{J_Z^\lambda(x_1) J_{Z\lambda}(x_2)}{\hat{c}^2(k^2 - M_Z^2)} + \frac{\hat{s}^2 J_\gamma^\lambda(x_1) J_{\gamma\lambda}(x_2)}{k^2} \right] \right\rangle \Bigg|_{q^2=0}$$

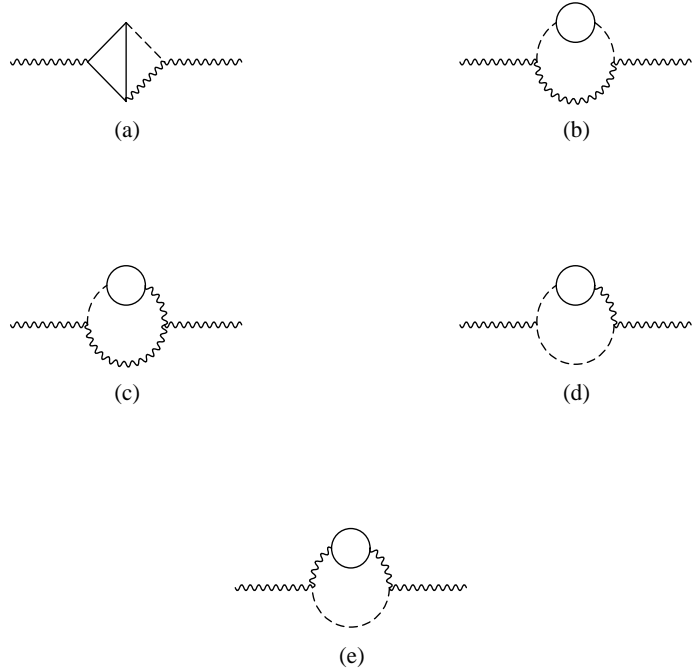

Figure C.1: Some of the self-energy diagrams contributing to $\mathcal{O}(G_\mu^2 m_t^2 M_Z^2)$. Solid lines represent fermions, dashed lines scalars, and wavy lines vector bosons (see text).

$$+ \frac{\hat{g}^4}{4} \frac{\partial}{\partial q^2} \int_{k,x,y} \frac{e^{-iqy+ikx}}{(k^2 - M_Z^2)[(k-q)^2 - M_W^2]} \, q_\alpha \, \langle \left( S^\dagger(y) J_W^\alpha(0) + \text{h.c.} \right) S_2(x) \rangle \Big|_{q^2=0}$$

$$- i\frac{\hat{g}^4}{2} \frac{\partial}{\partial q^2} \int_{k,x,y} \frac{e^{-iqy+ikx}}{[(k-q)^2 - M_W^2]} q_\alpha \langle \left( S^\dagger(y) S(0) - \text{h.c.} \right) \left[ \frac{(\hat{c}^2 - \hat{s}^2) J_Z^\alpha(x)}{\hat{c}^2(k^2 - M_Z^2)} + \frac{2\hat{s}^2 J_\gamma^\alpha(x)}{k^2} \right] \rangle \Big|_{q^2=0}$$

$$+ i\frac{\hat{g}^4}{2} \frac{\partial}{\partial q^2} \int_k \int_x \int_y \frac{e^{ik\cdot x - iq\cdot y}}{[(k-q)^2 - M_W^2]} \langle \left( S^\dagger(y) J_W^\alpha(0) + \text{ h.c.} \right) \times$$
$$\times \left[ \frac{(M_Z^2 - M_W^2)}{(k^2 - M_Z^2)} J_{Z\alpha}(x) - \frac{M_W^2 \hat{s}^2}{k^2} J_{\gamma\alpha}(x) \right] \rangle \Big|_{q^2=0}$$

$$+ 2\frac{\hat{g}^4}{n} \int_x \langle J_W^{\dagger\alpha}(x) J_{W\alpha}(0) \rangle \int_k \frac{1}{(k^2 - M_W^2)} \left( \frac{\hat{c}^2}{(k^2 - M_Z^2)} + \frac{\hat{s}^2}{k^2} \right)$$

$$+ \frac{\hat{g}^4}{2n} \int_k \int_x e^{ik\cdot x} \langle J_W^{\dagger\alpha}(x) J_{W\alpha}(0) \rangle \left[ \hat{s}^2 \left( \frac{(4-n)k^2 + n\lambda^2}{(k^2 - \lambda^2)^3} + \frac{2(n-5)}{k^2(k^2 - M_W^2)} \right. \right.$$
$$\left. + \frac{6-n}{(k^2 - M_W^2)^2} \right) + \hat{c}^2 \left( \frac{4M_Z^2(M_Z^2 - M_W^2)^2}{(k^2 - M_Z^2)^3(k^2 - M_W^2)^2} + \frac{6-n}{(k^2 - M_W^2)^2} + \frac{4-n}{(k^2 - M_Z^2)^2} \right.$$
$$\left. \left. + \frac{2(n-5)}{(k^2 - M_Z^2)(k^2 - M_W^2)} + \frac{M_Z^2}{(k^2 - M_Z^2)(k^2 - M_W^2)^2} \right) \right]$$

$$+ \frac{\hat{g}^4}{n} \int_k \int_x \frac{e^{ik\cdot x} \langle J_Z^\alpha(x) J_{Z\alpha}(0) \rangle}{\left(k^2 - M_Z^2\right)^2 \left(k^2 - M_W^2\right)} \times$$

$$\times \left[ \frac{(M_Z^2 - M_W^2)^2 (nM_W^2 + (4-n)k^2)}{\left(k^2 - M_W^2\right)^2} + 2M_Z^2 + M_W^2 \frac{\hat{c}^2 - \hat{s}^2 - 2\hat{c}^4}{\hat{c}^4} \right]$$

$$+ \frac{\hat{g}^4}{2n} \int_k \frac{\int_x e^{ik\cdot x} \langle S_2(x) S_2(0) \rangle - i \langle S_1(0) \rangle}{(k^2 - M_W^2)(k^2 - M_Z^2)} \left[ \frac{(n-6)\hat{c}^2 + 2\hat{s}^2}{2\hat{c}^2 \left(k^2 - M_Z^2\right)} + \frac{4\hat{c}^2 - 1}{\hat{c}^2 \left(k^2 - M_W^2\right)} \right]$$

$$+ \frac{\hat{g}^4}{n} \int_k \frac{\int_x e^{ik\cdot x} \langle S^\dagger(x) S(0) \rangle - \frac{i}{2} \langle S_1(0) \rangle}{(k^2 - M_W^2)} \left[ \frac{nk^2 + (6-n)M_Z^2 - 6M_W^2}{2\left(k^2 - M_Z^2\right)^2 \left(k^2 - M_W^2\right)} \right.$$

$$\left. + \hat{s}^2 \left( \frac{4}{\left(k^2 - \lambda^2\right)^2} - \frac{4}{\left(k^2 - M_Z^2\right)^2} - \frac{M_Z^2(3n-8)}{2k^2(k^2 - M_Z^2)(k^2 - M_W^2)} \right) \right]$$

$$+ \frac{\hat{g}^4 \hat{s}^2}{2} \int_k \int_x \frac{e^{ik\cdot x} \langle S^\dagger(x) S(0) \rangle}{\left(k^2 - M_W^2\right)^2} \left( \frac{1}{k^2} + \frac{\hat{s}^2}{\hat{c}^2 \left(k^2 - M_Z^2\right)} \right)$$

$$- 2\hat{s}^2 \frac{\hat{g}^4}{n} (M_Z^2 - M_W^2) \int_k \int_x \frac{e^{ik\cdot x} \langle J_{Z\alpha}(x) J_\gamma^\alpha(0) \rangle}{(k^2 - M_Z^2)(k^2 - M_W^2)} \left[ \frac{n-1}{k^2} - \frac{n}{(k^2 - M_W^2)} \right]$$

$$- M_W^2 \frac{2n+3}{n} \hat{s}^4 \hat{g}^4 \int_k \int_x \frac{e^{ik\cdot x} \langle J_{\gamma\alpha}(x) J_\gamma^\alpha(0) \rangle}{k^2 (k^2 - M_W^2)^2}. \tag{C5}$$